\documentclass[preprint]{aastex}

\newcommand{\kms}{km~s$^{-1}$}

\newcommand{\subsun}{\mbox{$_{\odot}$}}
\newcommand{\teff}{$T_{eff}$}
\newcommand{\grav}{log($g$)}
\newcommand{\etal}{{\it et al.\/}}
\newcommand{\eqw}{$W_{\lambda}$}
\newcommand{\fe}{[Fe/H]}

\begin{document}

\title{Abundances in a Large Sample of Stars in M3 and M13\altaffilmark{1}}

\author{Judith G. Cohen\altaffilmark{2} and Jorge Melendez\altaffilmark{2}}

\altaffiltext{1}{Based in part on observations obtained at the
W.M. Keck Observatory, which is operated jointly by the California 
Institute of Technology, the University of California, and the
National Aeronautics and Space Administration.}

\altaffiltext{2}{Palomar Observatory, Mail Stop 105-24,
California Institute of Technology, Pasadena, Ca., 91125, jlc(jorge)@astro.caltech.edu}

\begin{abstract}

We have carried out a detailed abundance analysis for 21 elements 
in a sample of 25 stars with a wide range in 
luminosity from luminous giants to stars near the main sequence turnoff
in the globular cluster M13 ([Fe/H] $-1.50$ dex)
and in a sample of 13 stars distributed 
from the tip to the base of the RGB in the globular cluster M3
([Fe/H] $-1.39$ dex).
The analyzed spectra, obtained with HIRES at the Keck Observatory,
are of high dispersion (R=$\lambda / \Delta \lambda$=35,000).
Most elements, including Fe, show no trend with \teff, 
and low scatter around the mean between the
top of the RGB and near the main sequence turnoff, 
suggesting that 
at this metallicity, non-LTE effects and
gravitationally induced heavy element diffusion
are not important for this set of elements over the range
of stellar parameters spanned by our sample.

We have detected an anti-correlation between O and Na abundances, 
observed previously
among the most luminous RGB stars in both of these clusters, 
in both M3 and in M13 over the full range of luminosity of our
samples, i.e. in the case of M13
to near the main sequence turnoff.  M13 shows a larger range
in both O and Na abundance than does M3 at all luminosities, 
in particular having a few stars at 
its RGB tip with unusually strongly depleted O.

We detect a correlation between Mg abundance and O abundance among the
stars in the M13 sample. We also find a decrease in the mean Mg abundance
as one moves towards lower luminosity, which we tentatively suggest is due
to our ignoring non-LTE effects in Mg.

Although CN burning must be occurring in both M3 and in M13, and ON
burning is required for M13,
we combine our new O abundances with published C and N abundances 
to confirm with quite high precision that 
the sum of C+N+O is constant near the tip of the giant
branch, and we extend this down to
the bump in the luminosity function.
The same holds true for a smaller
sample in M3, with somewhat larger variance.

Star I--5 in M13 has large excesses of Y and of Ba, with no strong
enhancement of Eu, suggesting an
$s$-process event contributed to its heavy element abundances.

The mean abundance ratios for M3 and for M13 are identical to within the errors.
They show the typical pattern for metal-poor
globular clusters of scatter among the light elements, with
the odd atomic number elements appearing in the mean enhanced.
The Fe-peak elements, where the odd atomic number elements are excessively
depleted, do not show any detectable star-to-star variations in either cluster.

The abundance ratios for 13 Galactic globular clusters with recent detailed
abundance analyses, obtained by combining our samples
with published data, are compared to those of published large surveys
of metal-poor halo field stars.  For most elements, the agreement is
very good, suggesting a common chemical history for the halo field
and cluster stars.

\end{abstract}

\keywords{globular clusters: general --- 
globular clusters: individual (M3, M13) --- 
stars: evolution -- stars: abundances}

\section{Introduction}

Abundance determinations of stars in Galactic globular clusters can provide 
valuable information about important astrophysical processes such as
stellar evolution, stellar structure, Galactic chemical evolution and
the formation of the Milky Way. Surface stellar abundances of C, N, O,
and often Na, Mg, and Al are found to be variable among red giants within 
a globular cluster. 
The physical process responsible for these star-to-star element variations 
is still uncertain (see the reviews of Kraft 1994 and
Pinsonneault 1997, as well as Cohen, Briley \& Stetson 2002 
and Ventura \etal\ 2001).  

In order to study the origin of the star-to-star abundance variations,
we started a program to determine chemical abundances of the nearer
Galactic globular cluster stars. 
In previous papers in this series, we studied 
a sample of 25 stars in M71, the nearest globular 
cluster reachable from the northern hemisphere \citep{cohen01,ramirez01,ramirez02},
followed by a similar sized sample in M5,
the nearest intermediate metallicity globular cluster accessible
from a northern hemisphere site \citep{ramirez03}.
Our sample in each cluster
includes stars over a large range in luminosity,
in order to study in a consistent manner red giants, horizontal
branch stars, and stars at the main sequence turnoff.
We measured the iron abundance and the abundance ratios for $\sim$20 elements 
with respect to Fe in each case, using high dispersion 
(R=$\lambda / \Delta \lambda$=35,000) optical spectra obtained with 
HIRES at the Keck Observatory.
We found that the \fe\ abundances\footnote{The 
standard nomenclature is adopted; the abundance of
element $X$ is given by $\epsilon(X) = N(X)/N(H)$ on a scale where
$N(H) = 10^{12}$ H atoms.  Then
[X/H] = log$_{10}$[N(X)/N(H)] $-$ log$_{10}$[N(X)/N(H)]\subsun, and similarly
for [X/Fe].}
from both Fe I
and Fe II  lines agree with each other and with
earlier determinations and that the \fe\ obtained from Fe I and Fe II lines 
is constant within the rather small uncertainties over the full
range in effective temperature (\teff) and luminosity \citep{ramirez01}.
We also found that the neutron capture, the iron peak and the
$\alpha-$element abundance ratios
show no trend with \teff, and low scatter around the mean between the
top of the RGB and near the main sequence turnoff in each cluster.
We detected an anti-correlation between O and Na abundances in our sample
of members of M71 which extends to the main sequence.
We observed a statistically significant correlation in M5 between Al and Na 
abundances extending to $M_V = +1.8$, fainter than the luminosity of the 
RGB bump in M5.  We merged our data with data compiled from the literature
for the Na -- O anticorrelation seen among globular cluster stars to find
that the slope of this relation is the same for all clusters studied to date, 
but the amplitude of the effect varies from cluster to cluster.

In the present paper, we study a sample of stars in
M3 and in M13, globular clusters of even lower metallicity than M5,
again covering a wide range in luminosity. 
There are many photometric studies of these two clusters
as they are the classic second parameter pair, having
similar abundances yet very different horizontal branch
morphologies.  Those which were utilized in the present work
are noted in \S\ref{section_photometry}.

The most important characteristic of the globular cluster M13 
as compared to M71 and M5
is that M13 is well known to show unusually large star-to-star
differences in abundance
of Al, Mg, Na and O among its red giants \citep*[see, e.g.][]{kraft97}.  
A study of C and N variations among a large sample
of M13 stars from the tip of the red giant branch
to the main sequence turnoff is given in \cite{briley02,briley04}; 
large star-to-star differences in C and N abundances were found.
M3, on the other hand,
resembles M5 and M71 in being relatively inactive in this regard,
although its metallicity is quite close to that
of M13.  

M3 and M13 have been the subject of several 
previous high dispersion analyses,
beginning with that of \cite{cohen78}.  \cite{sneden04} have
recently presented a high dispersion study of 28 red giants in M3 to
confirm the lower level of star-to-star variations in this cluster
in contrast to M13; they compare their results
with the older work of their group on both M3 and M13,
see, e.g. \cite{kraft97}.

M3 is more distant than M5 or M71, and we made no
effort to reach the main sequence turnoff in this cluster, although
our sample reaches far down the RGB.  For M13, which is at a
distance comparable to that of M5,
we reach close to the main sequence turnoff.

\section{Observations}

To the maximum extent possible,
the observing strategy, the atomic data and the analysis procedures
used here are identical to those developed in our earlier papers on M71 and on M5
(Cohen, Behr \& Briley 2001; Ram\'{\i}rez \etal\ 2001;
Ram\'{\i}rez \& Cohen 2002; Ram\'{\i}rez \& Cohen 2003)

\subsection{The Stellar Sample}

Stars were chosen to span the range from the tip of the red giant branch
to the base of the subgiant branch in M13 and to well below the RGB
bump in M3.  Both of these
clusters lie far from the Galactic plane, hence field star contamination
is minimal.  
The photometric database of \cite{stetson98} and \cite{stetson00}, which is
described in considerable detail in \cite{cohen02}, 
was used to verify that the  
selected stars lie on the cluster locus in various color-magnitude
diagrams.  This database, in the form available at the time,
only included small sections of the 
solid angle required to fully cover each of these two clusters.

Pairs were selected in M3 for which both stars
lie below the luminosity of the RGB bump and appear, based on
their broad band colors, to be members.  A separation of less than
9 arcsec was required. Only reasonably isolated stars were selected. 
To this we added a sample of bright giants, most of which
were selected to span the full range
in Na and/or O abundances from \cite{kraft92}.
Throughout this paper, 
the star names are from 
\cite{sandage53} for the brightest stars in M3, from \cite{vonzeipel08}
for the two bright stars near the center of M3 not included in the former work,
or, for the stars previously not cataloged,
are assigned based on the object's 
J2000 coordinates, so that a star with RA, Dec of 
13 rm rs.s +28 dm ds is identified in this paper with
the name Crmrss\_dmds.  For M13, the primary source of star identifications
is \cite{arp55} \citep*[see also][]{kadla66}.
For the M13 stars not previously cataloged, names
are assigned based on the object's 
J2000 coordinates, so that a star with RA, Dec of 
16 rm rs.s +36 dm ds is identified in this paper with
the name Crmrss\_dmds.  

\begin{figure}
\epsscale{0.9}
\figurenum{1a}
\plotone{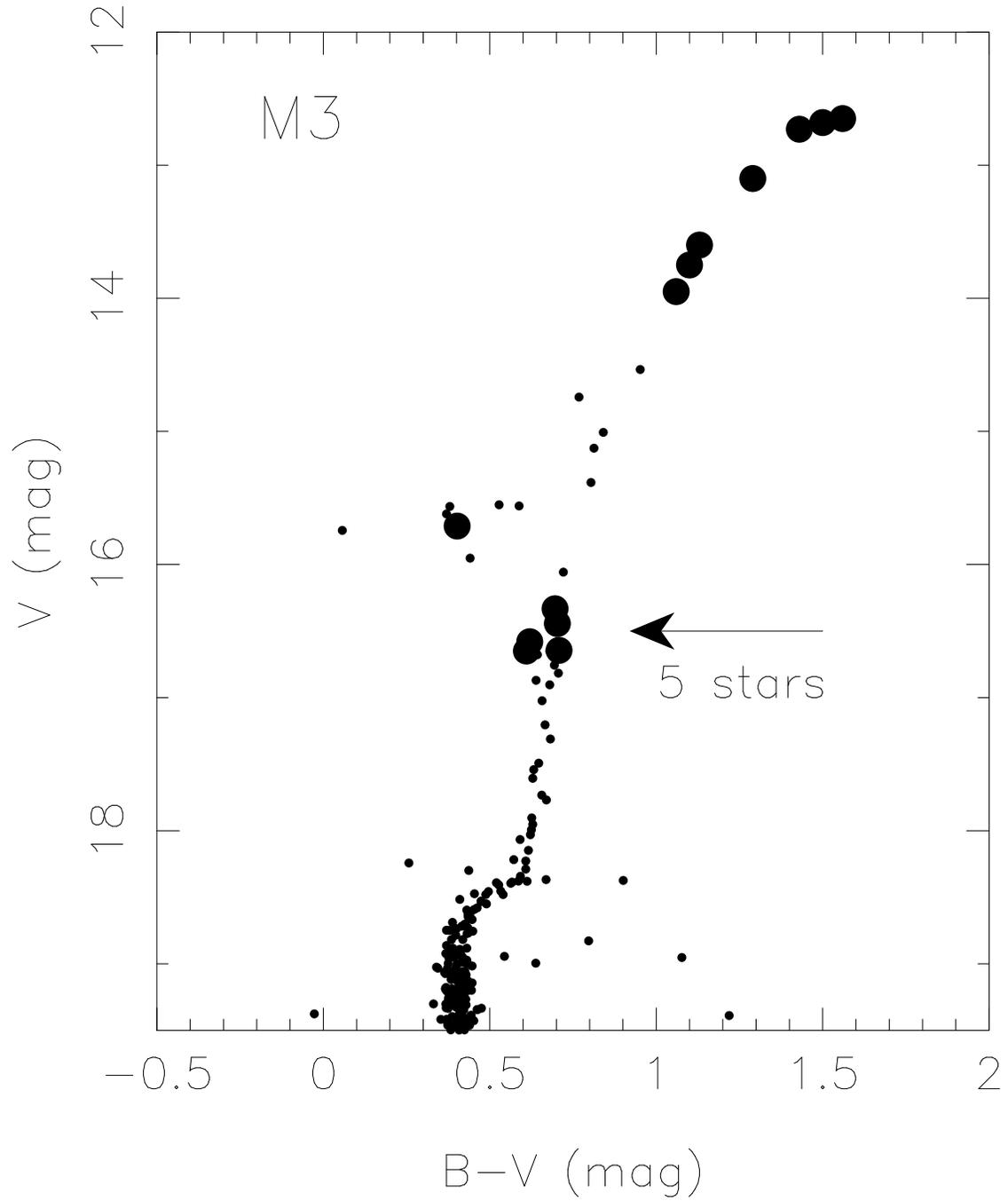}
\caption[]{The $B-V$ CMD for a sector of M3 is shown with
data from the database of \cite{stetson00}.  The 13 stars observed with HIRES
are indicated by large filled circles.
\label{figure_m3cmd}}
\end{figure}

Our sample in M3 totals 13 stars, including 5 on the RGB well below the 
luminosity of the HB and one RHB star.
Figure~\ref{figure_m3cmd} shows the sample
in M3 superposed on a color-magnitude diagram of this globular cluster.

\begin{figure}
\epsscale{0.9}
\figurenum{1b}
\plotone{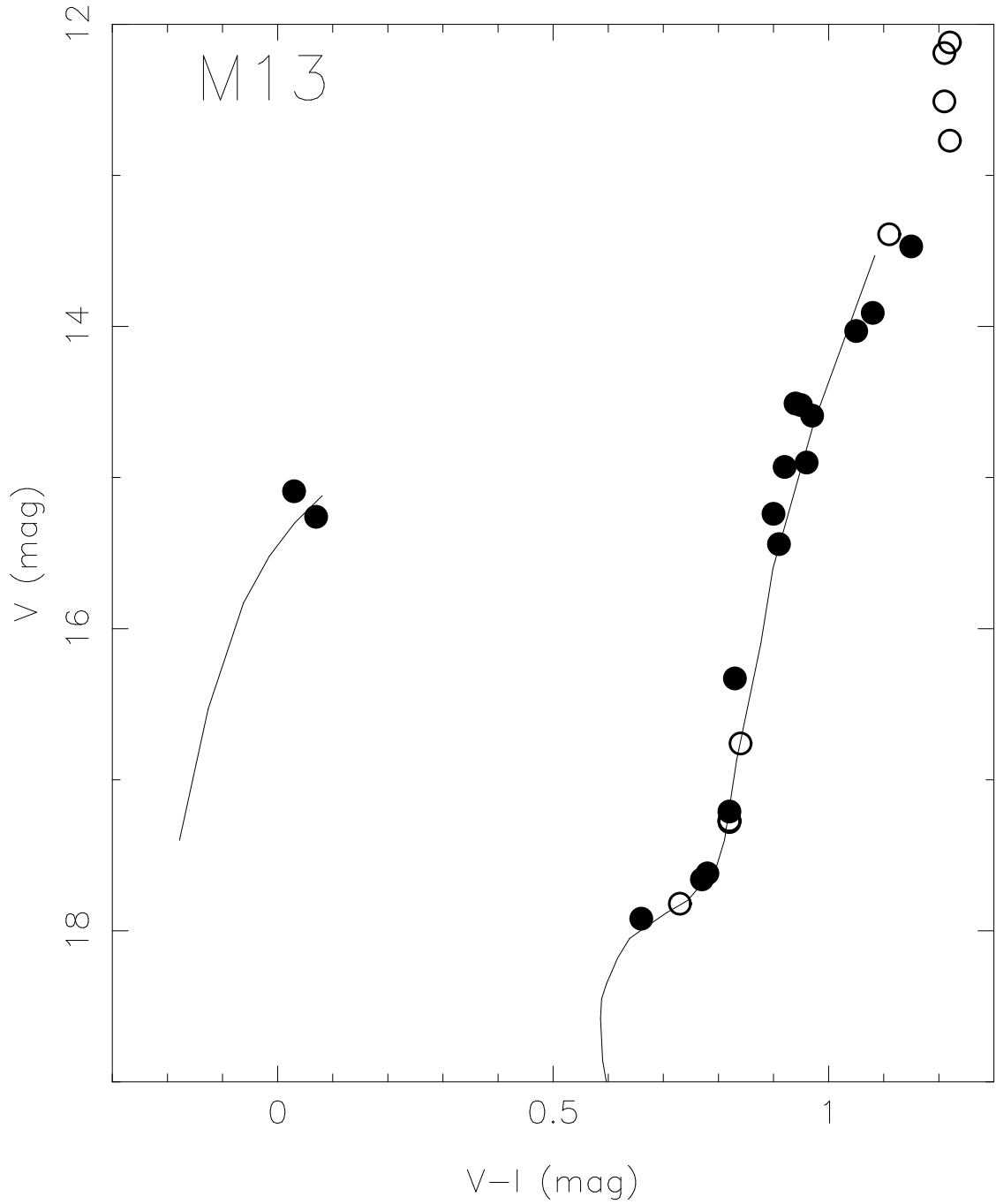}
\caption[]{The $V-I$ locus for M13  from \cite{johnson98} is shown 
with the 27 cluster members observed with HIRES superposed (large
circles).
Open circles
denote those sample stars without $I$ photometry from the
database of \cite{stetson00}.
\label{figure_m13cmd}}
\end{figure}

Eight pairs of stars 
were selected for observation in M13 using  criteria similar
to those used for the M3 pairs.  Two of the 16 stars turned out
not to be members of M13, and two others are horizontal branch stars
too hot to analyze.  These members of M13 reach to V=17.9 mag.
In addition, 13 bright giants were
observed, chosen to have previous observations of
[C/Fe] and of [N/Fe] from \cite{suntzeff81} or from
\cite{smith96}.  This sample include
stars covering the full range of
Na, Mg and Al abundances found in M13 by \cite{kraft97}.
Figure~\ref{figure_m13cmd} shows the sample
in M13 superposed on a color-magnitude diagram of this globular cluster.

Our sample in M13 totals 25 stars reaching from the tip
of the RGB to almost the main sequence turnoff, plus two hot HB
stars which we have not attempted to analyze.

\subsection{Data  Acquisition and Reduction}

All spectra were obtained with HIRES \citep{vogt94} at 
the Keck Observatory.  In May 2001, we observed with an instrumental
configuration similar to that used in our earlier globular cluster work.
The wavelength range from 5500 to 7800~\AA\ was covered with gaps
between the orders due to the undersized HIRES detector.
We wanted to include key lines of critical elements,
specifically the 6300, 6363 [OI] lines, the 7770 O triplet, the
Na doublet at 6154, 6160\AA, and the 6696, 6698\AA\ Al I lines.  
However, it was impossible to create a single instrumental
configuration which included all
the desired spectral features in the wavelength range 6000 to 8000 \AA,
and a single compromise configuration had to be adopted.
In particular, although the 6696, 6698\AA\ Al I doublet is the most useful 
feature of that element in this spectral region, we could not get it
to fit into a single HIRES setting together with the O lines.  
A 1.1 arcsec wide slit, corresponding to a spectral resolution
of 34,000, was used.  A maximum
slit length of 14 arc sec can be used with this instrumental configuration
without orders overlapping. However, we found that covering beyond 7000~\AA,
while highly desirable to reach the O~I triplet, added little else of
interest in the spectra of 
these low metallicity stars.  Thus for the June 2003 observations
(the spring 2002 run being totally lost to weather), we shifted the
instrument configuration toward the
blue, covering the range 4650 to 7010~\AA, again with small gaps
between the orders.  The
maximum slit length without orders overlapping at the blue
end of the spectrum is then reduced to 7 arcsec.  We call these two instrument
configurations the ``yellow'' and the ``red'' configurations.

Since an image rotator for HIRES is available
(built under the leadership of David Tytler), if we can find pairs of program stars
with suitable separations, they can be observed together on
a single exposure.  Pairs were pre-selected as described above 
to contain two members of the fainter M3 stars.
In May 2001, three (6) pairs of M3 (M13) 
stars\footnote{When values of a parameter
are given simultaneously for both M3 and M13, the value for M3 is given, with
that for M13 following in parentheses.}
were observed with HIRES using
the ``red'' setting.  One of the six stars in the M3 pairs turned out to be
a red horizontal branch star.  
In June 2003, two of the pairs in M3 as well as seven
bright giants were reobserved at higher SNR using
the ``yellow'' setting; the third pair was too far apart for this HIRES
configuration.   For M13, two of the pairs were reobserved in June or
August 2003, as were two new pairs  as well as several bright giants
with the ``yellow'' HIRES configuration.

The desired minimum SNR was 75 over a 4 pixel resolution element for
a wavelength near the center of the HIRES detector.
This is calculated strictly from the counts in the object spectrum, and 
excludes noise from cosmic ray hits, sky subtraction, flattening problems, etc. 
Since the nights were relatively dark, sky subtraction is not an issue except at
the specific wavelengths corresponding to strong night sky emission lines, 
such as the Na D doublet.  This SNR goal was
achieved for most of the stars reported here; see Table~\ref{table_sample_m13}
for details regarding the faintest stars in the M13 sample.
Note that for a fixed 
SNR in the continuum, for a star of a given luminosity,
the lower metallicity of these clusters
leads to weaker absorption lines, making it difficult to 
maintain the desired precision of the analysis.  It is this which
led to the repeat spectra with the ``yellow'' HIRES configuration.

The magnitude of the cluster radial velocity
is large for both M3 and M13, and the cluster abundances are low. It was
easy to tell after one integration whether or not
a star is a member of the cluster.
Approximate measurements of the radial velocity were made on line,
and if a star was determined to be a non-member, the observations were 
terminated.  Very few non-members turned up in this way.
If the probable non-member was the second component in a pair, an attempt
was made to switch to another position angle to pick up a different 
second star, when a possible candidate that was bright enough 
was available within the maximum allowed separation.

These data were reduced using a combination of Figaro scripts and
the software package MAKEE \footnote{MAKEE was developed
by T.A. Barlow specifically for reduction of Keck HIRES data.  It is
freely available on the world wide web at the
Keck Observatory home page, http://www2.keck.hawaii.edu:3636/.}.

Table~\ref{table_sample} gives details of the HIRES exposures for each star, with the
total exposure time for each object.
All long integrations were broken 
up into separate exposures, each 1200 sec long, to optimize cosmic
ray removal.
The last column of the table gives
the heliocentric radial velocity for each star,
measured from the HIRES spectra; see \cite{ramirez03} for
the details of the procedure used to determine $v_r$.
Based on their measured $v_r$, the 13 stars of our
sample are all members of M3.  They have
a mean $v_r$ of $-147.6$ \kms, which
agrees exactly with  the value of \citet{harris96}. 
The velocity dispersion of our sample in M3 is  $\sigma$ = 4.7$\pm$1 \kms,
which when the instrumental contribution is removed is about
4.4 \kms.  This is comparable to the value of 5.6~\kms\
obtained by \cite{gunn79} and by \cite{pryor88}.
For the 27 stars in M13 in our sample, the mean $v_r$ is $-245.7$ \kms,
with $\sigma = 7.1$ \kms.  This is identical to the velocity dispersion
obtained by \cite{lupton89} from a sample of more than 130 bright giants.

Based on their HIRES spectra,
stars C41262\_2248 (V=15.6) and C40560\_2847 are not members of M13.

\section{EQUIVALENT WIDTHS  \label{equiv_widths}}

The search for absorption features present in our HIRES data and the
measurement of their equivalent width (\eqw) was done automatically with
a FORTRAN code, EWDET, developed for our globular cluster project. 
Details of this code and its features are described in \citet{ramirez01}.
Because M3 and M13 are considerably more metal poor than M5 or M71, the determination
of the continuum level was easier, and the equivalent widths
measured automatically should be more reliable.

The initial list of unblended atomic lines
and their atomic parameters was adopted from our work on M5,
where it was created
by inspection of the 
spectra of M5 stars, as well as the online Solar spectrum taken with 
the FTS at the National Solar Observatory of \cite{wallace98}
and the set of Solar line identifications of 
\citet{moore66}.
The list of lines identified and measured by EWDET was then correlated,
taking the radial velocity into account, 
to the template list of suitable unblended lines
to specifically identify the various atomic lines. 
The automatic identifications were accepted as valid for
lines with \eqw\ $\ge 15$ m\AA.  They were checked by hand
for all lines with smaller \eqw\ and for all the rare earths.
The equivalent widths of the O I lines were measured by hand, since 
they were generally very weak.
The resulting \eqw\ for $\sim$320 lines in the spectra of the 13 
stars in M3 are
listed in Table~\ref{table_eqw_m3}.  \eqw\ for the 25 stars in M13 
are listed in Tables~\ref{table_eqw_m13_1} and 
Table~\ref{table_eqw_m13_2}.  Note that lines with \eqw $ > 200$~m\AA\
are not generally tabulated and are not used unless there are no other
available lines of that species.

\section{ATOMIC PARAMETERS}

The provenance of the $gf$ values and damping constants we adopt
in our analysis of M3 and M13 stars
is discussed below.  In general, the atomic data and the analysis procedures
used here are identical to those developed in our recent paper
on M5  \citep{ramirez03}.

\subsection{Transition Probabilities}

Transition probabilities for the Fe I lines were obtained from several
laboratory experiments, including studies of Fe I absorption
lines produced by iron vapor in a carbon tube furnace
\citep{bla79,bla82a,bla82b,bla86} (Oxford Group), measurement of 
radiative lifetimes of Fe I transitions by laser induced fluorescence 
\citep{obr91,bar91,bar94}, Fe I emission line spectroscopy from a low 
current arc \citep{may74}, and emission lines of Fe I from a shock 
tube \citep{wol71}.
We also considered solar $gf$ values from \citet{the89,the90} when needed.
The Fe I $gf$ values obtained by the different experiments were placed
into a common scale with respect to the results from
\citet{obr91} (see \cite{ramirez01} for details).
The $gf$ values for our Fe II lines were taken from the solar analysis of
\citet{bla80}, \citet{bie91}, and from the semi-empirical calculations of
\citet{kur93b}. For Fe I and Fe II $gf$ values, we used the same priority 
order for the $gf$ values from different experiments as in \citet{ramirez01}.

Transition probabilities for the lines of atomic species other than iron
were obtained from the NIST Atomic Spectra Database (NIST Standard
Reference Database \#78, see \citep{wei69,mar88,fuhr88,wei96}) when possible.
Nearly 80\% of the lines selected as suitable from the
HIRES spectra have transition probabilities from the NIST database.
For the remaining lines the $gf$ values come from the inverted solar
analysis of \citet{the89,the90},
with the exception of La~II, Nd~II and Eu~II lines, for which
we have updated our values to those of 
\cite{lawler01a}, \cite{lawler01b} and \cite{denhartog03}.

Six elements show hyperfine structure splitting (Sc II, V I, Mn I, Co I, Cu I,
and Ba II). The corresponding hyperfine structure constants were taken
from \citet{prochaska00}.
For Ba~II, we adopt the HFS from \cite{mcwilliam98}.
We use the laboratory spectroscopy of \cite{lawler01a}
and \cite{lawler01b} to calculate the HFS patterns
for La~II and for Eu~II.

We use the damping constants of \cite{barklem00} which were
calculated based on the
theory of Anstee, Barklem \& O'Mara \citep{barklem98} 
when available.  If there is no entry in their database,
as in our earlier work, the 
damping constants were set to twice
that of the Uns\"{o}ld approximation for van der Waals broadening
following \citet{hol91}.

\subsection{Solar Abundances \label{section_sun}}

The regime in which we are operating is so metal poor that we cannot 
in general
attempt to calculate Solar abundances corresponding to our particular choices of
atomic data because the lines seen in these metal poor globular
cluster stars are far too strong
in the Sun.  We  must therefore rely on the accuracy of the
$gf$ values for each element across the large relevant range of
line strength and wavelength.
We adopt the Solar abundances of \cite{anders89} for most elements.
For Ti and for Sr, we adopt the slightly modified values
given in \cite{grevesse98}.  For the special cases 
of La~II, Nd~II and Eu~II we use the results found
by the respective recent laboratory studies
cited above.  For Mg, we adopt the slightly updated value
suggested by \cite{holweger01}, ignoring the small suggested non-LTE
and granulation corrections, since we do not implement such in our analyses.

There is considerable controversy regarding the abundances
of the CNO elements in the Sun, with
the recent results of 
\cite{allende02}, \cite{asplund03} and \cite{asplund04} being
considerably ($\sim$0.2 dex) lower than those of  \cite{anders89} and
somewhat lower than those of \cite{grevesse98}.  We have derived
an inverse Solar O abundance for our particular choice of atomic data
and of model atmospheres using both the forbidden doublet at
6300,6363~\AA\ and the triplet at 7770~\AA.  The equivalent widths
of these lines in the Sun
were measured from  the Solar atlas of \cite{kurucz84}, and checked
in the McDonald Observatory Solar spectrum used by \cite{allende04}.
Corrections for the contribution of the Ni~I line for the 
6300~\AA\ line \citep{allende01} and the CN lines for the 6363~\AA\ line
\citep{asplund04} were made. We adopted the O abundance from [OI]; after
applying a non-LTE correction interpolated from the calculations of \cite{gratton99},
the O abundance from the triplet lines is
only 0.05 dex smaller than derived from the forbidden lines.

We adopt log$\epsilon$(Fe) = 7.45 dex for iron  
following the revisions in the Solar photospheric abundances
suggested by \cite{asplund00} and by \cite{holweger01}.
This value is somewhat lower
than that given by \cite{grevesse98} and considerably lower
than that recommended by \cite{anders89}.
Some papers in the literature use the \cite{grevesse98} value
and some older ones use 7.67 dex, the value recommended by
\cite{anders89}. In such cases, their values of [Fe/H]
will be 0.1 to 0.2 dex smaller than ours while their
abundance ratios [X/Fe] the same amount
larger than ours.

Table~\ref{table_sun} gives the Solar abundances used here.

\section{Stellar Parameters \label{section_photometry}}

We follow the philosophy developed in our earlier work on
globular cluster stars and described in
\citet{cohen01}. \teff\ is derived by comparing reddening-corrected
broad band colors with the predictions of grids of model atmospheres.
We utilize here the grid of predicted broad band colors and
bolometric corrections of \citet{houdashelt00}
based on the MARCS stellar atmosphere code of \citet{gus75}.  
In \cite{cohen01} we demonstrated that the Kurucz and MARCS 
predicted colors are essentially
identical, at least for the specific colors used here.
We adopt current values
from the on-line database of \citet{harris96}
for the distances of 10.4 (7.5) kpc for M3 (M13)
with a reddening of E(B--V) = 0.01 (0.02) mag.
The relative extinction in various passbands is taken from
\citet{cohen81} \citep*[see also][]{schlegel98}.
Based on the earlier high dispersion analyses of 
\cite{kraft97} and \cite{sneden04}, 
we adopt as an initial guess  
[Fe/H] = $-$1.5 dex for these two globular clusters.

Our primary source for optical photometry is the
BVI database of Stetson \citep{stetson98,stetson00}.
All the bright giants in M3 except VZ~1000 and VZ~1397, which
are too close to the center of this cluster, are included there.
All components of the pairs except V-30 + V-31, which are too far from
the center of M3, are included as well.  Photometry
for the missing stars was obtained from
\cite{sandage53}, \cite{johnson56}, \cite{buonanno94} or
\cite{ferraro97}, with preference given to the most recent study.

For M13, we used (in order of preference) the optical
photometry of 
\cite{cudworth79}, \cite{rey01}
\cite{buonanno94}, and Stetson \citep{stetson98,stetson00}.
After a small adjustment in the zero point of the (uncalibrated)
photometry of \cite{buonanno94}, the agreement between the
various datasets is reasonable.  The area in M13 covered
by the work of \cite{johnson98}, 
\cite{rosenberg00} and by \cite{piotto02} 
does not overlap with our stars, nor does that of the
IR photometry of \cite{davidge99} and of \cite{valenti04}.  

We only used V and I magnitudes from these sources,
combining them with J and K photometry from 2MASS \citep{2mass1,2mass2}.
This worked extremely well for the bright giants, with a scatter
in deduced \teff\ from V-I, V-J and V-K of under 40~K.
The fainter stars in our sample (i.e. the pairs)
show much larger scatter in their
deduced \teff\ from the various colors.  These stars are too
crowded (and in some cases rather faint) for 2MASS, which uses
an aperture size of 2 arcsec.  
This problem is  more serious in M13, as our sample reaches fainter
there than in M3; a large fraction of the
pairs observed in M13 had no entry or only a single entry in the 2MASS database.

To get around this problem, 
in April, 2004 we observed the fields of the fainter
stars in our samples in both M3 and M13
with the Wide Field Infrared Camera \citep{wilson03}
at the 5-m Hale Telescope for the purpose of establishing reliable J,K magnitudes
for the fainter stars in our sample.  The 2MASS colors of nearby isolated
somewhat brighter stars were used to calibrate our WIRC photometry.  
This new infrared photometry for
only the pairs in M3 and M13
is listed in Table~\ref{table_irphot}.  The agreement with 2MASS is very
good for the brighter, wider pairs.

With improved infrared photometry in hand,
the observed broad band colors V-I, V-J and V-K for
each program star from the sources listed above,
corrected for extinction, were used to determine
\teff.  The set of models with metallicity of $-$1.5 dex, nearest to our
initial estimate of [Fe/H], is used. Table~\ref{table_teff} 
(Table~\ref{table_teff_m13})
lists the \teff\ thus deduced for the sample in M3 and in M13.
The reddening to each of these clusters is small, making possible 
extinction variations across the cluster irrelevant.
We assume a random photometric error of 0.02 mag applies to
$V-I$ from \cite{stetson00}.  Following \cite{cohen01},
this translates into a total uncertainty in \teff\ of 75 K for giants rising
to 150 K for main sequence stars using only $V-I$.  The errors in
\teff\ deduced from V-J or V-K are about a factor of two smaller.

We have slightly smoothed the \teff\ for the fainter stars in our
sample by small amounts to ensure that stars at the 
approximately same evolutionary stage have
approximately the same stellar parameters.

Once an initial guess at \teff\ has been established from a broad
band color, it is possible with minimal assumptions
to evaluate \grav\ using  observational data.
The adopted distance modulus for the cluster, initial
guess at \teff\ for the star, and an assumed stellar mass 
(we adopt 0.8 $M$\subsun\
for the stars in M3 and in M13)  are combined with 
the known interstellar absorption, the
bolometric corrections predicted by the 
model atmosphere grid, as well as a broad band observed $V$ mag to
calculate \grav. 
An iterative scheme is used to correct for the small
dependence of the predictions of the model atmosphere grid on
\grav\ itself.  Rapid convergence is achieved.

It is important to note that because of the constraint of
a known distance to each cluster, the
uncertainty in \grav\ for any star in our sample is small, 
$\le0.1$ dex when comparing
two members of the same cluster.  Propagating an uncertainty of 15\% in the cluster
distance, 5\% in the stellar mass, and a generous 3\% in \teff, 
and ignoring any covariance, leads to
a potential systematic error of $\pm$0.2 dex for \grav.

\section{Abundance Analysis \label{section_anal}}

We rely heavily in the present work on the procedures and atomic data
for abundance analyses of metal-poor stars
described in our earlier papers referenced above reporting analyses of globular
cluster stars.

Given the derived stellar parameters from Table~\ref{table_teff}
(\ref{table_teff_m13}), we 
determined the abundances using the equivalent widths obtained as 
described above.
The abundance analysis is carried out using a current version of the LTE
spectral synthesis program MOOG \citep{sneden73}.
We employ the grid of stellar atmospheres from \cite{kurucz93a} 
without convective overshoot, when available. We compute the
abundances of the species observed in each star using 
the four stellar atmosphere
models with the closest \teff\ and log($g$) to each star's parameters.
The abundances were interpolated using results from the closest stellar model
atmospheres to the appropriate \teff\ and log($g$) for each star given
in Table~\ref{table_teff} (\ref{table_teff_m13}).

The microturbulent velocity ($v_t$) of a star is determined 
spectroscopically by requiring the abundance to be independent of the 
strength of the lines.  
The uncertainty
in our derived $v_t$ is estimated to be +0.4,$-$0.2 \kms\
based on repeated trials with the same line list for several stars
varying $v_t$. 
We apply this technique here to the large sample of detected
Fe~I lines in each star; the results are listed with the
stellar parameters in Table~\ref{table_teff} (\ref{table_teff_m13}).

\begin{figure}
\epsscale{0.9}
\figurenum{2}
\plotone{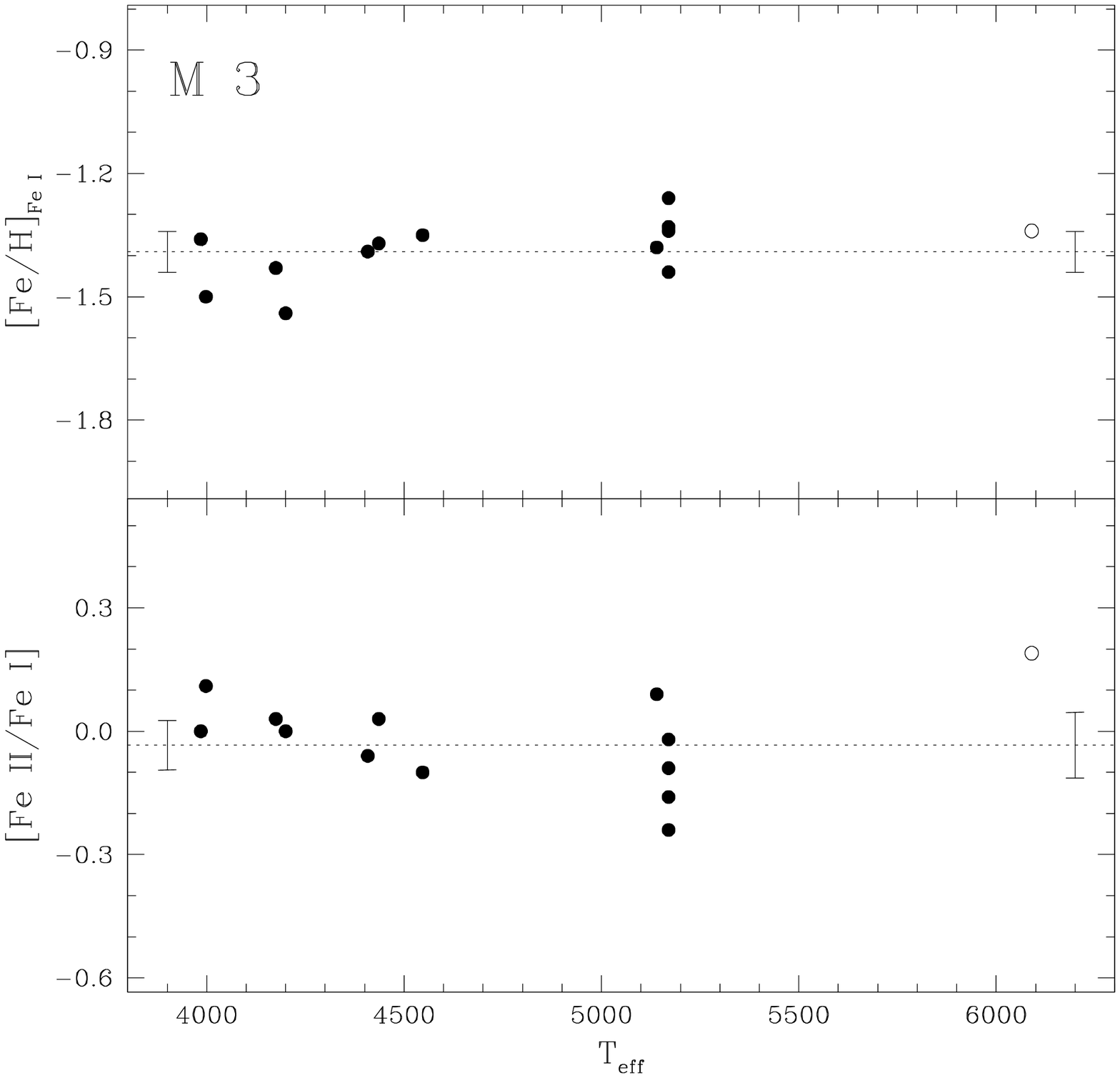}
\caption[]{The [Fe/H] from lines of Fe~I is shown as a function of \teff in
the upper panel, while the lower panel shows the ionization equilibrium of
Fe for our sample of 13 stars in M3. The open circle indicates the HB star.
The error bars on the left margin are those of the most luminous giants, while 
the error bars on the right margin are those of the faintest stars 
in our sample.  The dotted horizontal line indicates the mean value
for our sample in this globular cluster.
\label{figure_m3_feion}}
\end{figure}

\begin{figure}
\epsscale{0.9}
\figurenum{9}
\plotone{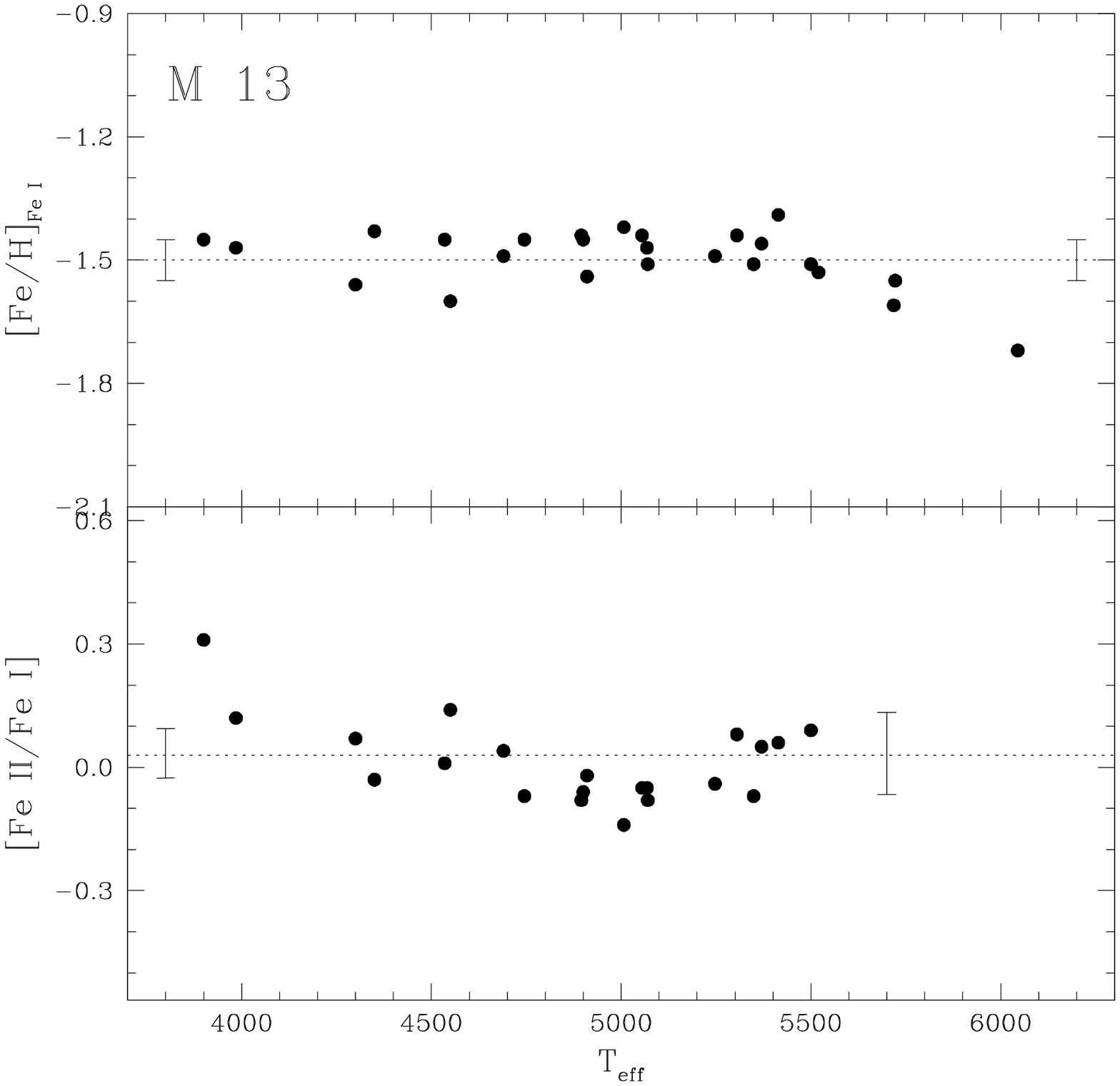}
\caption[]{The [Fe/H] from lines of Fe~I is shown as a function of \teff in
the upper panel, while the lower panel shows the ionization equilibrium of
Fe for our sample of 25 stars in M13.
The error bars on the left margin are those of the most luminous giants, while 
the error bars on the right margin are those of the faintest stars 
in our sample.  The dotted horizontal line indicates the mean value
for our sample in this globular cluster.
\label{figure_m13_feion}}
\end{figure}

Iron is the only element with more than a few detected lines in 
each of two
different stages of ionization, hence useful for determining the
ionization equilibrium. 
Figure~\ref{figure_m3_feion} (\ref{figure_m13_feion})
shows the Fe abundance as inferred from
lines of Fe~I as well as the Fe ionization equilibrium for the
stars in our sample in M3 (M13).
The ionization equilibrium for Fe~I versus Fe~II 
is satisfactory in each of these globular clusters.
The average difference 
between [Fe/H] as inferred from Fe~II lines
and from Fe~I lines for the 8 luminous RGB stars in M3 
is 0.00 dex, while the
largest difference (in absolute value) is only 0.11 dex.  
For the 5 low luminosity giants in M3, the range is somewhat larger, but
the average difference is only 0.01 dex.  Over the full M13 sample,
the average difference is only 0.04 dex.
The Fe ionization
equilibrium shifts by 0.2 dex for a 100 K change in \teff\ in this
temperature regime, so our $\pm50$~K uncertainty in \teff\
is capable of producing the observed dispersion in Fe ionization equilibrium.
For Ti, which has far fewer detected lines of the neutral species
than does Fe, the mean difference in [Ti/Fe] as deduced from the
neutral lines and from the ionized lines for the stars in our sample
in M3 (M13) is only 0.10 (0.15) dex. 

Following upon our previous work, no non-LTE corrections have been applied
for the specific ions studied in the M3 and M13 stars, with the exception
of O, where we rely on the calculations of \cite{gratton99}.
The detailed non-LTE
calculations of \cite{gratton99} and of \cite{takeda03} for the two Na~I doublets
we use suggest that for this regime of \teff, the
non-LTE correction is about $+0.15$ dex.  For Ba~II, the 
non-LTE calculations of \cite{mashonkina99} and of
\cite{mashonkina00} suggest that a non-LTE correction
of $-$0.1 dex is appropriate for the metallicity of M3 and M13
and the set of Ba~II lines we used. 
In comparing with other
abundance analyses, the issue of implementing non-LTE
corrections and their adopted values must be considered.

The resulting abundance ratios for 13 (25) stars in M3 (M13) are given in
Tables~\ref{tab4a} to \ref{tab4e} (Tables~\ref{table_m13abund_a} to
\ref{table_m13abund_e}).  The abundance ratio for
a species with only one detected line 
in a particular star is assigned
an uncertainty of 0.10 dex.
Table~\ref{table_sens} indicates
the changes in derived abundance ratios for small changes in the 
adopted stellar parameters,
the [Fe/H] for the adopted model atmosphere, or the set of \eqw\
for the lines of each species.  Table~\ref{table_abundsig}
(Table~\ref{table_abundsig_m13})
gives the mean abundance and 1$\sigma$ variance for the species
observed in M3 (M13).

\subsection{Comments on Individual Elements \label{sec_individual}}

The oxygen abundance is derived from the forbidden lines
at 6300 and 6363~\AA\ and from the triplet lines at 7770~\AA\
when these were included within the wavelength range and were
detected.  The subtraction of the night sky emission
lines for the forbidden lines was reasonably straightforward
given that the radial
velocities of M3 and of M13 sufficiently different from 0 \kms\ 
that their \eqw\ can be reliably measured.
The C/O ratio was assumed to be Solar.
Small non-LTE corrections, calculated from \cite{gratton99}, 
were applied for abundances deduced from the O~I triplet.
Since $N$(CN)/$N$(H) is roughly 
$\propto {\{}N$(Fe)/$N$(H)${\}}^2$,
CN lines are much weaker relative to O lines in metal poor stars,  so that
the correction for CN contamination to the 6363~\AA\
[OI] line for the M3 and M13 stars is negligible.
Finally, the contribution of the Ni~I line to the 6300~\AA\
forbidden line of O can be ignored in M3 and M13, as we will see
that O/Fe (and O/Ni) is larger than the Solar 
value\footnote{\cite{sneden04} estimate that \eqw\ for this
Ni~I lines is less than 0.5~m\AA.}.
The O abundance from the triplet lines is given with respect to [Fe/H] deduced
from lines
of Fe~I, while that from the forbidden lines is given
with respect to Fe~II; the mean [O/Fe] becomes 0.04 dex 
larger if expressed using the  Fe~II lines instead.

The usual lines of Al in this wavelength region, in particular
the 6696,6698~\AA\ doublet, can not be reached with our HIRES configuration
as they fall in an interorder gap.  We have detected the much weaker
Al~I line at 5557~\AA\ in the coolest M13 giants only. There
is a difference of 0.4 dex between the Al abundance we deduce
and that obtained by 
\cite{sneden04} \citep*[see also][]{kraft97,shetrone96}
 for the two M13 luminous giants
in common. The $gf$ value adopted 
for this rarely used line ($-1.67$ dex) may be wrong, or non-LTE may play a role;
\cite{baumuller97} have demonstrated that non-LTE effects can be very
strong for Al~I features in this \teff\ range, and the effect varies
a lot from multiplet to multiplet.

The Na abundance was obtained from the 5680~\AA\ doublet in general.
For the faintest, hottest stars in M13, these lines become very weak,
and so the Na abundance was checked using the D lines (after a small
empirical correction not exceeding 0.08 dex 
to put them on the same abundance scale as
for the weak Na lines).  

\begin{figure}
\epsscale{0.9}
\figurenum{3}
\plotone{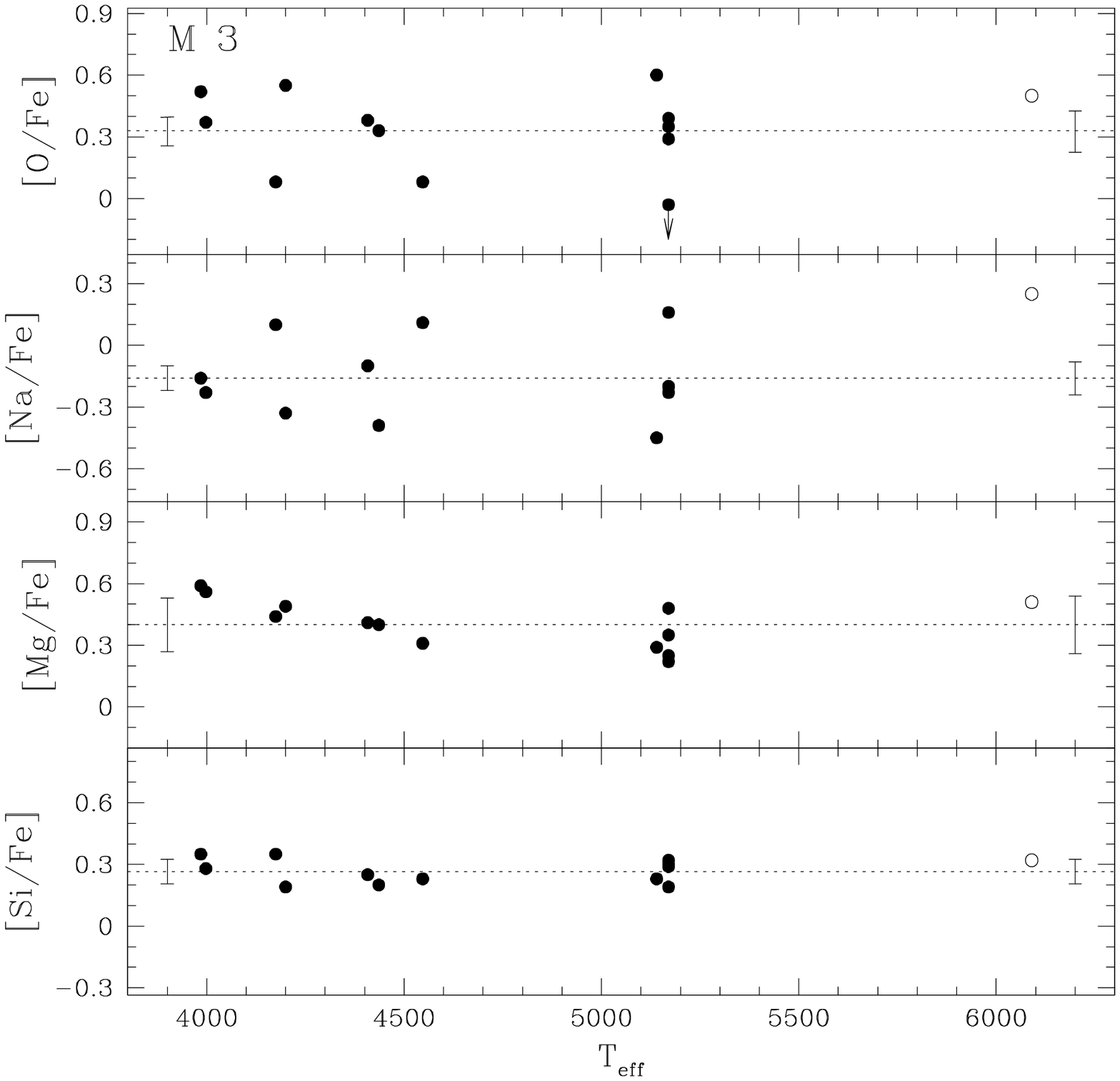}
\caption[]{[X/Fe] for the elements O, Na, Mg and Si are shown
as a function of \teff\
for our sample of 13 stars in M3. The open circle indicates the HB star.
The error bars for the most luminous and least luminous stars,
as well as the cluster mean, are indicated as in Figure~\ref{figure_m3_feion}.
\label{figure_m3_o_si}}
\end{figure}

\begin{figure}
\epsscale{0.9}
\figurenum{4}
\plotone{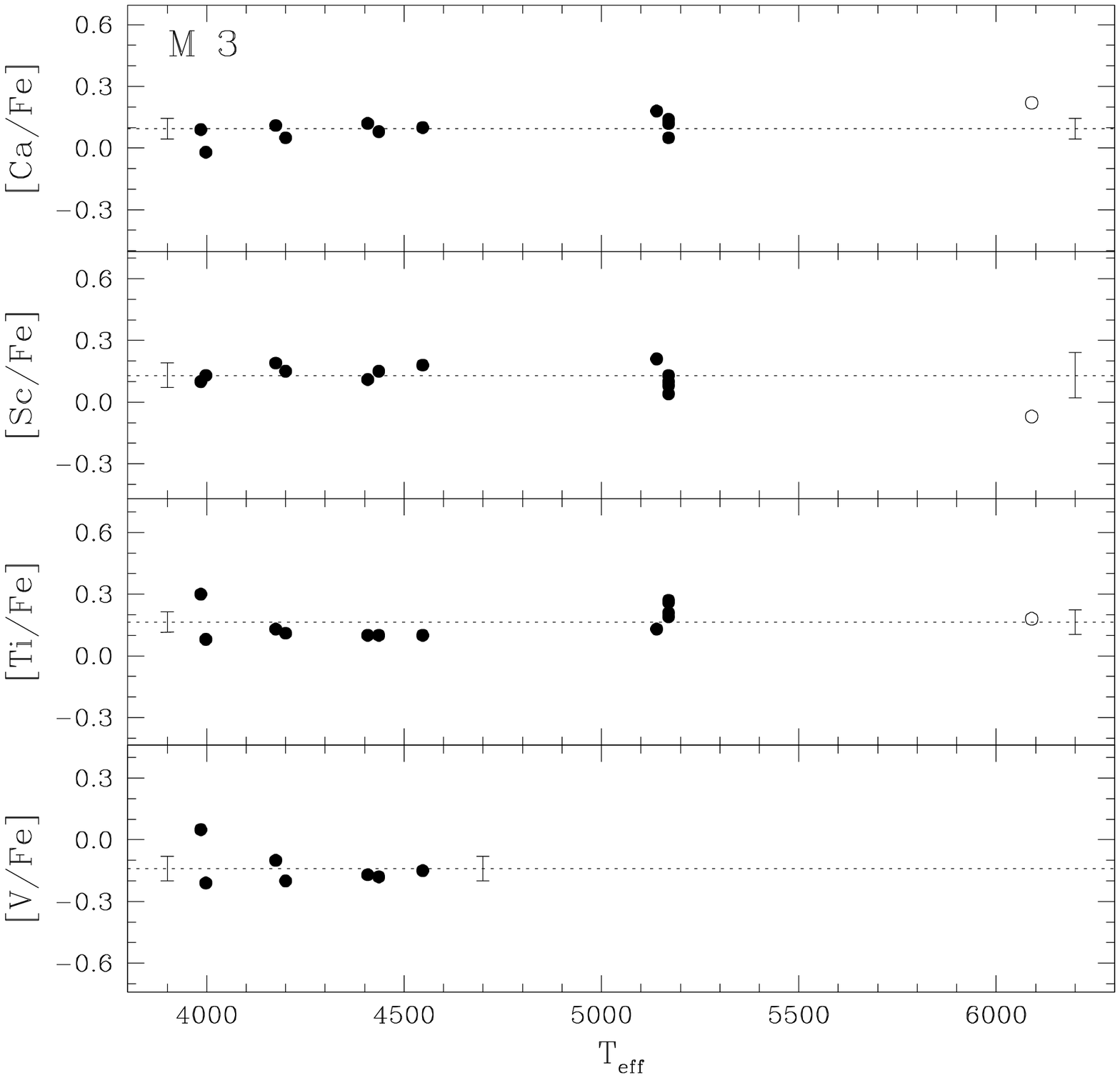}
\caption[]{Same as Figure~\ref{figure_m3_o_si} for the elements
Ca, Sc, Ti and V in M3.
\label{figure_m3_ca_v}}
\end{figure}

\begin{figure}
\epsscale{0.9}
\figurenum{5}
\plotone{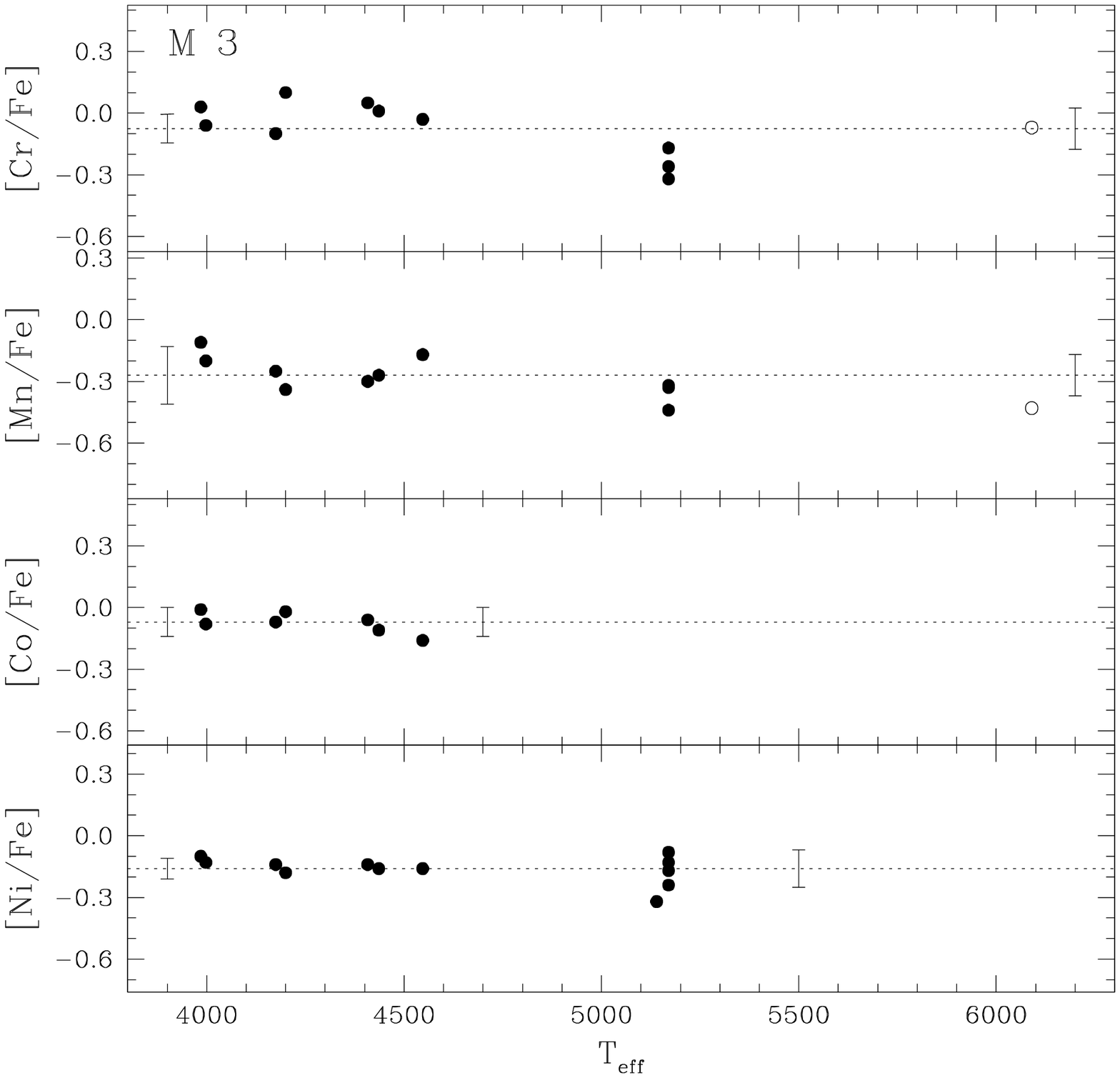}
\caption[]{Same as Figure~\ref{figure_m3_o_si} for the elements
Cr, Mn, Co and Ni in M3.
\label{figure_m3_cr_ni}}
\end{figure}

\begin{figure}
\epsscale{0.9}
\figurenum{6}
\plotone{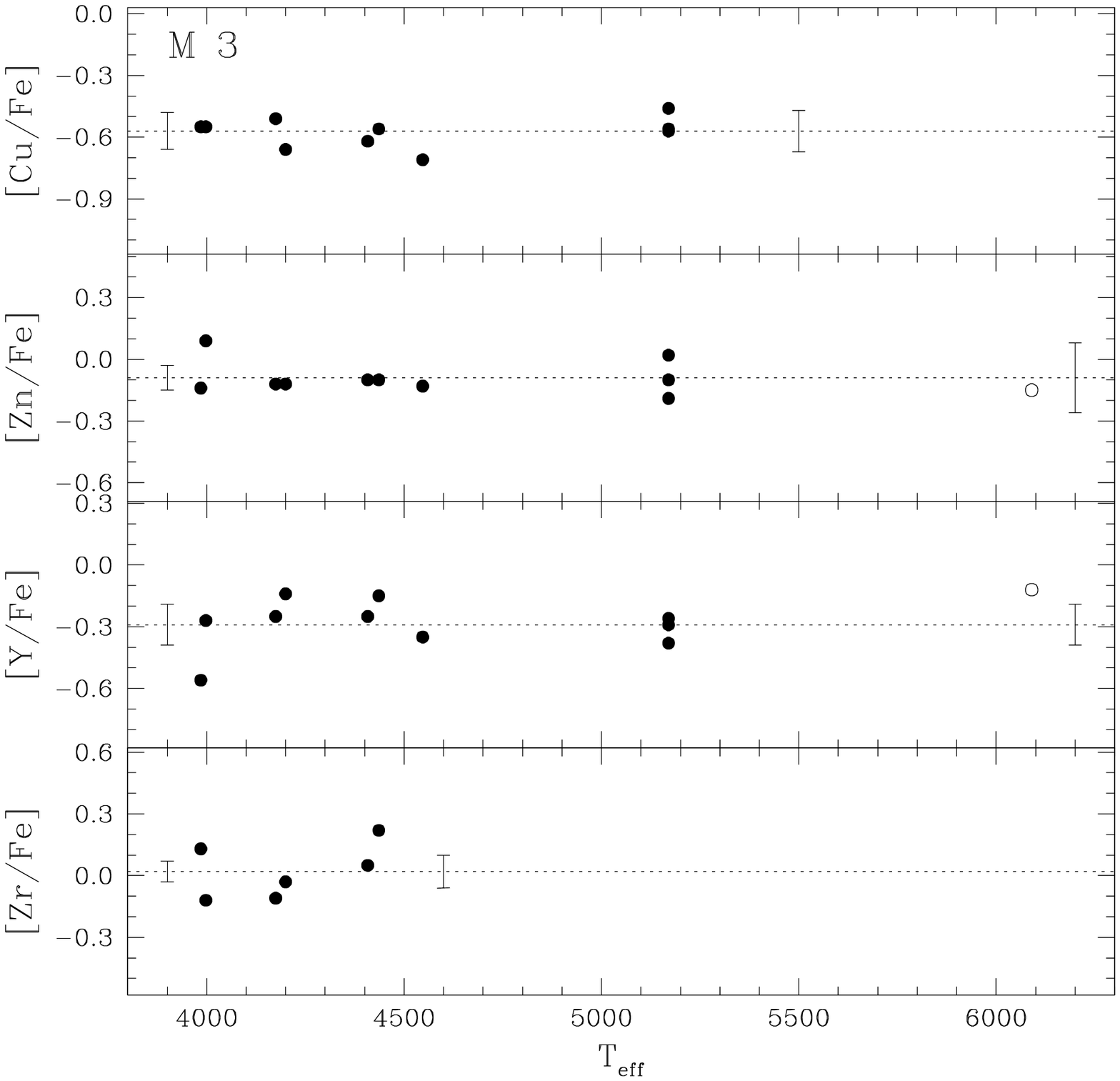}
\caption[]{Same as Figure~\ref{figure_m3_o_si} for the elements
Cu, Zn, Y and Zr in M3.
\label{figure_m3_cu_zr}}
\end{figure}

\begin{figure}
\epsscale{0.9}
\figurenum{7}
\plotone{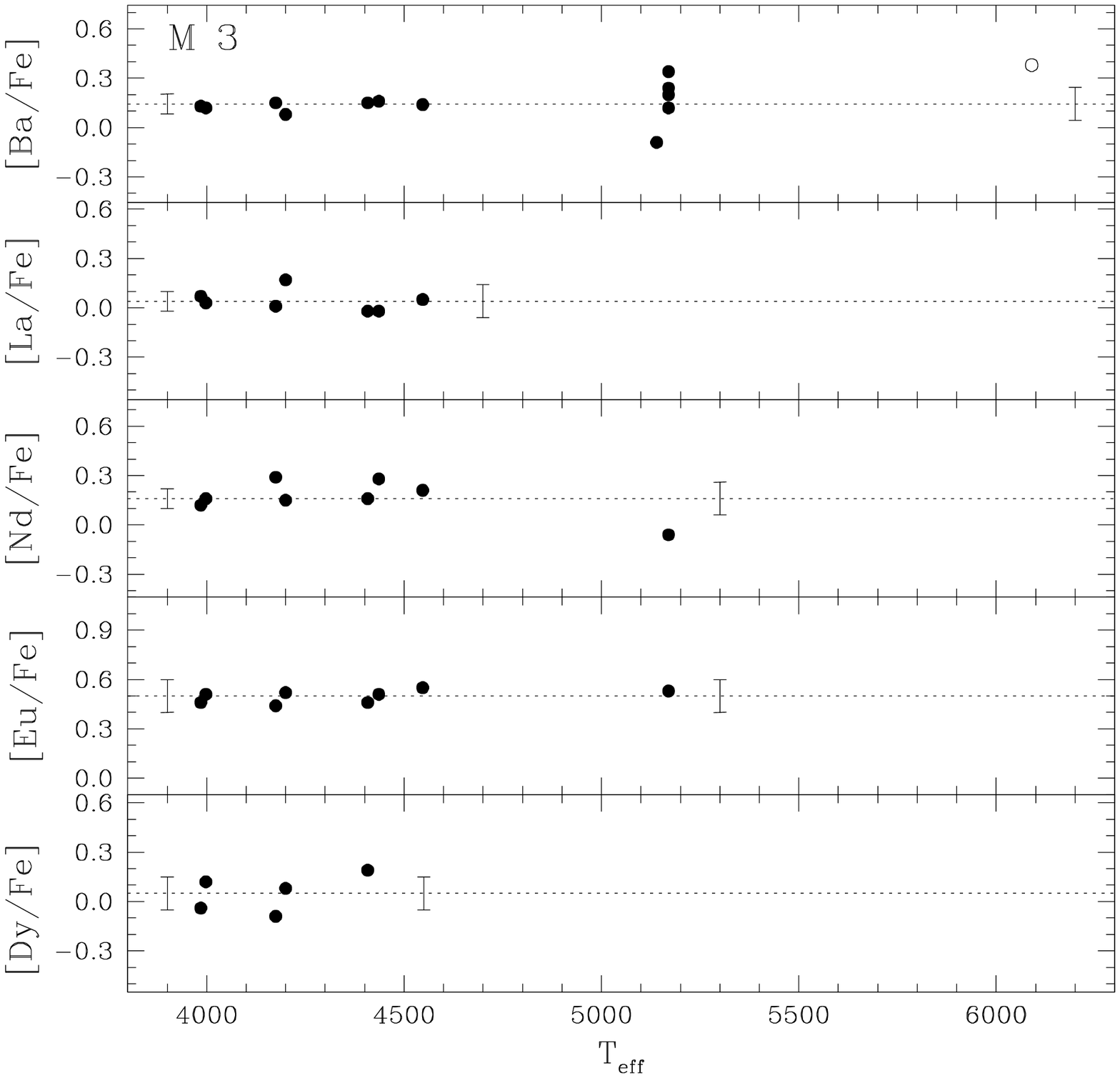}
\caption[]{Same as Figure~\ref{figure_m3_o_si} for the elements
Ba, La, Nd, Eu and Dy in M3.
\label{figure_m3_ba_dy}}
\end{figure}

\begin{figure}
\epsscale{0.9}
\figurenum{10}
\plotone{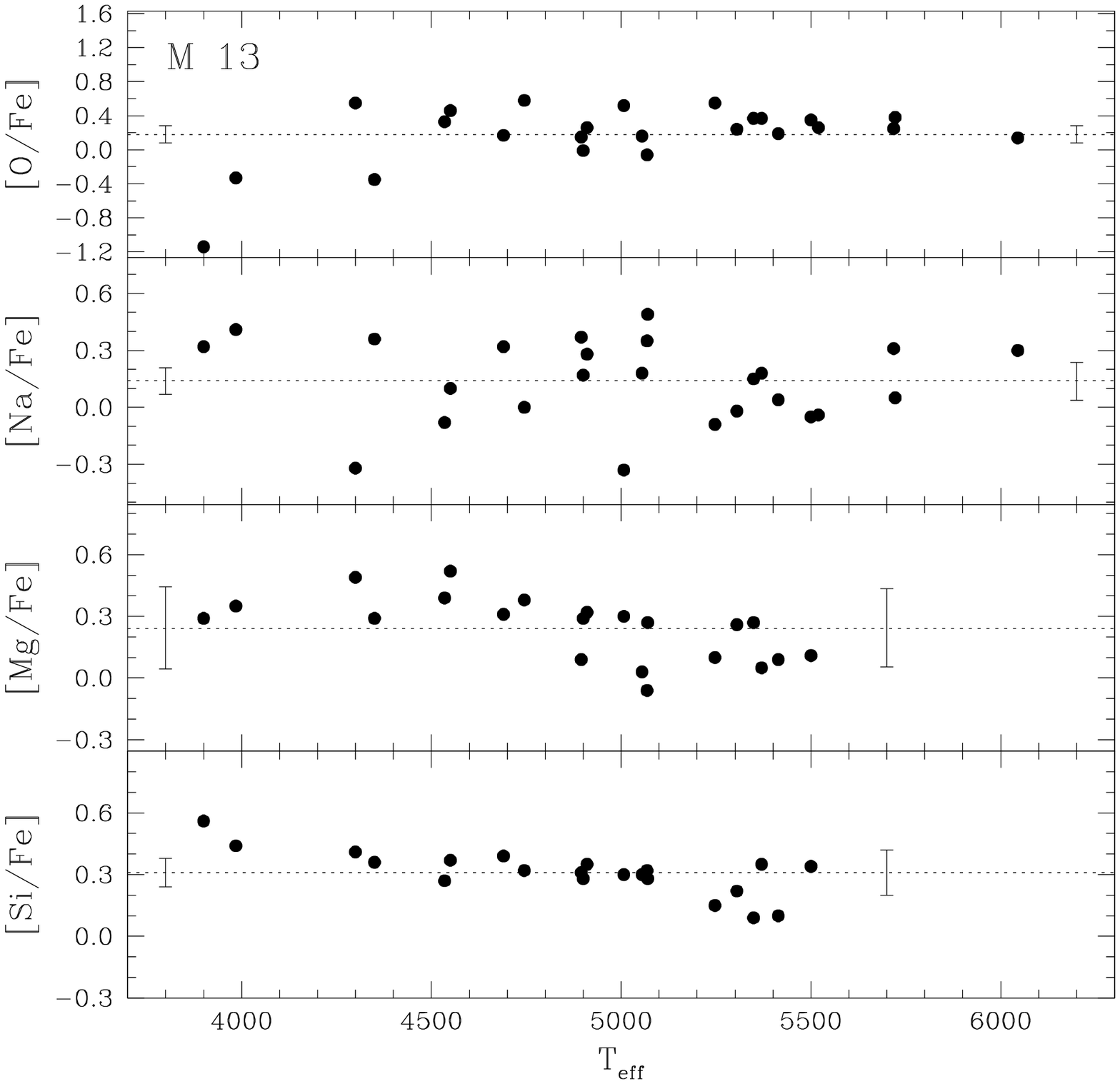}
\caption[]{[X/Fe] for the elements O, Na, Mg and Si are shown
as a function of \teff\
for our sample of 13 stars in M13. 
The error bars for the most luminous and least luminous stars,
as well as the cluster mean, are indicated as in Figure~\ref{figure_m3_feion}.
\label{figure_m13_o_si}}
\end{figure}

\begin{figure}
\epsscale{0.9}
\figurenum{11}
\plotone{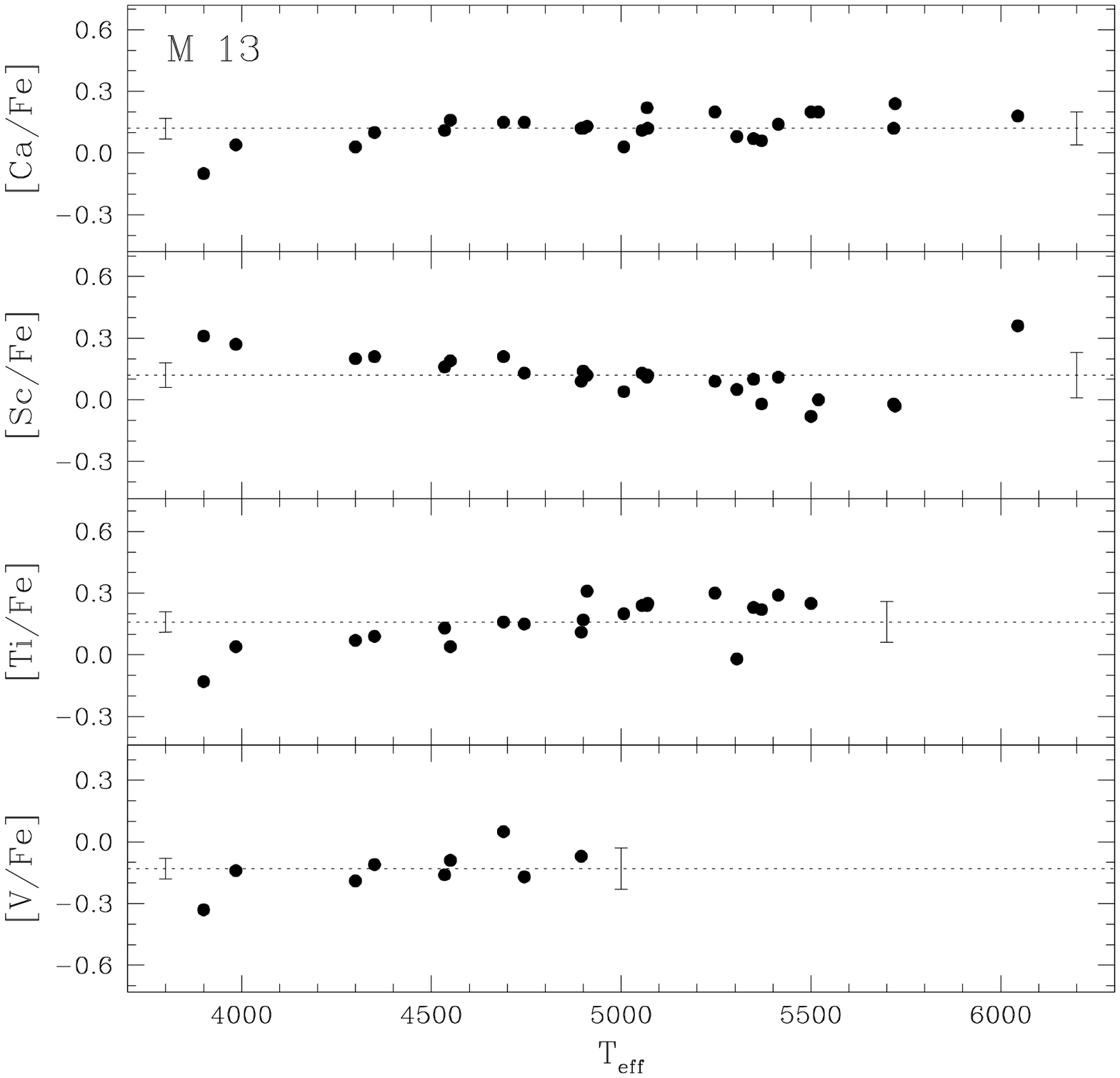}
\caption[]{Same as Figure~\ref{figure_m13_o_si} for the elements
Ca, Sc, Ti and V in M13.
\label{figure_m13_ca_v}}
\end{figure}

\begin{figure}
\epsscale{0.9}
\figurenum{12}
\plotone{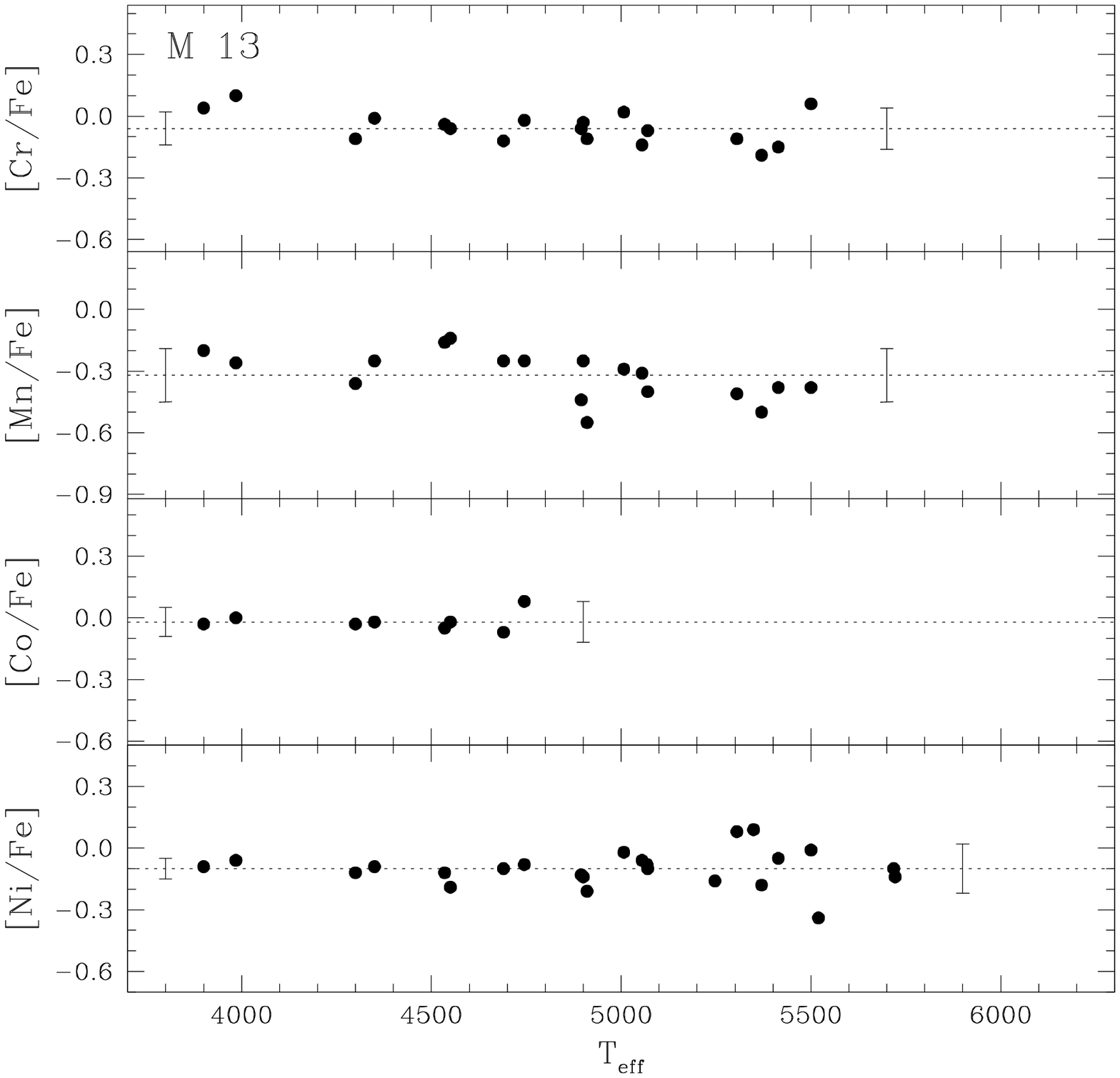}
\caption[]{Same as Figure~\ref{figure_m13_o_si} for the elements
Cr, Mn, Co and Ni in M13.
\label{figure_m13_cr_ni}}
\end{figure}

\begin{figure}
\epsscale{0.9}
\figurenum{13}
\plotone{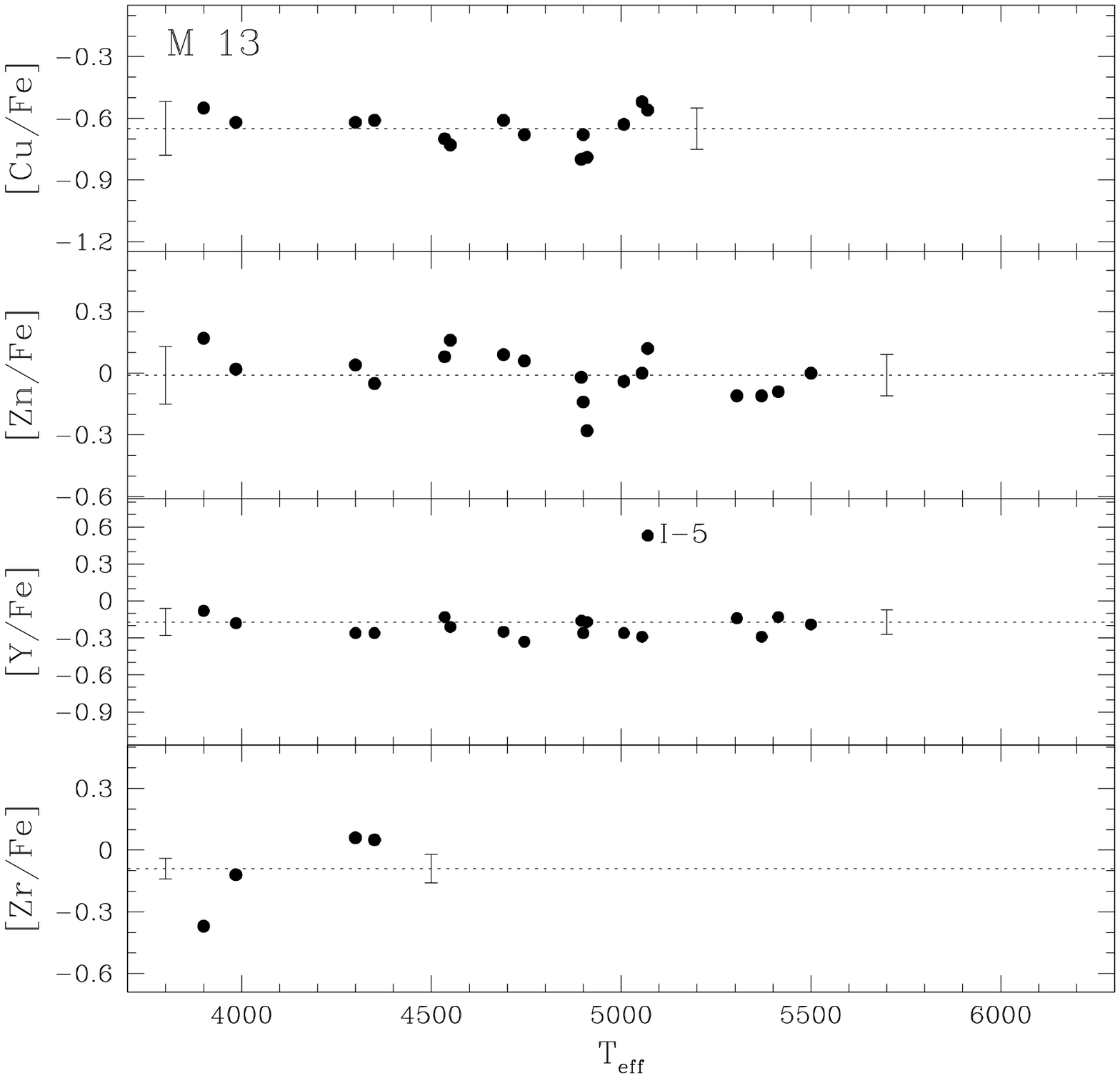}
\caption[]{Same as Figure~\ref{figure_m13_o_si} for the elements
Cu, Zn, Y and Zr for M13.
\label{figure_m13_cu_zr}}
\end{figure}

\begin{figure}
\epsscale{0.9}
\figurenum{14}
\plotone{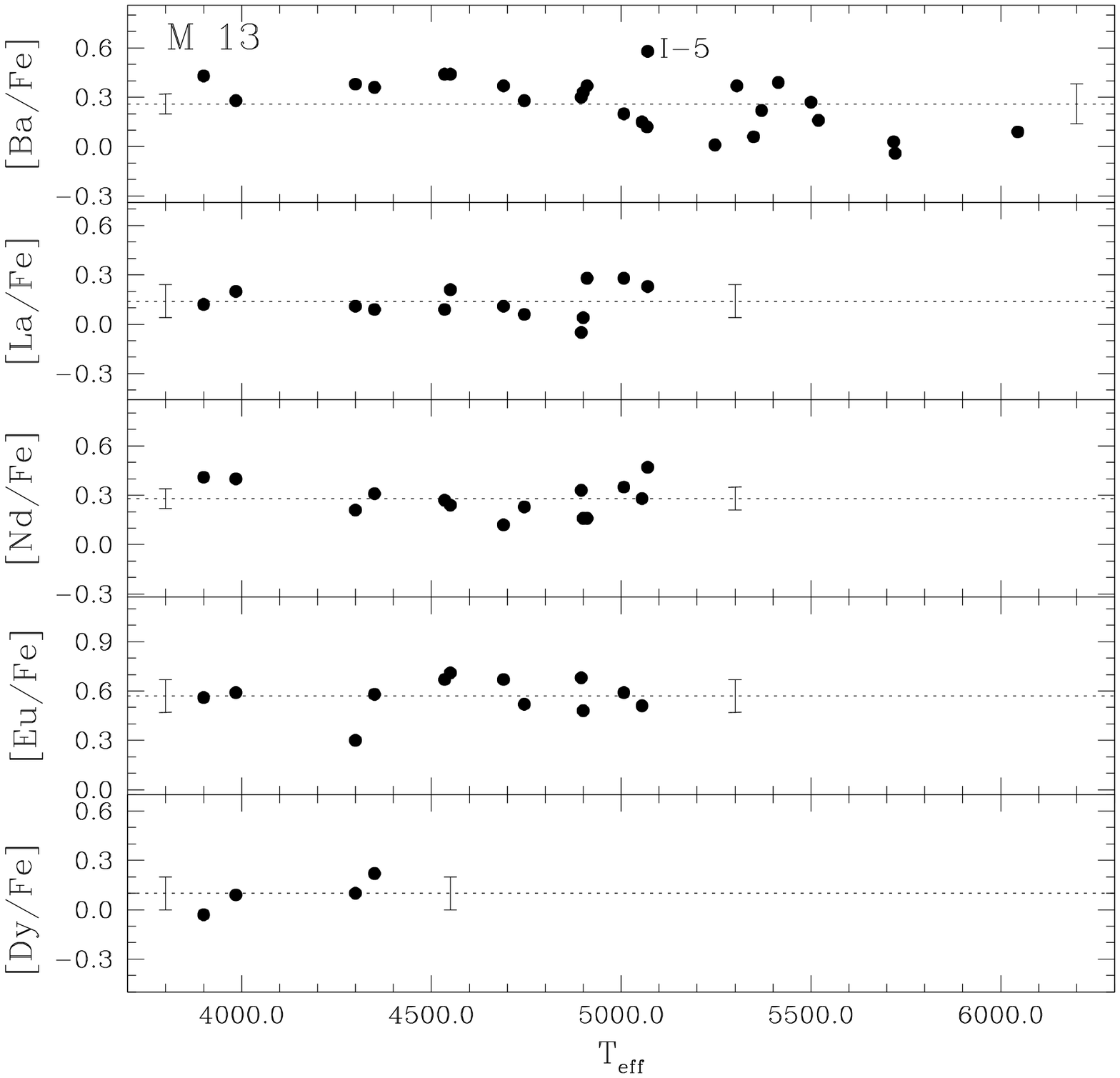}
\caption[]{Same as Figure~\ref{figure_m13_o_si} for the elements
Ba, La, Nd, Eu and Dy in M13.
\label{figure_m13_ba_dy}}
\end{figure}

The abundances of the elements with respect to Fe, [X/Fe],
as a function of \teff\ are shown in 
Figures~\ref{figure_m3_o_si} (\ref{figure_m13_o_si}),
covering O, Na, Mg and Si, 
Figures~\ref{figure_m3_ca_v} (\ref{figure_m13_ca_v}), which include 
Ca, Sc, Ti and V, 
Figures~\ref{figure_m3_cr_ni} (\ref{figure_m13_cr_ni}), which include Cr, Mn,
Co and Ni, 
Figures~\ref{figure_m3_cu_zr} (\ref{figure_m13_cu_zr}), which include 
Cu, Zn, Y and Zr,
and Figure~\ref{figure_m3_ba_dy} (\ref{figure_m13_ba_dy}), for
Ba, La, Nd, Eu and Dy.
Note the apparent star-to-star variation in [O/Fe] and in [Na/Fe],
which becomes undetectably small, if it exists at all, for the
elements heavier than Na. The scatter for Ni, which is detected with
several lines in every star, is remarkably small.  Even the
rare earths, for which in general only a few weak lines are
detected only in the more luminous stars, show very
small variations in general.

\subsection{Abundance Spreads \label{section_spread}}

Detection and quantitative measurement of
star-to-star variations in abundance ratios within a single Galactic
globular cluster is one of the primary goals of this effort.
As a global indicator of the presence of such effects we use
a parameter we call the ``spread ratio'' ($SR$).  The numerator of
$SR$ is the 1$\sigma$ rms variance for the sample of 13 (25) stars in M3 (M13)
about that mean
abundance for each atomic species ($X$) with detected
absorption lines,
denoted $\sigma$.  The mean abundance and $\sigma$ for
each species observed are given in the first three columns  
of Table~\ref{table_abundsig} (Table~\ref{table_abundsig_m13}).  
The denominator of $SR$ is the total
expected uncertainty, $\sigma(tot)$, which is the sum in quadrature of 
the known contributing terms.  Included are  a term corresponding to an uncertainty
of 50~K in \teff, the same for an uncertainty of 0.2 dex in \grav,
and for an uncertainty of 0.2~\kms\ in $v_t$, and
the observed uncertainty [$\sigma(obs)$],
The parameter $\sigma(obs)$, which is calculated from data given 
in Tables~\ref{tab4a} to \ref{tab4e} 
(Tables~\ref{table_m13abund_a} to \ref{table_m13abund_e}),
is taken as the variance about the mean abundance for a given species
in a given star, i.e. the  1$\sigma$ rms value about the mean
abundance of species $X$ in a given star/$\sqrt{N}$, where $N$ is the number of
observed lines of species $X$.  It includes contributions from
errors in the measured \eqw, random errors (i.e. between
lines of a given species) in the adopted $gf$ values, etc.
Some species, an example being Fe~I
with its very large value of $N$, 
have unrealistically small values of $\sigma(obs)$;
we adopt a minimum of 0.05 dex for this parameter.

The ratio $\sigma/\sigma(tot)$ is an indication of whether
there is any intrinsic star-to-star variation in [X/Fe].  A high value
of this ``spread ratio'', tabulated in the fifth column of
this table, suggests a high probability of intrinsic scatter for
the abundance of the species $X$. Ideally the mean $SR$ for those
elements with no star-to-star variation should be unity.  However, 
we use the entries in Table~\ref{table_sens} for the lowest \teff,
i.e. for giants near the RGB tip, to calculate $\sigma(tot)$.
Since the abundance sensitivities in general decrease as \teff\ increases,
we will slightly overestimate $\sigma(tot)$, and hence underestimate
$SR$, thus explaining why the deduced values for $SR$ tend
to be slightly less than 1. A second indication of the reality
of the star-to-star variation in [X/Fe] for any species $X$ is deduced
using a $\chi^2$ analysis, and evaluating the probability of exceeding
by chance
the measured value of $\chi^2$ from our sample of stars in a globular cluster.
Only O~I and Na~I have  values of $\chi^2$ much higher 
than the number of degrees of freedom ($N(star)-1$), where $N(star)$ is the
number of stars in which that species was detected.

Inspection of Table~\ref{table_abundsig} shows that for all but two species the
$SR$ in M3 ranges from 0.5 to 1.1, indicating little 
sign of an intrinsic star-to-star range
in abundance.  O~I and Na~I, however, have $SR$ exceeding 2.0 and their
$\chi^2$ are very large.
Note that $SR$  for Mg~I is 0.7, suggesting no real 
star-to-star abundance variations for this element in M3.  For M13,
O~I and Na~I again have by far the largest values of $SR$, exceeding 2.5
in both cases, with no other species having a value exceeding 1.5.
We therefore assume that the range of abundances seen in our M3 and in our M13
samples for Na~I and O~I represent real star-to-star abundance variations;
while no other element shows definite evidence for such variations
from this simple analysis, but see \S\ref{section_correlated}.

We have also examined whether one can discern a difference in the mean 
$<$[Fe/H$]>$ between stars with high and low O abundances in  M13
and between stars with high and low Na abundances in M13. 
No statistically significant difference was found.

\subsection{The Peculiar Star M13 I--5}

\begin{figure}
\epsscale{1.0}
\figurenum{15}
\plotone{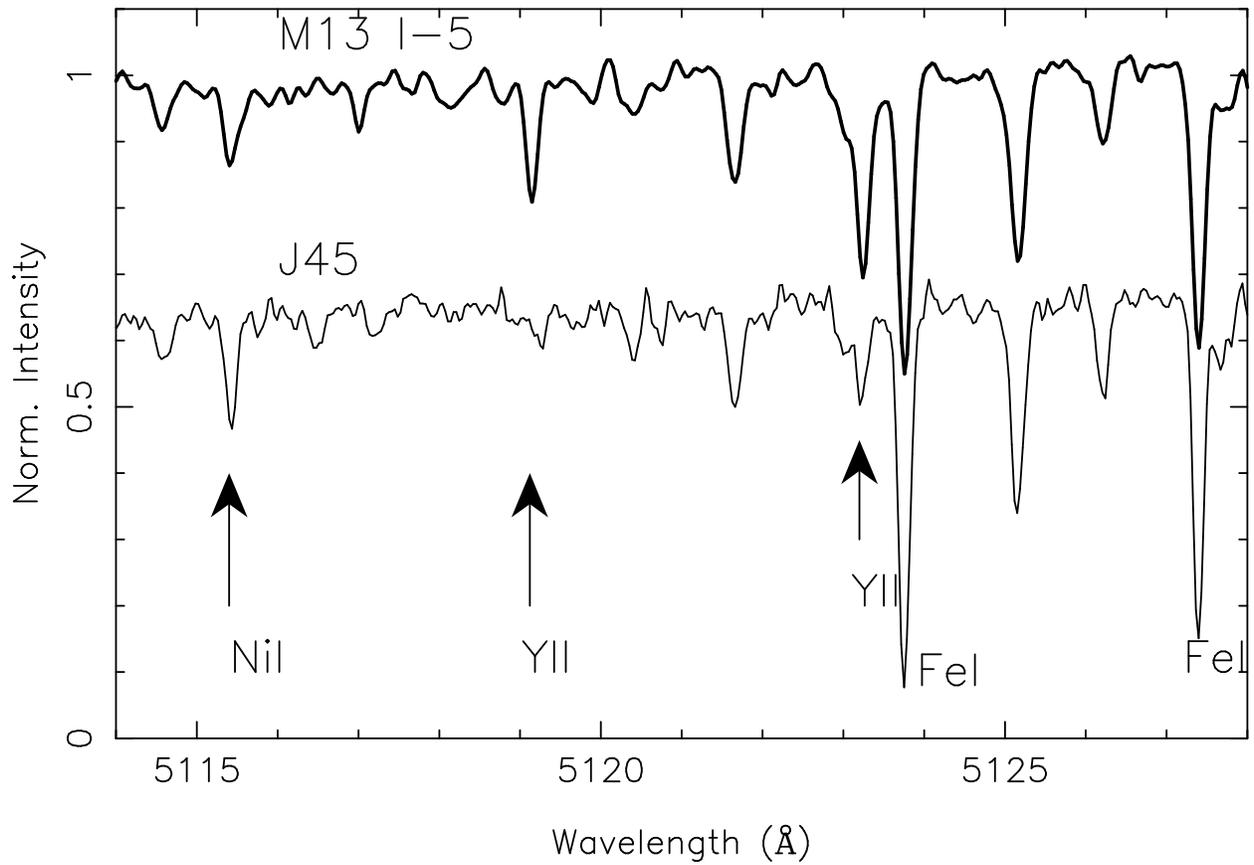}
\caption[]{A section of the spectrum of the Y-rich star M13 I--5 is shown
in the region of several Y~II lines.  The same region in the spectrum of star
M13 J45, a star 0.5 mag brighter in V along the RGB, hence
slightly cooler than star I--5, is shown for comparison.
\label{figure_m13_i-5}}
\end{figure}

With regard to the heavy elements, astute readers
will have noticed that there is one star (M13 I--5)
with anomalously high Y and Ba (see Table~\ref{table_m13abund_d}).  
This star, which is too
faint to have been included in any previous high dispersion analyses
in this globular cluster, stands out (and is marked)
in Figure~\ref{figure_m13_cu_zr} (Y panel) and in 
Figure~\ref{figure_m13_ba_dy} (Ba panel).  It has the highest abundance of 
each of these two elements in the entire M13 sample; 
its deduced $\epsilon$(Y) is
five times the mean of the remaining 24 stars in the M13 sample.
Y~II has the fourth largest
values of $SR$ ($SR = 1.27$) in M13, but eliminating this star would reduce it
to below 0.5.  Figure~\ref{figure_m13_i-5} 
shows a section of the spectrum of this star compared with one of
similar luminosity and \teff\ in M13.  This figure demonstrates
convincingly that the very high abundance of Y measured for star
M13 I--5 is definitely 
real. 

With regard to the other heavy elements, in the spectrum
of star M13 I--5 
there is one weak and uncertain detection for La~II, and several weak lines
ascribed to Nd~II; these give [La/Fe] at the upper end of those for the
other M13 stars and the largest (but only by $\sim$0.05 dex)
[Nd/Fe] for any star in our sample in M13. 
An upper limit to the 6645~\AA\ line of Eu~II of 10.5 m\AA\ yields an
upper limit to [Eu/Fe] of +0.7 dex.  Since the M13 mean for [Eu/Fe] is  +0.57 dex,
Eu is not significantly enhanced in this star, consistent with a $s$-process
enhancement.  A better spectrum for this
peculiar star has just been acquired, and results will be reported in
a future publication.

\subsection{Comparison With Previous Analyses}

Since M3 and M13 are key globular clusters, there have been several previous
high dispersion analyses of  the most luminous stars in them.
A comparison of the \teff\ determined in the present work 
with those from previous
investigations (most of which relied on B--V colors) 
for the stars in common in M3 and in M13 with the analyses of 
\cite{cohen78}, \cite{kraft92}, \cite{kraft97},
\cite{cavallo00} and \cite{sneden04} 
is shown in Table~\ref{table_teff_comp}.  The agreement is very gratifying.
No difference exceeds 100~K, and most are 50~K or less.  

A comparison
of the mean abundance ratios of large sample of bright giants 
in M3 (M13) analyzed
by \cite{sneden04} with our determinations is given in 
Table~\ref{table_compare}.  The results are encouraging.  
The differences
are given in the last column of the table. The deduced [Fe/H](Fe~I) between
the current work and that of \cite{sneden04} 
differ by 0.23 (0.13) dex, while the difference using 
Fe~II is only 0.12 (0.03) dex.
The largest difference in abundance ratios 
[X/Fe] for M3 is 0.24 dex (for Sc~II), with only 3 species (O~I, Mg~I and Sc~II) 
having differences exceeding 0.15 dex.
For M13, [O/Fe] and [Mg/Fe] both have differences exceeding 0.15 dex.
Part of the difference in O
arises as \cite{sneden04} adopt a Solar O abundance of 8.93 dex, 0.08 dex
higher than we do.  It could also in principle,
at least partially, be produced if the amplitude of the
star-to-star variations of O were a function of stellar luminosity
or equivalently \teff, given the differences in mean luminosity of the
two samples.  Other
differences in the details of the analysis may enter as well.
For example, the difference in Ca appears to be due to difference in the absolute
scaling of the Ca~I $gf$ values adopted by the Lick-Texas group and by us.
This may also play a role for Sc~II, but there are not sufficient
lines in common to be certain.  The HFS corrections for Sc~II in these stars
are probably not large enough to contribute significantly to this difference.

Because we are interested in star-to-star variations in abundance, we
also carry out a comparison of our derived
O and Na abundances for the individual stars in
common with the sample of bright giants of 
\cite{sneden04}, rather than
comparing the mean of the samples in M3 as was done in 
Table~\ref{table_compare}.   We choose to compare only to the most recent
detailed abundance analysis for stars in these two clusters, ignoring earlier
work, in the hopes of demonstrating good agreement.
The results are shown in Table~\ref{table_ab_comp_m3} (Table~\ref{table_ab_comp_m13});
the mean differences for each species have been removed, thus removing
the systematic differences in the analyses. 
The table reveals the scatter about the mean, i.e. the
non-systematic differences, whose variance is given as the final column
in each table. 
There are three stars in common in M3 and five in M13. 
For these stars in common, the Na abundances as analyzed by 
\cite{sneden04} and by
us are in extremely good agreement, ignoring a constant offset
between us and the Lick-Texas group, but the
O abundances are not.  Overall,
for M3, the agreement is pretty good, with $\sigma \le 0.15$ dex
for 9 of the 12 species;
only for [O/Fe], [Sc/Fe] and [Eu/Fe] is it larger.  O has only a few weak lines
in the relevant wavelength region, and difficulties in determining the O abundance
are notorious.  
[Eu/Fe] has essentially the same mean in our analysis of 
M3 and of M13 as was
found by \cite{sneden04}.  There is only one line used by both
analyses, that of Eu~II at 6645~\AA, which is very weak.
The derived Sc abundances might be affected by differences in the treatment of HFS.
The largest $\sigma$ is 0.22 dex, for [Eu/Fe].  In M13
two of the 8 differences have $\sigma > 0.3$ dex ([O/Fe] and [La/Fe]),
again two species with only a few weak lines.  (Sc is not included in
Sneden {\etal}'s analysis for M13, while there are no stars in common in M3
with [La/Fe] measurements.)

Having already removed the systematic differences, one might admit
``random'' variations in deduced abundance ratios 
of up to 0.15 dex as resulting from different assumptions made in these
two independent analyses.  However,
the larger differences found for  M13, with  $\sigma > 0.3$ dex
for two species, suggest that 
the abundance errors in one or both of these analyses are being underestimated.  
We have compared our measured \eqw\ with those of the Lick-Texas group, when
available.  The agreement is extremely good; \eqw(us: Keck) -- \eqw(Sneden et al 2004,
Keck) has a mean of 7\% with $\sigma$ = 7\%, while 
\eqw(us: Keck) -- \eqw(Kraft et al 1992, Lick)
has a mean of $-5$\% with $\sigma$ = 8\%\footnote{This good agreement suggests
that our adoption of the uncertainty in \eqw\ as 10\% is reasonable.}.

To summarize, there is very good agreement on a star-by-star basis for the derived
[Na/Fe] in M3 and in M13 for the sample of stars in common with \cite{sneden04}.  
The agreement in derived [O/Fe] is not as good,
but fortunately the star-to-star variations in O are large.  Furthermore, we
believe we understand the origin of most of the discrepancies, and believe
that our analysis is sound.
Hence we have some confidence
that we may proceed to analyze the abundance spreads in terms
of star-to-star variations in [X/Fe].

\section{Correlated Abundance Variations of the Light Elements 
\label{section_correlated}}

C, N, O, Na, Mg, and Al are known to show correlated abundance
variations from star-to-star among the most luminous stars 
in globular clusters;
see, e.g. the review of \cite{kraft94}.
Our simple spread ratio analysis (see \S\ref{section_spread}) shows 
definite star-to-star
variations in abundance of both O and Na in both M3 and in M13.
Variations in Mg, if present
are smaller and subtle.  Al is not effectively covered in 
our spectra\footnote{Our spectra only include the line at 5557~\AA,
which is very weak and only detected in a few stars; they 
do not include the most commonly used Al lines.}.

It is well established that O and Na are anti-correlated among luminous
giants in globular clusters, see, e.g. \cite{kraft94}.  Furthermore,
\cite{ramirez02} compiled the data from the literature, combined it
with their own, and showed that the same linear relation can be used
to fit the O and Na data for all globular clusters studied in detail
thus far.  The latest addition to the clusters studied in detail,
NGC~2808, by \cite{carretta04b}, does so as well.  The question
of interest is what happens when we look at lower luminosity stars.

\begin{figure}
\epsscale{0.9}
\figurenum{8}
\plotone{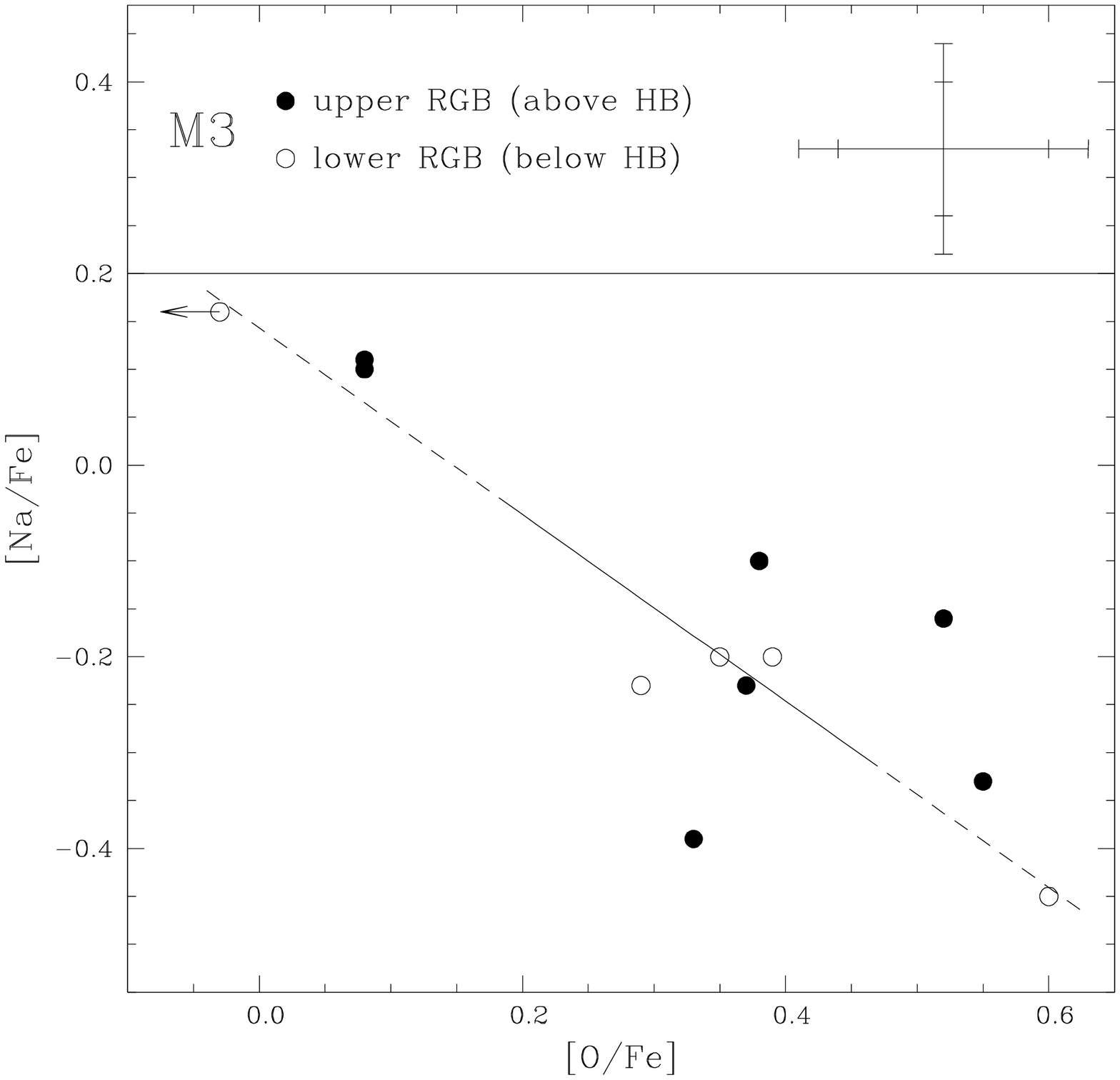}
\caption[]{The ratio [Na/Fe] is shown as a function of [O/Fe]
for our sample of 13 stars in M3. The filled circles denote
the luminous RGB stars; the open circles the lower luminosity giants.
The error bars typical of the most luminous and least luminous stars 
in our sample are indicated.
The line represents the relationship found by \cite{sneden04}, with a shift
of +0.07 dex in [O/Fe] applied;  the line is solid between the first and
third quartiles of his sample and is dashed outside that regime.
\label{figure_m3_ona}}
\end{figure}

Figure~\ref{figure_m3_ona} shows
the relationship between Na and O abundances (both with respect to Fe)
for our sample in M3.  Also superposed is the line representing 
the fit for this anti-correlation determined by \cite{sneden04} for the
luminous giants in M13, shifted by +0.07 dex in [O/Fe].
The first and last
quartiles of the O--Na anti-correlation seen by \cite{sneden04} in their
sample of luminous giants in M3 are indicated.
There is a reasonably clear anti-correlation even for the faintest stars
in our M3 sample, which agrees well
with that of \cite{sneden04}, given a small shift in O abundance scale. 
Thus the anti-correlation between O and Na persists in M3 from the RGB to 
at least V = 16.6 mag, which is about 2 mag fainter than the regime
over which \cite{sneden04} established this relationship. 
The anti-correlation between O and Na has approximately the same slope
and extends over approximately the same range among the fainter stars
in M3 as it does among the most luminous M3 giants.

Although \cite{sneden04}
claim a marginal detection of variations in [Mg/Fe] among their sample
of luminous giants in M3, with
increasing Mg abundance being correlated with increasing Na abundance,
we fail to find any credible evidence of such in M3; it might
be only slightly larger than $\sigma(tot)$.

\begin{figure}
\epsscale{0.9}
\figurenum{16}
\plotone{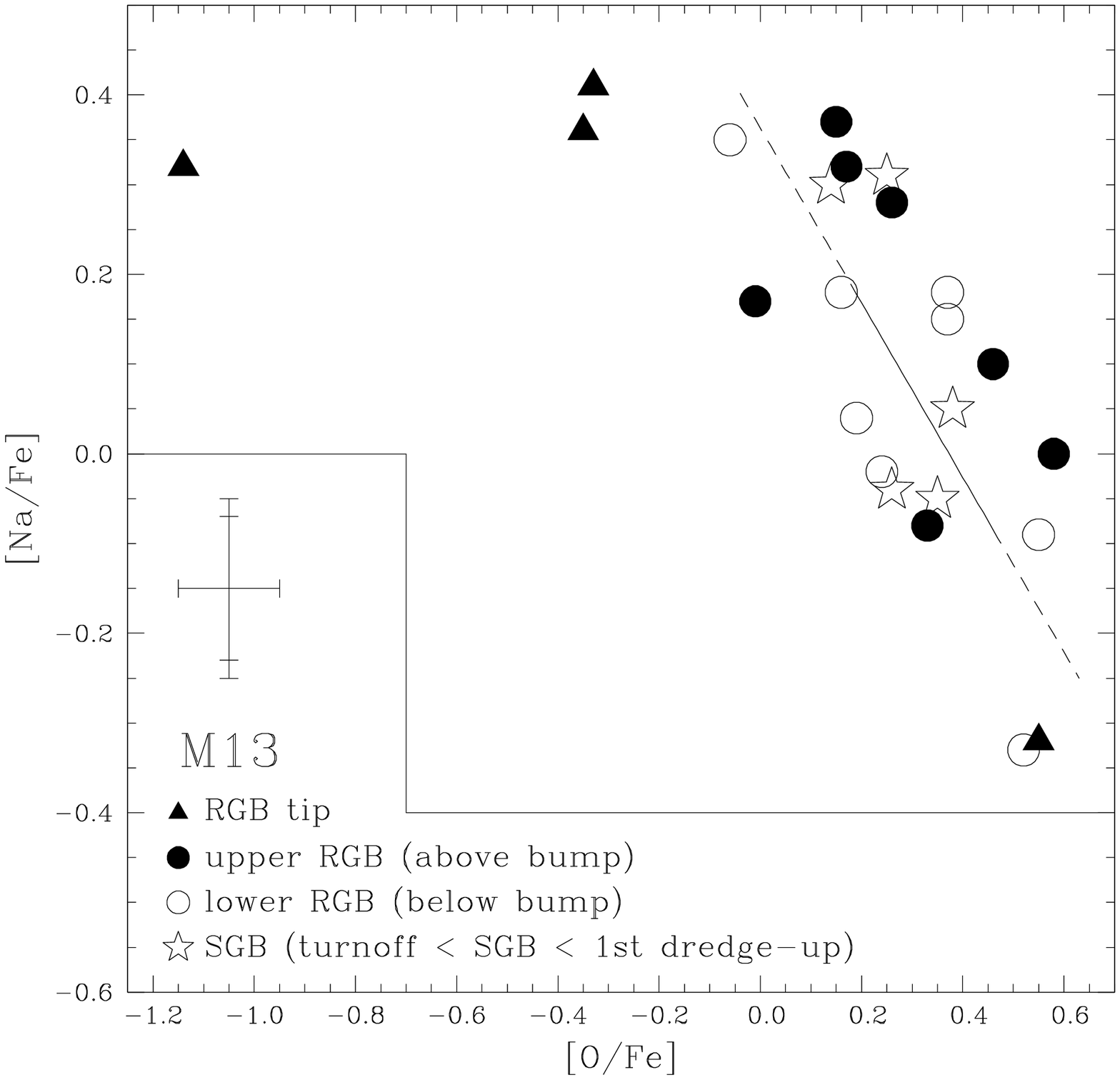}
\caption[]{The ratio [Na/Fe] is shown as a function of [O/Fe]
for 24 of our sample of 25 stars in M13. 
The error bars typical of the most luminous and least luminous stars 
in our sample are indicated.
The line represents the relationship found for M3, with a vertical offset of 
of +0.22 dex ;  the line extends over the range covered
by the sample of \cite{sneden04}, it is solid between the first and
third quartiles of their sample and is dashed outside that regime.
The different symbols denote the luminosity of the star.
\label{figure_m13_ona}}
\end{figure}

For M13, there is a much richer phenomenology.  Figure~\ref{figure_m13_ona} 
shows
the relationship between Na and O abundances (both with respect to Fe)
for our sample.  Also superposed is the line representing 
the fit for this anti-correlation we determined  for the
luminous giants in M3, shifted vertically by +0.22 dex.
The anti-correlation is immediately apparent.  It contains
a regime extending to very strong depletion of O at the highest
values of [Na/Fe], which is not seen in M3, but was seen by
\cite{kraft97} in M13.  This regime is populated only by 
the most luminous giants in M13.
It is also immediately apparent that the anti-correlation  holds
to the lowest luminosities probed in the present dataset, i.e. to
the regime just above the main sequence turnoff in M13.  
With the exception of the extremely O-depleted RGB tip stars, the amplitude
of the Na-O anti-correlation in M13 appears to be constant over the full range
of luminosities in the present sample.

\begin{figure}
\epsscale{0.9}
\figurenum{17}
\plotone{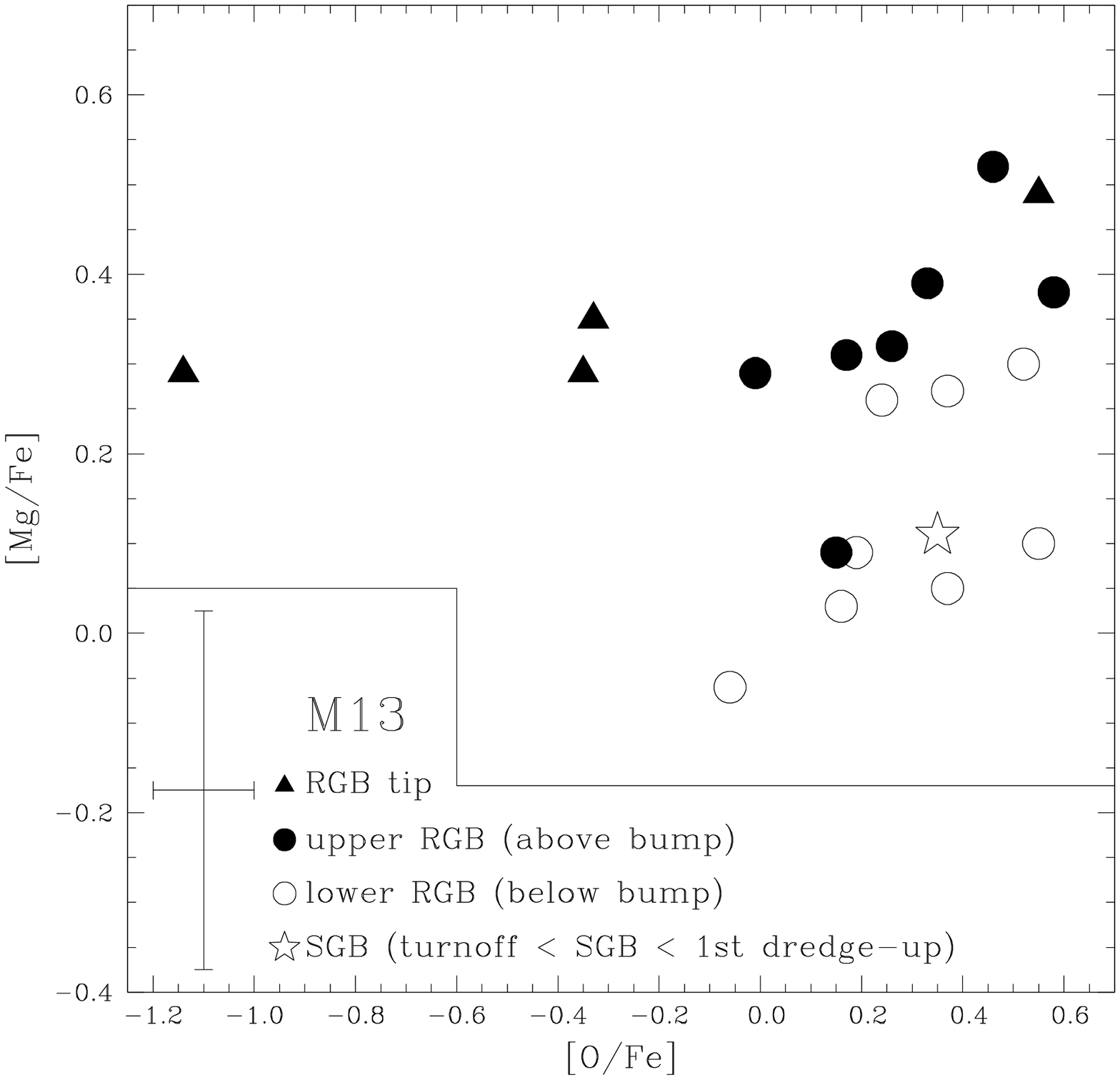}
\caption[]{The ratio [Mg/Fe] is shown as a function of [O/Fe]
for 20 of our sample of 25 stars in M13. 
The different symbols denote the luminosity of the star.
The error bars typical of the most luminous and least luminous stars 
in our sample are indicated.
\label{figure_m13_mgo}}
\end{figure}

\begin{figure}
\epsscale{0.9}
\figurenum{18}
\plotone{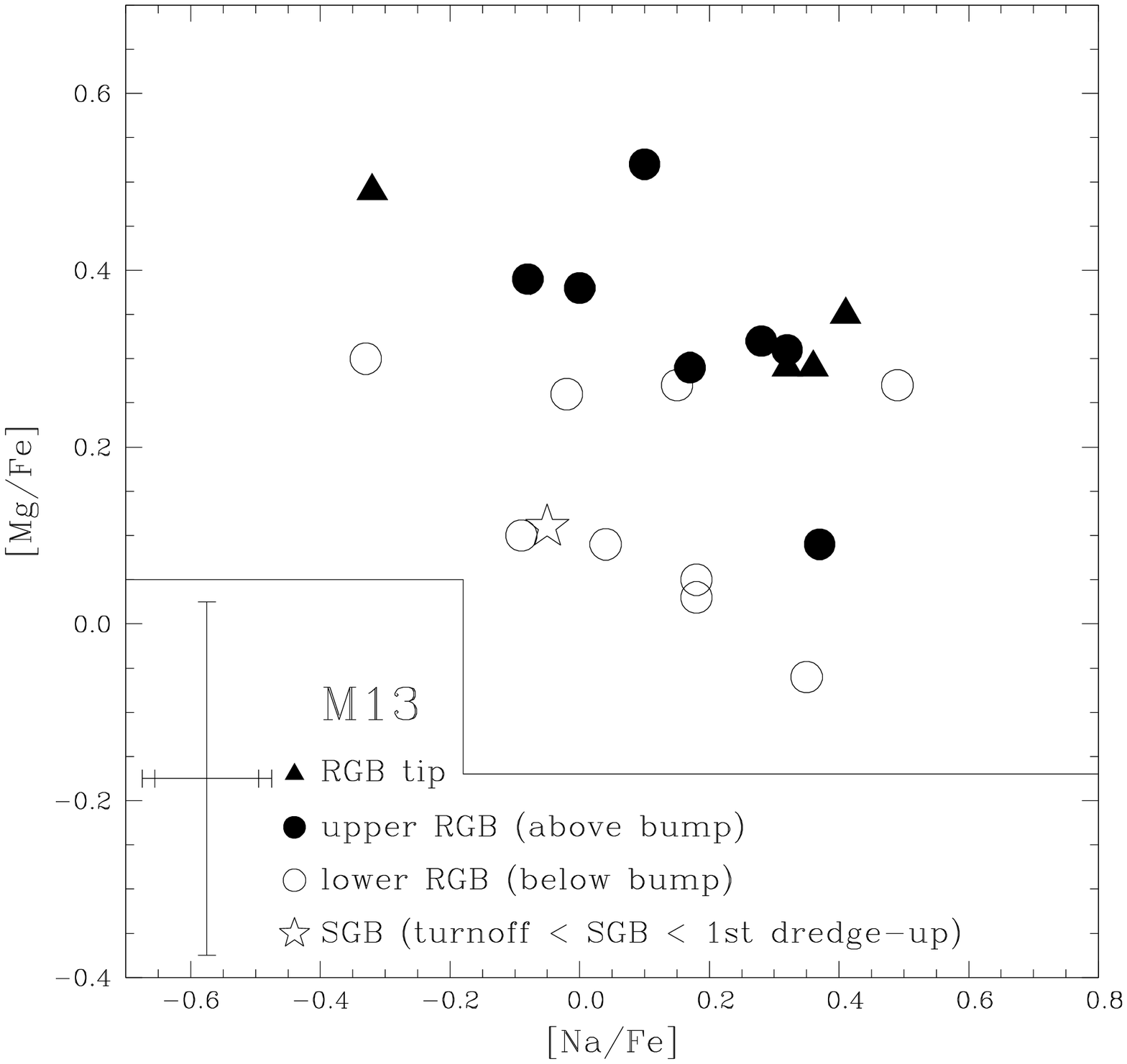}
\caption[]{The ratio [Mg/Fe] is shown as a function of [Na/Fe]
for 21 of our sample of 25 stars in M13. 
The different symbols denote the luminosity of the star.
The error bars typical of the most luminous and least luminous stars 
in our sample are indicated.
\label{figure_m13_mgna}}
\end{figure}

Figure~\ref{figure_m13_mgo} shows the behavior of Mg as a function
of [O/Fe].  The general previously observed small trend of increasing
Mg as O increases or Na decreases is present with an
amplitude of $\sim$0.4 dex,
as was found by \cite{kraft97}.  
This trend is present at all luminosities probed.  However
the mean abundance ratio [Mg/Fe] appears to decrease  with 
decreasing luminosity along the RGB.  The effect is small,
$\sim$0.2 dex in [Mg/Fe] over the range of our sample, but is
clearly seen in this figure and also in Figure~\ref{figure_m13_mgna},
which shows [Mg/Fe] as a function of [Na/Fe].  A similar difference
between [Mg/Fe] for dwarfs and for subgiants for stars in metal-poor
GCs was noted by
\cite{gratton01}.
We are using the same three Mg lines for each star
(the Mg triplet lines are not used), and in all
but the three faintest stars, all three lines are detected. We use 
the same atomic parameters throughout. Given the strong increase in
Mg~I line strength with luminosity along the RGB, a systematic
error in the choice of $v_t$ with \teff\ could in principle
produce 
this increase in Mg abundance with luminosity
along the M13 RGB.  However, a more likely culprit is non-LTE
effects in Mg.  These, as calculated by \cite{zhao00}
\citep*[see also][]{gehren04}  increase
with \teff\ and with decreasing [Fe/H] and have the right
amplitude and sign to produce this effect.

\subsection{Correlations of C, N and O}

Correlated C and N variations in M13 have been studied in detail
by \cite{briley02,briley04}. 
They determined C abundances
from the G band of CH for a large sample of stars reaching 
from the RGB to below the main sequence turnoff in M13.
Combining their results with the earlier work of
\cite{smith96} and \cite{suntzeff81} who cover the most luminous
RGB stars in M13, they evaluated
the variation from star-to-star of [C/Fe] from the RGB tip
to very low luminosities.
They  find big spreads in [C/Fe] at all luminosities probed, with 
[C/Fe] ranging from $-0.8$ to $-0.2$ dex among the lower luminosity giants.
There is a marked decline in [C/Fe] towards higher luminosities
among the upper RGB M13 stars, found originally by 
\cite{suntzeff81}.
Briley, Cohen \& Stetson also see a large range in [N/Fe] of
0.0 to +1.5 dex. 

\begin{figure}
\epsscale{0.9}
\figurenum{19}
\plotone{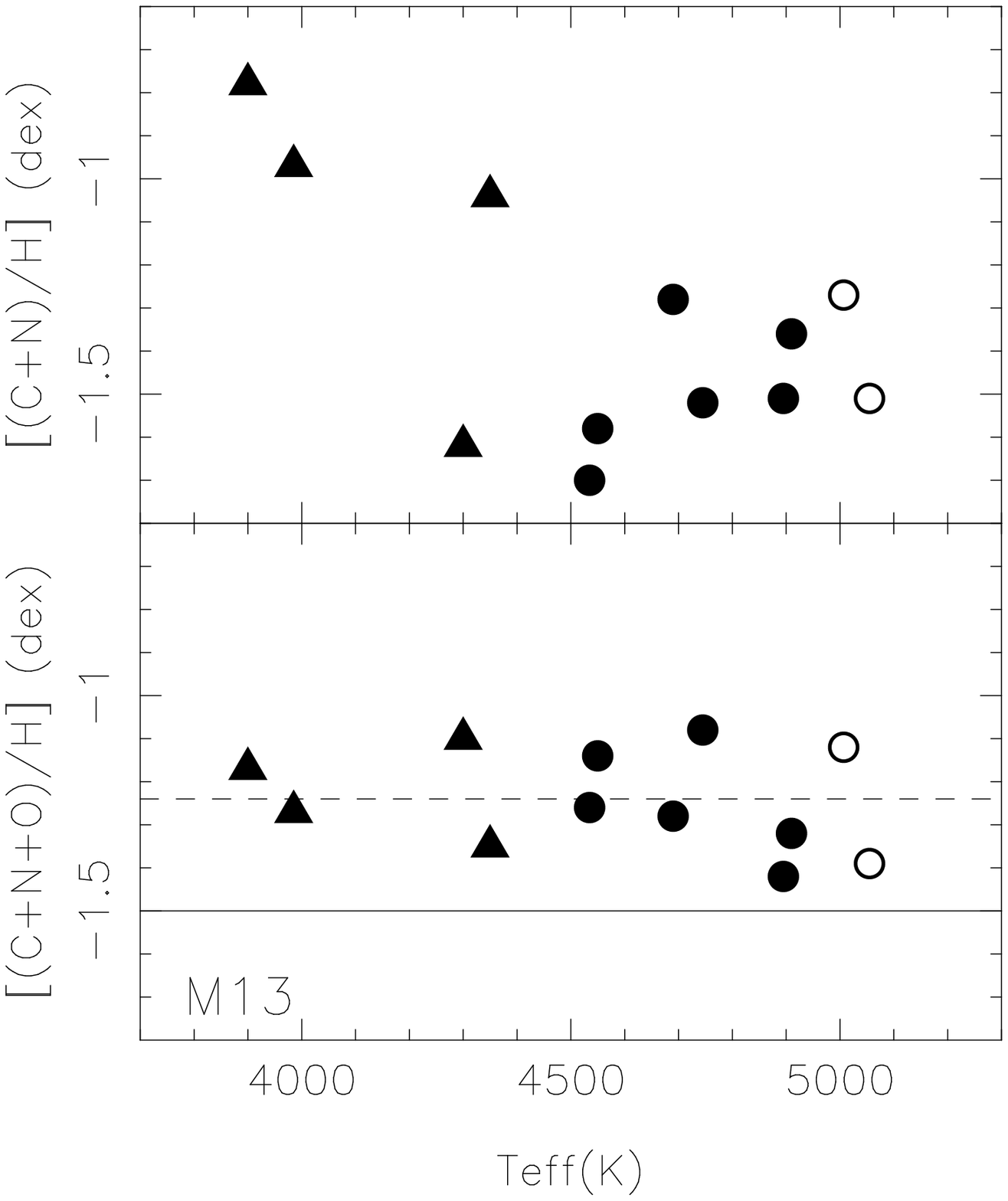}
\caption[]{The sum of C+N+O (lower panel) and of C+N (upper panel)
is shown as a function of \teff\ for 12 stars in M13.
The different symbols denote the luminosity of the star 
(as in Figure~\ref{figure_m13_ona}).  The dashed 
horizontal line in the lower panel is the mean of (C+N+O)/H, while
the solid horizontal line is the cluster mean for [Fe/H].
\label{figure_m13_cno}}
\end{figure}

A crucial test is to determine whether just CN
burning is occurring, C being transformed into N, in which case 
$\epsilon$(C+N) will be constant,
or if ON burning is also occurring.  In either case, the sum 
$\epsilon$(C+N+O) should be approximately constant, as nucleosynthesis
of even heavier elements proceeds only at very high temperatures.
Such a test was carried out by \cite{brown91} for a small sample of
stars; we carry it out for a considerably larger
sample of 5 (12) stars using our new
O abundances and the C and N abundances of \cite{smith96}
or \cite{suntzeff81}. 
Figure~\ref{figure_m13_cno} shows the result for M13;
log[$\epsilon$(C+N+O)] is constant at $-1.24$ dex  with a 
relatively small $\sigma$ of $0.12$ dex; this value is
$\sim0.3$ dex higher than [Fe/H] for M13.  The sum of C+N is not
constant, and O burning, especially near the RGB tip of M13, is required.
For M3, the necessary data exists for five stars, 
combining our new determination of [O/Fe] with the C,N
measurements of \cite{smith96}, \cite{suntzeff81}
or of \cite{bell80}.
We find a value of $-1.2$ dex for the mean of sum of
log[$\epsilon$(C+N+O)], identical to that of M13.

\subsection{Comments on Nucleosynthesis}

\begin{figure}
\epsscale{0.9}
\figurenum{20}
\plotone{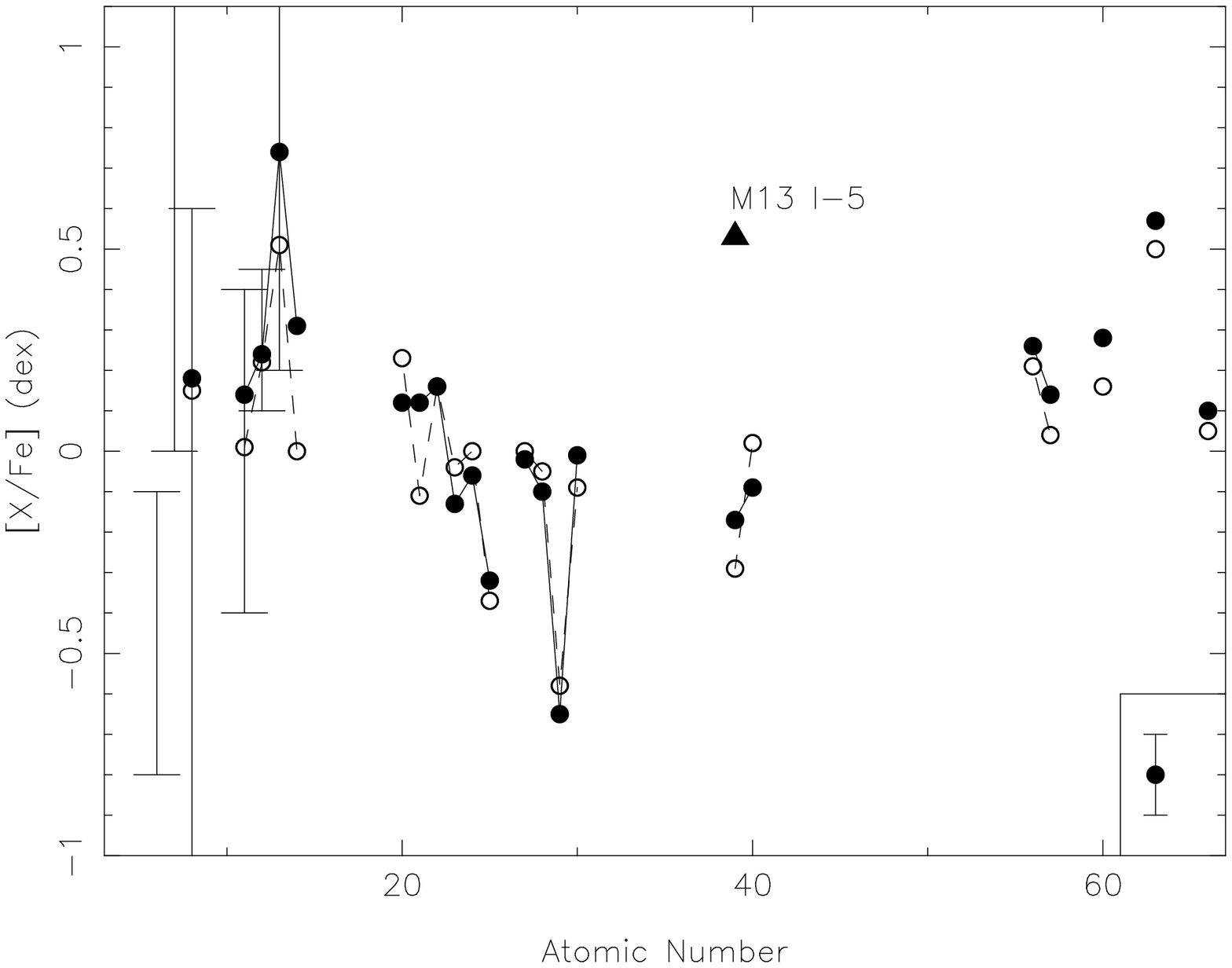}
\caption[]{The abundance ratios [X/Fe] are shown as a function of
atomic number for M3 and for M13.  Filled circles are used for M13 and open
circles for M3, while a filled triangle denotes the [Y/Fe]
for the peculiar star M13 I--5.  Elements with consecutive 
atomic numbers whose abundances
have been determined are indicated by solid (dashed) lines.  For those
elements which show star-to-star variation,
ranges are indicated for C, N (both from Briley, Cohen \& Stetson 2004),
O, Na, Mg, and Al, where the Al range is from \cite{sneden04}.  A typical
error bar is shown at the lower right.
\label{figure_atomic_number}}
\end{figure}

We now turn to what we can learn about the chemical history of
the two globular clusters M3 and M13 from our work.  
Figure~\ref{figure_atomic_number} shows the abundance [X/Fe]
we have derived for these two populous Galactic GCs. 
The light
elements are characterized by very large ranges of
star-to-star abundance variations.
Since the odd atomic number
elements among the light elements normally are of lower abundance
(in terms of $X$/H), a small increment of these elements
can lead to large enhancements in [X/Fe].  Such behavior is
seen throughout this range of atomic number.
The correlations among the elements showing such
star-to-star variations suggest that 
CN and, especially for M13, ON burning, as well NaMg burning
are required.
These spreads have been widely discussed in the literature for
the luminous giants in
these and other Galactic GCs.  

Our main contributions  are twofold.  
We have confirmed with quite high precision that 
the sum of C+N+O is constant  near the tip of the giant
branch as shown by \cite{smith96} but we extend this down to
the bump in the luminosity function.
Second we have demonstrated
that the correlations and anti-correlations among these light
elements extend with the same amplitude seen on the upper RGB
(but not with the extreme ratios characteristic of the tip of the RGB
in M13) 
to the faintest luminosities probed, which for our
sample in M13, is near the main sequence turnoff. 
The range of luminosities for our sample in
each of these two GCs reaches well below
their RGB bump, which is at V = 15.45 for M3 and at V=14.75 for M13
\citep{ferraro99}, and is where the first dredge
up is believed to begin.  Stars less luminous than this cannot
have mixed significantly according to current theory
\citep*[see, e.g.][]{charbonnel95,pinsonneault97,palacios03}.

This behavior
demonstrates yet again that intrinsic nuclear processes within
these low mass, and in some cases relatively unevolved, stars is
not capable of explaining the observed phenomena.
In this context, we note the work of \cite{gratton00} on 
metal-poor field stars, which do not show these phenomena,
but rather a much more orderly and smaller amplitude change
in abundance ratios with luminosity.   Thus these phenomena
seem restricted to globular clusters, where the stellar density
is high, and the gas density presumably was also quite high in
the past.  Since internal nucleosynthesis has been ruled out,
some form
of external pollution onto the low mass stars we 
currently see in GCs is
required.  This could be from a companion AGB star or from 
mass loss into the ISM of the cluster itself
by higher mass stars presumably in the AGB phase, followed by
accretion as the low mass star passes through the denser regions near the
cluster core, or by incomplete mixing of the cluster ISM
prior to the formation of the generation of stars we see today.
The problems with this general type of mechanism have been
discussed at length by many, see, e.g. \cite{cohen02}, and
basically involve whether enough mass can be accreted,
whether with the additional accreted material the observed abundance
ratios can be reproduced
\citep*[see, e.g.][for a discussion regarding massive AGB stars]{fenner04}, and over what mass zones
and how deeply down from the surface does the 
accreting star mix and hence dilute the accreted material.

All this exotic behavior 
ceases with Si, whose abundance appears to be constant, with a small range
consistent with the observational and modeling uncertainties.
The regime from Si through the Fe-peak is characterized by no detectable 
star-to-star variations, and by strong over-depletion of the
odd atomic number elements, expected for Fe-peak nuclei
from considerations  of explosive nucleosynthesis
\citep{arnett71,arnett96}.  (Recall that among the lightest 
elements, the odd atomic numbers show large enhancements in [X/Fe].)

The heaviest elements are not well sampled by our data.  However,
the [Ba/Eu] ratio is $-0.35$ ($-0.31$) dex in M3 (M13), which value is intermediate
between the Solar ratio and that of the pure $r$-process, presumably
reflecting the increased dominance of the $r$-process contribution
at low metallicities.

The heaviest elements also show the first signs of star-to-star
variations again, not surprising since their abundances
$X$/H are so low that any small addition of material could
raise [X/Fe].  In particular, M13 I--5 shows strong excesses of Y and of Ba,
presumably from a $s$-process event, but we need, and have already
obtained, better spectra
with more complete wavelength coverage to verify this.  Ignoring
$\omega$ Cen\footnote{We ignore 
$\omega$ Cen in discussing star-to-star variations among
globular clusters
for the rest of this paper as it has
been known for more than 20 years to have a wide range of
heavy element abundance 
\citep*[see, e.g. the latest such work,][]{pancino02} and has
been repeatedly suggested recently as the remnant of the nucleus
of an accreted dwarf galaxy.}, star-to-star variations among
the heavy elements
have been previously detected only in M15 \citep{sneden97}, and 
there they appear quite different in character.

\section{The Evolution of Abundances Within the GC System}

While it is clear that the GCs differ from the field stars
in the amplitude of their star-to-star variation of the light elements,
we need  to establish whether this difference persists 
in their average abundance ratios.  If so, this would provide
evidence for a difference in the chemical history and/or
formation mechanisms for globular
clusters from those of the halo field stars.
With the advent of our program at the Keck Observatory and similar
programs at ESO using UVES, there are now a substantial
number of Galactic GCs for which detailed abundance analyses
using high precision, high resolution spectra have been published.  
We collect those carried out at Keck and at the VLT, and add in
only recent analyses of relatively nearby
GCs using 4-m telescopes.  We impose a minimum of 4 stars per cluster. 
Ignoring the GCs associated with the Sgr dwarf galaxy,
we find 13 clusters with suitable analyses,
which in order of decreasing metallicity are 
NGC 6528 \citep{carretta01},
NGC~6553 \citep{cohen99,carretta01}, 
47 Tuc \citep{carretta04a,james04b},
M71 \citep{ramirez02,ramirez03}, 
M5 \citep{ramirez03}, 
NGC~288 \citep{shetrone00},
NGC~362 \citep{shetrone00}, 
NGC~6752 \citep{james04a},
M3 \citep*[this paper, see also][]{sneden04}, 
M13 \citep*[this paper, see also][]{sneden04},
NGC~7492 (Cohen \& Melendez, in preparation),
NGC~6397 \citep{thevenin01,gratton01,james04b} and
M15 \citep{sneden97}.
There are at least 10 more Galactic GCs which have been observed
at Keck or at the VLT within the past two years with analyses in
progress, so a significant fraction of the total Galactic population
of GCs has been covered.

This strict selection of GC analyses
guarantees the maximum possible accuracy of the abundance ratios, without
of course guaranteeing consistency between the various analyses.
Although the first author has been associated with seven of the 13,
we have not tried to homogenize the details of the procedures
adopted by other groups. 

To characterize the behavior of the metal-poor halo field stars, we
adopt abundance ratios from recent large surveys of such
by \cite{gratton91},
\cite{mcwilliam95}, \cite{fulbright00}, \cite{nissen00} and
by \cite{johnson02}.  No effort has been made to homogenize
these analyses either, but since they were carried out over
the course of more than a decade, we have corrected for the difference in the Solar
Fe abundance adopted by each.

For both the field star surveys and the GCs, we have 
looked at the differences
between the transition probabilities used for each of these analyses,
and set them to the same absolute scale.   Such differences are
small for most species, but are large (up to 0.2 dex) for Ca~I,
with substantial scatter in the difference from line to line.

\begin{figure}
\epsscale{0.9}
\figurenum{21}
\plotone{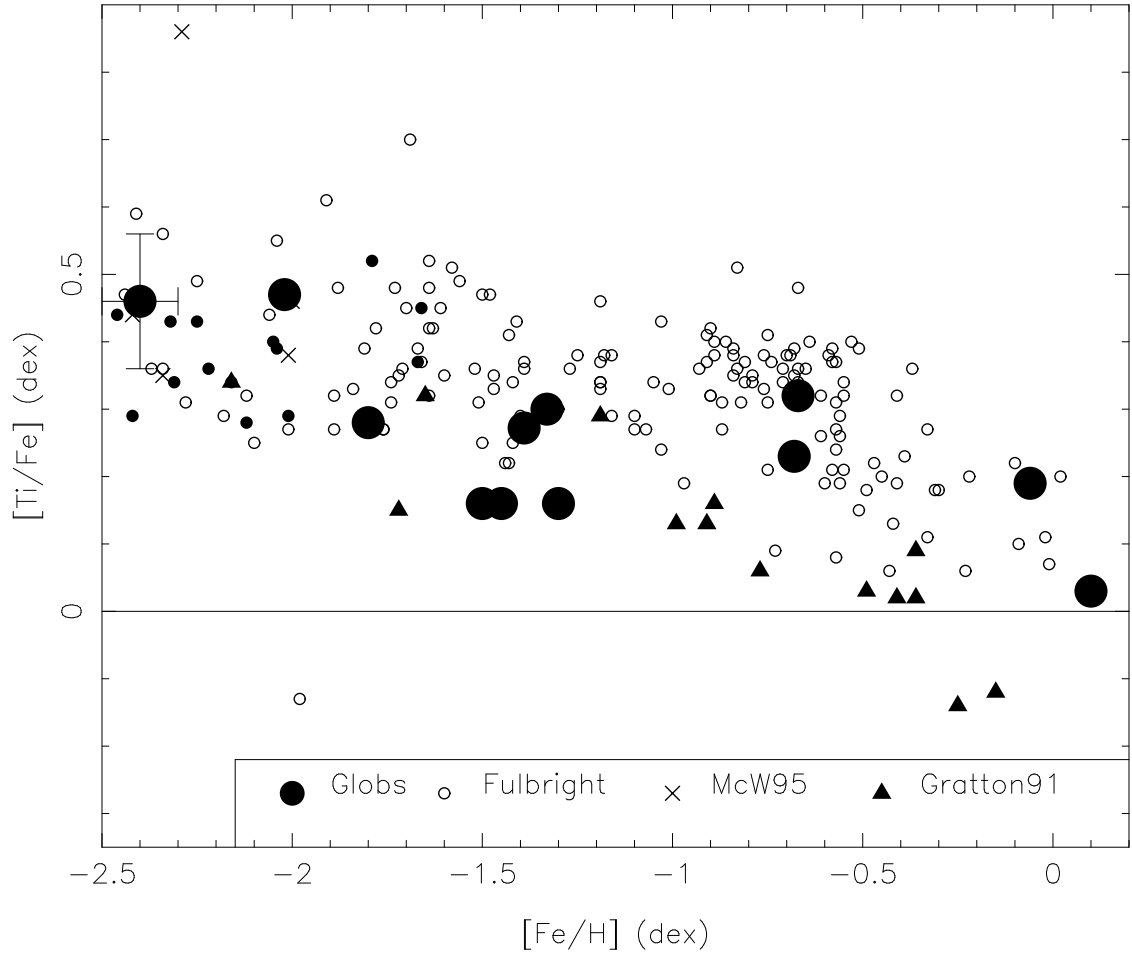}
\caption[]{The abundance ratio [Ti/Fe] is shown as a function of
[Fe/H] for a sample of 13 Galactic GCs (see text for references),
indicated as large filled circles.  This
is compared to the same relationship for halo field stars (sources
and symbols indicated on the figure). An 
error bar typical of the GCs is shown for the lowest metallicity GC.
\label{figure_ti_all}}
\end{figure}

\begin{figure}
\epsscale{0.9}
\figurenum{22}
\plotone{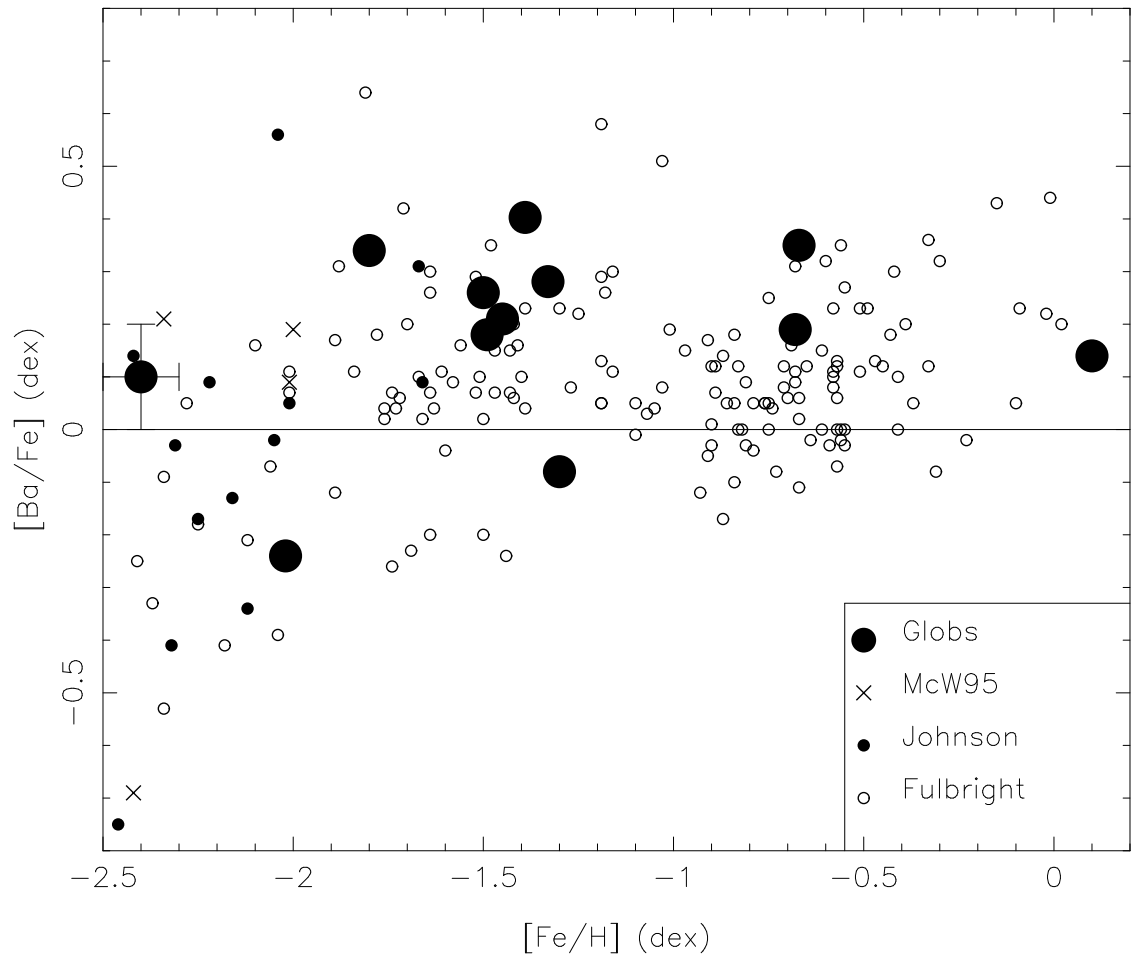}
\caption[]{The same as Figure~\ref{figure_ti_all}, but for [Ba/Fe].
\label{figure_ba_all}}
\end{figure}

When one examines the resulting relations between [X/H] and [Fe/H]
for the Galactic globular clusters and for the halo field stars, one
is struck by the similarity for many elements of the trends between the two
systems, present in spite of completely independent analyses on
very different samples (the halo stars being mostly dwarfs, while
the GC samples focus on giants in many cases).  Two typical cases
are illustrated; for Ti (Figure~\ref{figure_ti_all}) and for
Ba (Figure~\ref{figure_ba_all}), as for most
elements, the relations of abundance ratio with metallicity of the GCs
and of the halo field are indistinguishable.  

\begin{figure}
\epsscale{0.9}
\figurenum{23}
\plotone{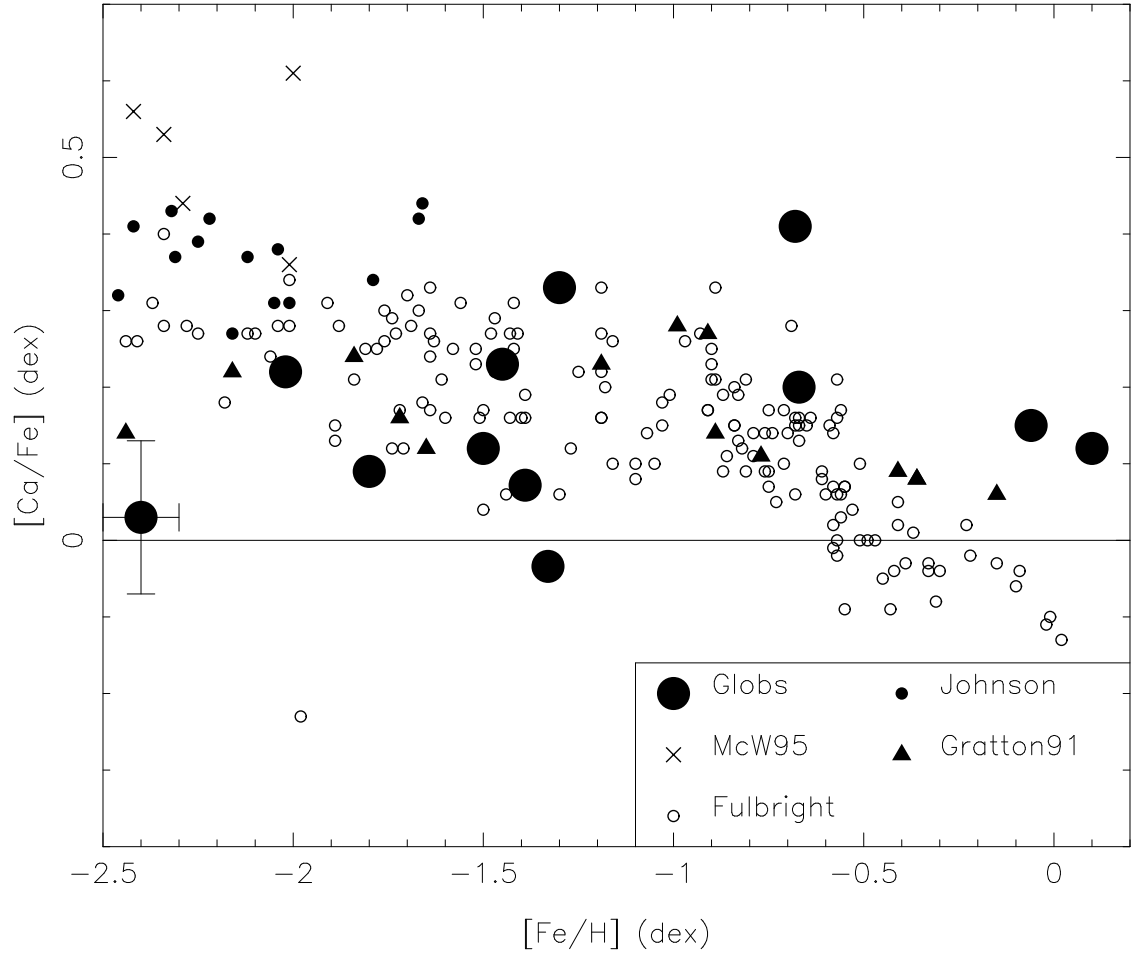}
\caption[]{The same as Figure~\ref{figure_ti_all}, but for [Ca/Fe].  The $gf$ values
adopted by the various groups have been adjusted to a common absolute scale.
\label{figure_ca_all}}
\end{figure}

The most credible difference we notice is shown by the $\alpha$-elements Mg
and Ca (we show that of [Ca/Fe] versus [Fe/H] in Figure~\ref{figure_ca_all}). 
The most metal poor GC appears to be deficient in Mg and in Ca
at [Fe/H] $\le -1.5$ dex by $\sim$0.3 dex as compared to the field.
One explanation for this could be that the luminous giants usually
analyzed in the GCs are slightly depleted in Mg, an element
which shows modest star-to-star variations in GCs, or perhaps
non-LTE effects in Mg are playing a role.  The notoriously uncertain
Mg $gf$ values could also be relevant if the same lines are not
used by all the groups, and there are also problems in the $gf$ values for
Ca~I at the level of  $\sim$0.2 dex which may not have been completely
removed.
This difference in behavior between the GCs and the halo field
is at present smaller than 2$\sigma$ and involves just the most metal poor GC.
We await 
publication of analyses for more of the most metal deficient halo
GCs to confirm the reality of this potential difference.

\section{Summary}

We have carried a detailed abundance analysis for 21 elements 
in a sample of 25 stars with a wide range in 
luminosity from luminous giants to stars near the main sequence turnoff
in the globular cluster M13 ([Fe/H] $-1.50$ dex)
and in a sample of 13 stars distributed 
from the tip to the base of the RGB in the globular cluster M3
([Fe/H]$ -1.39$ dex).
The analyzed spectra, obtained with HIRES at the Keck Observatory,
are of high dispersion (R=$\lambda / \Delta \lambda$=35,000).
Most elements, including Fe, but excluding the elements
lighter than Si, show no trend in abundance
ratio [X/Fe] with \teff, 
and scatter around the mean between the
top of the RGB and near the main sequence turnoff consistent
with observational uncertainties.   This suggests
that at this metallicity, non-LTE effects and
gravitationally induced heavy element diffusion
are not important for this set of elements over the range
of stellar parameters spanned by our sample.

The elements lighter than Si that have been studied in detail, i.e. C and N
\citep*[by][]{briley02,briley04}, O, Na, Mg, and Al 
\citep*[see][]{sneden04} all show strong star-to-star variations and
correlations among each other.
We have detected an anti-correlation between O and Na abundances, 
observed previously only
among the most luminous RGB stars in both of these clusters.  We find
these anti-correlations to persist 
in both M3 and in M13 over the full range of luminosity of our
samples, i.e. in the case of M13
to near the main sequence turnoff.  M13 shows a larger range
in both O and Na abundance than does M3 at all luminosities, 
in particular having a few stars at 
its RGB tip with very strongly depleted O.

We detect a correlation between Mg abundance and O abundance among the
stars in the M13 sample, but no credible star-to-star variation in [Mg/Fe]
within the M3 sample.  We also find a decrease in the mean Mg abundance
as one moves towards lower luminosity, which we tentatively suggest is due
to ignoring non-LTE effects in Mg.

Although CN burning must be occurring in both M3 and in M13, and ON
burning is required for M13,
we have confirmed with quite high precision that 
the sum of C+N+O is constant,
log[$\epsilon$(C+N+O)] = $-1.24$ dex  with a 
relatively small $\sigma$ of $0.12$ dex, previously shown
near the tip of the giant
branch by \cite{smith96} for luminous giants, but we extend this down to
the bump in the luminosity function.
The same holds true for a smaller
sample in M3, with somewhat larger variance.

We have shown that these star-to-star abundance variations 
among the light elements continue
well below the RGB bump in both M3 and M13.  The low luminosity at which
these phenomena are now detected in M3 and M13, and from our previous
work, in M71 and in M5 \citep{ramirez03}, as in other published
analyses \citep*[see, e.g.][]{gratton01,carretta04a} has effectively ruled out
the possibility of generating the spreads through
internal nucleosynthesis and mixing within the stars we observe today.  Instead some
form of external pollution involving a previous generation of stars, combined
with binaries or accretion of gas from the cluster ISM, must be involved.
But as discussed most recently by \cite{fenner04}, the details do not fit (yet).

Star I--5 in M13 has large excesses of Y and of Ba, with no strong
enhancement of Eu, suggesting an
$s$-process event contributed to its heavy element abundances.  This is
the first star we have found in the present long term effort
that shows any credible deviation from the cluster mean
for any heavy element.  

The mean abundance ratios we derive for M3 and for M13 are identical 
to within the errors.
They show the typical pattern of scatter among the light elements, with
the odd atomic number elements appearing enhanced, then no star-to-star 
variations
among the Fe-peak elements, where the odd atomic number elements are excessively
depleted.  The mean [Eu/Ba] ratio is essentially the same in both clusters;
it is intermediate between the Solar ratio and that of the $r$-process,
suggesting the additional $r$-process contribution characteristic of metal-poor
populations.  It does not appear possible to explain the significant differences
in horizontal branch characteristics of M3 and M13, 
the classic second parameter pair,
through differences in abundances (unless He is the culprit), since
we have shown that these two
clusters have essentially
identical values of [Fe/H] and of mean [X/Fe], for all elements studied here.

The abundance ratios for 13 Galactic globular clusters with recent detailed
abundance analyses, obtained by combining our samples
with published data, are compared to those of published large surveys
of metal-poor halo field stars.  For most elements, the agreement is
very good, suggesting a common chemical history for the halo field and for
the Galactic globular clusters.  

We see yet again that the abundances of the 
Fe-peak elements in M3 and in M13, which are rather massive globular
clusters ($\sim6 \times 10^5 M$\subsun)
are single valued, with
extremely narrow peaks. 
It is ironic that such intense theoretical and observational effort has in recent
years focused on the correlations and anti-correlations among the light
elements C, N, O, Na, Mg and Al in globular clusters, and the mixing and
nuclear processes that might produce these, and so little has focused
on the amazingly narrow range of abundances of the Fe-peak elements
characteristic of the Galactic globular clusters.   

\acknowledgements
The entire Keck/HIRES user communities owes a huge debt to 
Jerry Nelson, Gerry Smith, Steve Vogt, and many other 
people who have worked to make the Keck Telescope and HIRES  
a reality and to operate and maintain the Keck Observatory. 
We are grateful to the W. M.  Keck Foundation for the vision to fund
the construction of the W. M. Keck Observatory. 
The authors wish to extend special thanks to those of Hawaiian ancestry
on whose sacred mountain we are privileged to be guests. 
Without their generous hospitality, none of the observations presented
herein would have been possible.
This publication makes use of data from the Two Micron All-Sky Survey,
which is a joint project of the University of Massachusetts and the 
Infrared Processing and Analysis Center, funded by the 
National Aeronautics and Space Administration and the
National Science Foundation.
We are grateful to the National Science Foundation for partial support under
grant AST-0205951 to JGC.  

\clearpage


%
%
\clearpage
\begin{deluxetable}{lrrrrr}
\tablenum{1a}
\tablewidth{0pt}
\tablecaption{The Sample of Stars in M3 \label{table_sample}}
\tablehead{
\colhead{ID\tablenotemark{a}} & 
\colhead{V\tablenotemark{b}} &
\colhead{Date Obs.} & 
\colhead{Exp. Time} &
\colhead{SNR\tablenotemark{c}} &
\colhead{$v_{r}$} \\ 
\colhead{} & 
\colhead{(mag)} &   
\colhead{} & 
\colhead{(sec)} &
\colhead{} & \colhead{(\kms)} }
\startdata 
VZ1397    & 12.65 & 6/2003 & 600  & $>$100  & -146.1 \\
II-46     & 12.68 & 6/2003 & 600  & $>$100  & -151.4 \\
III-28    & 12.73 & 6/2003 & 800  & $>$100  & -154.0 \\
VZ1000    & 13.01 & 6/2003 & 400  & $>$100  & -151.8 \\
IV-25     & 13.60 & 6/2003 & 1000 & $>$100  & -150.4 \\
C41303\_2217 & 13.75 & 6/2003 & 1000 & $>$100  & -144.5 \\
IV-27     & 13.95 & 6/2003 & 1000 & $>$100  & -151.4 \\
C41543\_2334 & 15.71 & 5/2001 & 4800 &  $>$100  & -139.9 \\
C41543\_2334 & 15.71 & 6/2003 & 7200 & 100  & -145.7 \\
III-61    & 16.33 & 5/2001 & 3600 &  80      & -148.7 \\
III-60    & 16.44 & 5/2001 & 3600 &  75    & -141.4 \\
V-30      & 16.58 & 5/2001 & 4800 & 85      & -143.6 \\
V-30      & 16.58 & 6/2003 & 8400 & 90     & -143.7 \\
C41544\_2336 & 16.65 & 5/2001 & 4800 & 80   & -150.4 \\
C41544\_2336 & 16.65 & 6/2003 & 7200 & 90  & -153.8 \\
V-31      & 16.65 & 5/2001 & 4800 & 80       & -138.6 \\
V-31      & 16.65 & 6/2003 & 8400 & 90     & -149.8 \\
\enddata
\tablenotetext{a}{Identifications are from \cite{sandage53} or from
\cite{vonzeipel08}, or are
based on J2000 coordinates (Cmmsss\_mmsss means R.A. 13 mm ss.s, Dec. +28 mm ss.s)}
\tablenotetext{b}{V photometry from \cite{sandage53}, \cite{buonanno94}, 
\cite{ferraro97}, \cite{stetson00}}
\tablenotetext{c}{Signal to noise ratio in the continuum near
6380~\AA\ for the 2001 spectra and near 5865~\AA\ 
for the 2003 spectra per 4 pixel spectral resolution element.}
\end{deluxetable}


%
%
\clearpage
\begin{deluxetable}{lrrrrr}
\tablenum{1b}
\tablewidth{0pt}
\tablecaption{The Sample of Stars in M13 \label{table_sample_m13}}
\tablehead{
\colhead{ID\tablenotemark{a}} & 
\colhead{V\tablenotemark{b}} &
\colhead{Date Obs.} & 
\colhead{Exp. Time} &
\colhead{SNR\tablenotemark{c}} &
\colhead{$v_{r}$} \\ 
\colhead{} & 
\colhead{(mag)} &   
\colhead{} & 
\colhead{(sec)} &
\colhead{} & \colhead{(\kms)} }
\startdata 
II$-$67         & 12.12 & 8/2003 & 200  & $>$100  & $-$244.0 \\
IV$-$25         & 12.19 & 8/2003 & 200  & $>$100  & $-$252.3 \\
II$-$76         & 12.51 & 6/2003 & 400  & $>$100  & $-$240.7 \\
III$-$18        & 12.77 & 6/2003 & 800  & $>$100  & $-$234.9 \\
K188          & 13.39 & 6/2003 & 800  & 100  & $-$246.3 \\
III$-$7         & 13.47 & 8/2003 & 1200 & $>$100  & $-$261.5 \\
I$-$18          & 13.91 & 6/2003 & 1000 & $>$100  & $-$247.1 \\
I$-$49          & 14.03 & 8/2003 & 1200 & $>$100   & $-$261.2 \\
\enddata
\tablenotetext{a}{Identifications are from \cite{arp55}, 
\cite{kadla66}
or, if not included in the above,
based on J2000 coordinates 
(Cmmsss\_mmsss means R.A. 16 mm ss.s, Dec. +36 mm ss.s)}
\tablenotetext{b}{V photometry from 
Stetson's database \citep{stetson98,stetson00},
from a 1990 update of \cite{cudworth79b} (private communication from K.~Cudworth),
or from \cite{rey01}.}
\tablenotetext{c}{Signal to noise ratio in the continuum near
6380~\AA\ for the 2001 spectra and near 5865~\AA\ 
for the 2003 spectra per 4 pixel spectral resolution element.}
\tablenotetext{d}{$v_r$ for this star not measured due to technical issues.
Star is a member of M13.}
\end{deluxetable}

%
%
\clearpage
\begin{deluxetable}{lclc}
\tablenum{2}
\tablewidth{0pt}
\tablecaption{Adopted Solar Abundances
\label{table_sun}}
\tablehead{
\colhead{Element} & \colhead{[X/H]\tablenotemark{a}} & 
\colhead{Element} & \colhead{[X/H]\tablenotemark{a}} \\ 
}
\startdata
O & 8.85 & Fe & 7.45 \\
Na & 6.33 & Ni & 6.25 \\
Mg & 7.54 & Cu & 4.21 \\
Al & 6.47 & Zn & 4.60 \\
Si & 7.55 & Ba & 2.13 \\
Ca & 6.36 & Y & 2.24 \\
Sc & 3.10 & Zr & 2.60 \\
Ti & 4.99 & La & 1.14 \\
V & 4.00 & Nd & 1.45 \\
Cr & 5.67 & Eu & 0.51 \\
Mn  & 5.39 & Dy & 1.10 \\
\enddata
\tablenotetext{a}{Given on a scale where log(N(H))=12.0; values in dex.}
\end{deluxetable}

%
%
\clearpage
\begin{deluxetable}{lcc}
\tablenum{3}
\tablewidth{0pt}
\tablecaption{New IR Photometry for the Fainter Stars
\label{table_irphot}}
\tablehead{
\colhead{ID\tablenotemark{a}} &
\colhead{$J$ } & \colhead{$K$} \\
\colhead{} &
\colhead{(mag)\tablenotemark{b}} &
\colhead{(mag)\tablenotemark{b}} \\
}
\startdata
\multicolumn{3}{c}{\bf M 3} \\
C41544\_2336 & 15.16 & 14.68 \\
C41543\_2334 & 14.67 & 14.38 \\
III--60 & 15.03 & 14.52 \\
III--61 & 14.92 & 14.41 \\
V--31 & 15.33 & 14.74 \\
V-30 & 15.18 & 14.61 \\
~ \\
\multicolumn{3}{c}{\bf M 13} \\
C41134\_3056 & 15.73 & 15.32 \\
C41135\_3053 & 16.29 & 15.91 \\
C41101\_3050 & 15.88 & 15.40 \\
C41099\_3046 & 15.88 & 15.44 \\
C40539\_2813 & 16.49 & 16.08 \\
C40535\_2819 & 16.58 & 16.19 \\
C41135\_2753 & 16.87 & 16.44 \\
C41133\_2750 & 16.30 & 15.86 \\
C41148\_3103 & 14.87 & 14.30 \\
C41155\_3103 &  13.64 & 13.09 \\
C40559\_2839 & 12.84 & 12.20 \\
C41196\_2632 & 12.84 & 12.20 \\
C41195\_2635 & 14.72 & 14.82 \\
\enddata
\tablenotetext{a}{Identifications as in notes to Table~\ref{table_sample}
(\ref{table_sample_m13}).}
\tablenotetext{b}{1$\sigma$ uncertainties are $\pm$0.03 mag.}
\end{deluxetable}

%
%
\clearpage
\begin{deluxetable}{lccc}
\tablenum{4a}
\tablewidth{0pt}
\tablecaption{Stellar Parameters for the M3 Sample. 
\label{table_teff}}
\tablehead{
\colhead{ID\tablenotemark{a}} &
\colhead{\teff } &
\colhead{log($g$) } &
\colhead{$v_t$} \\
\colhead{} &
\colhead{(K)} &
\colhead{(dex)} &
\colhead{(km/s)} 
}
\startdata
VZ1397       & 3985 &  0.40 & 1.90 \\
II-46        & 3998 &  0.40 & 1.90 \\
VZ1000       & 4175 &  0.70 & 1.85 \\
III-28       & 4200 &  0.60 & 1.80 \\
IV-25        & 4408 &  1.10 & 1.60 \\
C41303\_2217 & 4436 &  1.20 & 1.60 \\
IV-27        & 4547 &  1.30 & 1.60 \\
III-61       & 5140 &  2.70 & 1.35 \\ 
III-60       & 5170 &  2.65 & 1.30 \\ 
C41544\_2336  & 5170 &  2.70 & 1.25 \\ 
V-30         & 5170 &  2.75 & 1.40 \\ 
V-31         & 5170 &  2.80 & 1.20 \\ 
C41543\_2334\tablenotemark{b} & 6090 &  2.70 & 2.05 \\ 
\enddata
\tablenotetext{a}{Identifications as in notes to Table~\ref{table_sample}.}
\tablenotetext{b}{This is a RHB star}
\end{deluxetable}

%
%
\clearpage
\begin{deluxetable}{lccc}
\tablenum{4b}
\tablewidth{0pt}
\tablecaption{Stellar Parameters for the M13 Sample. 
\label{table_teff_m13}}
\tablehead{
\colhead{ID\tablenotemark{a}} &
\colhead{\teff } &
\colhead{log($g$) } &
\colhead{$v_t$} \\
\colhead{} &
\colhead{(K)} &
\colhead{(dex)} &
\colhead{(km/s)} 
}
\startdata
II$-$67         & 3900 & 0.45 & 1.9 \\
IV$-$25         & 3985 & 0.50 & 1.95 \\
II$-$76         & 4300 & 0.85 & 1.95 \\
III$-$18        & 4350 & 1.00 & 1.65 \\
K188            & 4535 & 1.37 & 1.45 \\
III$-$7         & 4550 & 1.40 & 1.35 \\
I$-$18          & 4690 & 1.65 & 1.5 \\
I$-$49          & 4745 & 1.75 & 1.55 \\
J37             & 4895 & 2.00 & 1.45 \\
C41196\_2632    & 4900 & 2.00 & 1.5 \\
II$-$4          & 4910 & 2.05 & 1.5 \\
IV$-$29         & 5007 & 2.22 & 1.55 \\
J45             & 5055 & 2.25 & 1.55 \\
C41155\_3103    & 5068 & 2.37 & 1.55 \\
I$-$5           & 5070 & 2.45 & 1.5 \\
C41148\_3103    & 5247 & 2.89 & 1.55 \\
C40559\_2839    & 5349 & 3.10 & 1.6 \\
C41134\_3056    & 5305 & 3.25 & 1.0 \\
C41099\_3046    & 5414 & 3.35 & 1.05 \\
C40535\_2813    & 5718 & 3.60 & 2.3 \\
C41101\_3050    & 5370 & 3.30 & 1.2 \\
C41135\_3053    & 5500 & 3.50 & 1.15 \\
C41133\_2750    & 5520 & 3.50 & 1.65 \\
C40535\_2819    & 5722 & 3.65 & 2.05 \\
C41135\_2753    & 6045 & 3.80 & 2.3 \\
\enddata
\tablenotetext{a}{Identifications as in notes to Table~\ref{table_sample_m13}.}
\end{deluxetable}

%
%
\clearpage                                                                      
\begin{deluxetable}{llrrrrrrrrrrrrrrr}     
\tabletypesize
\scriptsize              
\tablenum{5a}  
\tablewidth{0pt}                                                                
\tablecaption{Equivalent Widths for the M3 Stars \label{table_eqw_m3}} 
\tablehead{                                                                     
\colhead{Ion} & \colhead{$\lambda$} & \colhead{$\chi$} & 
\colhead{log $gf$} & 
\colhead{VZ} & \colhead{II} &  \colhead{III} & \colhead{VZ} & 
\colhead{IV} & \colhead{C41303} & \colhead{IV} &
\colhead{41543} & \colhead{III} & \colhead{III} & 
\colhead{V} & \colhead{41544} & \colhead{V} \\
\colhead{} & \colhead{(\AA)} & \colhead{(eV)} & \colhead{} &
\colhead{1397} & \colhead{46} & \colhead{28} & \colhead{1000} &
\colhead{25} & \colhead{\_2217} & \colhead{27} & 
\colhead{\_2334} & \colhead{61} & \colhead{60} & 
\colhead{30} & \colhead{\_2336} & \colhead{31} \\
\colhead{} & \colhead{} & \colhead{} & \colhead{} &
\colhead{(m\AA)} & \colhead{(m\AA)} & \colhead{(m\AA)} & \colhead{(m\AA)} &
\colhead{(m\AA)} & \colhead{(m\AA)} & \colhead{(m\AA)} & 
\colhead{(m\AA)} & \colhead{(m\AA)} & \colhead{(m\AA)} & 
\colhead{(m\AA)} & \colhead{(m\AA)} & \colhead{(m\AA)}
}
\startdata
 OI  &  6300.30 & 0.00 & $-$9.78 &    77.2 &    66.5 &    55.0 &    28.5 &    39.0 &    
 37.0 &    16.0 & \nodata & \nodata & \nodata & \nodata & \nodata & \nodata  \\
 OI  &  6363.78 & 0.02 & $-$10.30&    41.0 &    27.5 &    28.1 &    14.6 &    14.0 &    
 15.7 &     7.0 & \nodata & \nodata & \nodata & \nodata & \nodata & \nodata  \\
 OI  &  7771.94 & 9.15 &    0.37 & \nodata & \nodata & \nodata & \nodata & \nodata & 
 \nodata & \nodata &   111.0 &    32.0 &      27.0 &    ${\le}14.0$ &    32.0 &    22.0  \\
 OI  &  7774.17 & 9.15 &    0.22 & \nodata & \nodata & \nodata & \nodata & \nodata & 
 \nodata & \nodata &    90.0 &    30.0 &  20.0 &   \nodata &    25.0 & \nodata  \\
 OI  &  7775.39 & 9.15 &    0.00 & \nodata & \nodata & \nodata & \nodata & \nodata & 
 \nodata & \nodata &    71.0 & \nodata & \nodata & \nodata & \nodata & \nodata  \\
 NaI &  5682.63 & 2.10 & $-$0.70 &    72.1 &    58.0 &    30.2 &    73.9 &    46.3 &   
  33.0 &    55.9 & \nodata & \nodata & \nodata &    41.0 &    23.0 &    19.0  \\
\enddata
\end{deluxetable}

\clearpage
  
%
%
\clearpage                                                                      
\begin{deluxetable}{llrrrrrrrrrrrrrrr}     
\rotate
\tabletypesize
\scriptsize              
\tablenum{5b}  
\tablewidth{0pt}                                                                
\tablecaption{Equivalent Widths for the Brighter M13 Stars \label{table_eqw_m13_1}} 
\tablehead{                                                                     
\colhead{Ion} & \colhead{$\lambda$} & \colhead{$\chi$} & 
\colhead{log $gf$} & 
\colhead{II} & \colhead{IV} &  \colhead{II} & \colhead{III} & 
\colhead{K} & \colhead{III} & \colhead{I} &
\colhead{I} & \colhead{J} & \colhead{41196} & 
\colhead{II} & \colhead{IV} & \colhead{J} \\
\colhead{} & \colhead{(\AA)} & \colhead{(eV)} & \colhead{} &
\colhead{67} & \colhead{25} & \colhead{76} & \colhead{18} &
\colhead{188} & \colhead{7} & \colhead{18} & 
\colhead{49} & \colhead{37} & \colhead{\_2632} & 
\colhead{4} & \colhead{29} & \colhead{45} \\
\colhead{} & \colhead{} & \colhead{} & \colhead{} &
\colhead{(m\AA)} & \colhead{(m\AA)} & \colhead{(m\AA)} & \colhead{(m\AA)} &
\colhead{(m\AA)} & \colhead{(m\AA)} & \colhead{(m\AA)} & 
\colhead{(m\AA)} & \colhead{(m\AA)} & \colhead{(m\AA)} & 
\colhead{(m\AA)} & \colhead{(m\AA)} & \colhead{(m\AA)}
}
\startdata
  OI  &  6300.30 &  0.00 & $-$9.78 &     6.1 &    20.0 &    55.0 &    10.0 &    26.0 &    30.0 &    16.0 &    30.0
 &     9.0 &     6.5 &    10.0 &    14.0 &     7.0 \\
  OI  &  6363.78 &  0.02 &$-$10.30 & \nodata &     7.0 &    23.8 & \nodata &    11.7 &    14.0 & \nodata &     9.0
 & \nodata & \nodata & \nodata & \nodata & \nodata \\
 NaI  &  5682.63 &  2.10 & $-$0.70 &   109.2 &   112.0 &    30.0 &    83.7 &    36.0 &    40.7 &    55.0 &    39.4
 &    56.0 &    42.3 &    43.5 &    15.5 &    41.0 \\
 NaI  &  5688.19 &  2.10 & $-$0.42 &   124.0 &   138.0 &    52.8 &   106.4 &    59.0 &    57.8 &    80.4 &    53.0
 &    75.6 &    61.7 &    63.4 &    29.4 &    57.3 \\
 NaI  &  6154.23 &  2.10 & $-$1.53 &    45.7 &    46.9 &     6.6 &    27.0 &    10.0 & \nodata &    14.5 & \nodata
 &    19.0 & \nodata &     8.5 & \nodata & \nodata \\
 NaI  &  6160.75 &  2.00 & $-$1.23 & \nodata &    67.9 &    11.0 &    44.0 &    18.1 & \nodata &    29.6 &    17.0
 &    23.5 &    20.0 &    21.0 & \nodata &    16.0 \\
\enddata
\end{deluxetable}

%
%
\clearpage                                                                      
\begin{deluxetable}{llrrrrrrrrrrrrrrr}     
\rotate
\tabletypesize
\scriptsize              
\tablenum{5c}  
\tablewidth{0pt}                                                                
\tablecaption{Equivalent Widths for the Fainter M13 Stars \label{table_eqw_m13_2}} 
\tablehead{                                                                     
\colhead{Ion} & \colhead{$\lambda$} & \colhead{$\chi$} & 
\colhead{log $gf$} & 
\colhead{I} & \colhead{41155} &  \colhead{41148} & \colhead{41134} & 
\colhead{40559} & \colhead{41101} & \colhead{41099} &
\colhead{41135} & \colhead{41133} & \colhead{40535} & 
\colhead{40539} & \colhead{41135} \\
\colhead{} & \colhead{(\AA)} & \colhead{(eV)} & \colhead{} &
\colhead{5} & \colhead{\_3103} & \colhead{\_3103} & \colhead{\_3056} &
\colhead{\_2839} & \colhead{\_3050} & \colhead{\_3046} & 
\colhead{\_3053} & \colhead{\_2750} & \colhead{\_2819} & 
\colhead{\_2813} & \colhead{\_2753}  \\
\colhead{} & \colhead{} & \colhead{} & \colhead{} &
\colhead{(m\AA)} & \colhead{(m\AA)} & \colhead{(m\AA)} & \colhead{(m\AA)} &
\colhead{(m\AA)} & \colhead{(m\AA)} & \colhead{(m\AA)} & 
\colhead{(m\AA)} & \colhead{(m\AA)} & \colhead{(m\AA)} & 
\colhead{(m\AA)} & \colhead{(m\AA)} 
}
\startdata
  OI  &  7771.94 &  9.15 &    0.37 & \nodata &    12.0 &    34.0 &    16.0 &    25.0 &    22.5 &    20.0 &    20.0
 &    20.0 &    25.0 &    19.0 &    16.5 \\
  OI  &  7774.17 &  9.15 &    0.22 & \nodata & \nodata &    25.0 &    14.0 &    17.0 & \nodata &    15.0 &    18.0
 &    14.0 &    25.0 &    19.0 &    20.0 \\
 NaI  &  5682.63 &  2.10 & $-$0.70 &    56.0 &    49.0 &    20.0 &    21.7 &    26.0 &    30.3 &    29.0 &    14.0
 & \nodata &    19.0 &    20.0 & \nodata \\
 NaI  &  5688.19 &  2.10 & $-$0.42 &    76.5 &    69.0 &    29.0 &    44.3 &    33.0 &    44.3 &    48.0 &    23.3
 &    27.0 &    26.0 &    35.0 &    27.0 \\
 NaI  &  6160.75 &  2.00 & $-$1.23 &    26.6 &    22.0 & \nodata & \nodata & \nodata & \nodata & \nodata & \nodata
 & \nodata & \nodata & \nodata & \nodata \\
\enddata
\end{deluxetable}

%
%
\begin{deluxetable}{lcrlrlcrcccccc}  
\tabletypesize                      
\footnotesize                  
\tablenum{6a}                    
\tablewidth{0pt}                
\tablecaption{Abundance Ratios for M3: O to Mg \label{tab4a}} 
\tablehead{                     
\colhead{Star} &                
\colhead{[Fe/H]$_{\rm I}$}  & \colhead{$N$} &
\colhead{[Fe/H]$_{\rm II}$} & \colhead{$N$} &
\colhead{[O/Fe]} & \colhead{$N$} &
\colhead{[Na/Fe]} & \colhead{$N$} &
\colhead{[Mg/Fe]} & \colhead{$N$} \\
\colhead{} &
\colhead{$\pm\sigma/ \sqrt{N}$} & \colhead{} & \colhead{$\pm\sigma/ \sqrt{N}$} & \colhead{
} &
\colhead{$\pm\sigma/ \sqrt{N}$} & \colhead{} & \colhead{$\pm\sigma/ \sqrt{N}$} & \colhead{
} & 
\colhead{$\pm\sigma/ \sqrt{N}$} & \colhead{} \\
\colhead{} & 
\colhead{(dex)} & \colhead{} & \colhead{(dex)} & \colhead{} &
\colhead{(dex)} & \colhead{} & \colhead{(dex)} & \colhead{} & 
\colhead{(dex)} & \colhead{} 
}
\startdata
VZ1397       & $-$1.36 $\pm$ 0.05*& 149 & $-$1.36 $\pm$ 0.06 & 13& 0.52 $\pm$0.05*& 2& $-$
0.16 $\pm$0.05~ & 4 & 0.59 $\pm$ 0.13 & 3 \\
II-46        & $-$1.50 $\pm$ 0.05*& 148 & $-$1.39 $\pm$ 0.05 & 13& 0.37 $\pm$0.05*& 2& $-$
0.23 $\pm$0.07~ & 4 & 0.56 $\pm$ 0.15 & 3 \\
VZ1000       & $-$1.43 $\pm$ 0.05*& 143 & $-$1.40 $\pm$ 0.05 & 13& 0.08 $\pm$0.10 & 2& 
+0.10 $\pm$0.06~ & 4 & 0.44 $\pm$ 0.10 & 3 \\
III-28       & $-$1.54 $\pm$ 0.05*& 134 & $-$1.54 $\pm$ 0.05 & 10& 0.55 $\pm$0.07 & 2& $-$
0.33 $\pm$0.06~ & 2 & 0.49 $\pm$ 0.15 & 3 \\
IV-25        & $-$1.39 $\pm$ 0.05*& 141 & $-$1.45 $\pm$ 0.05*& 14& 0.38 $\pm$0.05*& 2& $-$
0.10 $\pm$0.05~ & 4 & 0.41 $\pm$ 0.11 & 3 \\
C41303\_2217 & $-$1.37 $\pm$ 0.05*& 126 & $-$1.34 $\pm$ 0.05*& 14& 0.33 $\pm$0.05*& 2& $-$
0.39 $\pm$0.09~ & 3 & 0.40 $\pm$ 0.11 & 3 \\
IV-27        & $-$1.35 $\pm$ 0.05*& 129 & $-$1.45 $\pm$ 0.05 & 10& 0.08 $\pm$0.08 & 2&  
+0.11 $\pm$0.06~ & 4 & 0.31 $\pm$ 0.17 & 3 \\
III-61       & $-$1.38 $\pm$ 0.05*&  61 & $-$1.29 $\pm$ 0.07 &  3& 0.60 $\pm$0.05 & 2& $-$
0.45 $\pm$0.09~ & 2 & 0.29 $\pm$ 0.10 & 1 \\
\enddata                                          
\tablenotetext{*}{The minimum value of 0.05 dex has been adopted; the nominal 
calculated value is smaller.}
\end{deluxetable}    

%
\begin{deluxetable}{llrrrrrlrrrlr}  
\tabletypesize                      
\footnotesize                       
\tablenum{6b} 
\tablewidth{0pt}                
\tablecaption{Abundance Ratios for M3: Si to V \label{tab4b}} 
\tablehead{                     
\colhead{Star} &                
\colhead{[Si/Fe]} & \colhead{$N$} &
\colhead{[Ca/Fe]} & \colhead{$N$} &
\colhead{[Sc/Fe]} & \colhead{$N$} &
\colhead{[Ti/Fe]} & \colhead{$N$} &
\colhead{[V/Fe]} & \colhead{$N$} \\
\colhead{} &
\colhead{$\pm\sigma/ \sqrt{N}$} & \colhead{} & \colhead{$\pm\sigma/ \sqrt{N}$} & \colhead{
} &
\colhead{$\pm\sigma/ \sqrt{N}$} & \colhead{} & \colhead{$\pm\sigma/ \sqrt{N}$} & \colhead{
} & 
\colhead{$\pm\sigma/ \sqrt{N}$} & \colhead{} \\
\colhead{} & 
\colhead{(dex)} & \colhead{} & \colhead{(dex)} & \colhead{} &
\colhead{(dex)} & \colhead{} & \colhead{(dex)} & \colhead{} & 
\colhead{(dex)} & \colhead{} 
}
\startdata
VZ1397       &  0.35 $\pm$0.05*& 11 &  0.09$\pm$ 0.05* & 16 &  0.10 $\pm$ 0.06~ & 7  & 0.3
0 $\pm$ 0.05* &35 &  0.05 $\pm$ 0.06~ &  9  \\
II-46        &  0.28 $\pm$0.06 &  9 &$-$0.02$\pm$ 0.05* & 15 &  0.13 $\pm$ 0.05*& 7  & 0.0
8 $\pm$ 0.05* &33 & $-$0.21 $\pm$ 0.05~ &  9  \\
VZ1000       &  0.35 $\pm$0.05*&  9 &  0.11$\pm$ 0.05* & 15 &  0.19 $\pm$ 0.05~ & 7  & 0.1
3 $\pm$ 0.05* &32 & $-$0.10 $\pm$ 0.05* & 10  \\
III-28       &  0.19 $\pm$0.08 &  3 &  0.05$\pm$ 0.05* & 15 &  0.15 $\pm$ 0.07~ & 7  & 0.1
1 $\pm$ 0.05* &29 & $-$0.20 $\pm$ 0.05* & 10  \\
IV-25        &  0.25 $\pm$0.05*&  9 &  0.12$\pm$ 0.05* & 15 &  0.11 $\pm$ 0.05~ & 7  & 0.1
0 $\pm$ 0.05* &26 & $-$0.17 $\pm$ 0.05* &  8  \\
C41303\_2217 &  0.20 $\pm$0.05*&  9 &  0.08$\pm$ 0.05* & 15 &  0.15 $\pm$ 0.05~ & 7  & 0.1
0 $\pm$ 0.05* &25 & $-$0.18 $\pm$ 0.06~ &  8  \\
IV-27        &  0.23 $\pm$0.05*& 12 &  0.10$\pm$ 0.05* & 15 &  0.18 $\pm$ 0.06~ & 7  & 0.1
0 $\pm$ 0.05* &27 & $-$0.15 $\pm$ 0.05* &  8  \\
III-61       &  0.23 $\pm$0.13 &  4 &  0.18$\pm$ 0.05~ & 17 &  0.21 $\pm$ 0.11~ & 7  & 0.1
3 $\pm$ 0.06  & 3 & \nodata &   0  \\
\enddata     
\tablenotetext{*}{The minimum value of 0.05 dex has been adopted; the nominal 
calculated value is smaller.}                                     
\end{deluxetable}

%
\begin{deluxetable}{lrrrrrrrrrrr}  
\tabletypesize                      
\footnotesize 
\tablenum{6c} 
\tablewidth{0pt}                
\tablecaption{Abundance Ratios for M3: Cr to Cu \label{tab4c}} 
\tablehead{                     
\colhead{Star} &                
\colhead{[Cr/Fe]} & \colhead{$N$} &
\colhead{[Mn/Fe]} & \colhead{$N$} &
\colhead{[Co/Fe]} & \colhead{$N$} &
\colhead{[NiFe]} & \colhead{$N$} &
\colhead{[Cu/Fe]} & \colhead{$N$} \\
\colhead{} &
\colhead{$\pm\sigma/ \sqrt{N}$} & \colhead{} & \colhead{$\pm\sigma/ \sqrt{N}$} & \colhead{
} &
\colhead{$\pm\sigma/ \sqrt{N}$} & \colhead{} & \colhead{$\pm\sigma/ \sqrt{N}$} & \colhead{
} & 
\colhead{$\pm\sigma/ \sqrt{N}$} & \colhead{} \\
\colhead{} & 
\colhead{(dex)} & \colhead{} & \colhead{(dex)} & \colhead{} &
\colhead{(dex)} & \colhead{} & \colhead{(dex)} & \colhead{} & 
\colhead{(dex)} & \colhead{} 
}
\startdata
VZ1397       &  0.03   $\pm$ 0.06~ &  8 & $-$0.11 $\pm$ 0.14 & 5 & $-$0.01 $\pm$ 0.05 & 4 
& $-$0.10 $\pm$ 0.05*& 24 & $-$0.55 $\pm$ 0.09~ &  3 \\
II-46        & $-$0.06 $\pm$ 0.05~ &  8 & $-$0.20 $\pm$ 0.15 & 5 & $-$0.08 $\pm$ 0.06 & 4 
& $-$0.13 $\pm$ 0.05*& 21 & $-$0.55 $\pm$ 0.09~ &  3 \\
VZ1000       & $-$0.10 $\pm$ 0.09~ &  9 & $-$0.25 $\pm$ 0.09 & 5 & $-$0.07 $\pm$ 0.09 & 4 
& $-$0.14 $\pm$ 0.05*& 20 & $-$0.51 $\pm$ 0.06~ &  3 \\
III-28       &  0.10  $\pm$ 0.08~ &  8 & $-$0.34 $\pm$ 0.12 & 3 & $-$0.02 $\pm$ 0.09 & 4 &
 $-$0.18 $\pm$ 0.05*& 13 & $-$0.66 $\pm$ 0.05* &  3 \\
IV-25        &  0.05  $\pm$ 0.06~ & 10 & $-$0.30 $\pm$ 0.12 & 5 & $-$0.06 $\pm$ 0.08 & 4 &
 $-$0.14 $\pm$ 0.05*& 20 & $-$0.62 $\pm$ 0.11~ &  2 \\
C41303\_2217 &  0.01  $\pm$ 0.06~ &  9 & $-$0.27 $\pm$ 0.14 & 4 & $-$0.11 $\pm$ 0.03 & 3 &
 $-$0.16 $\pm$ 0.05*& 15 & $-$0.56 $\pm$ 0.21~ &  2 \\
IV-27        & $-$0.03 $\pm$ 0.05* &  9 & $-$0.17 $\pm$ 0.05 & 3 & $-$0.16 $\pm$ 0.06 & 3 
& $-$0.16 $\pm$ 0.05*& 14 & $-$0.71 $\pm$ 0.05*&  2 \\
III-61       & \nodata             &  0 & \nodata          & 0 & \nodata           & 0 & 
$-$0.32 $\pm$ 0.08~ &  6 & \nodata &  0 \\
\enddata 
\tablenotetext{*}{The minimum value of 0.05 dex has been adopted; the nominal 
calculated value is smaller.}
\end{deluxetable}

%
\begin{deluxetable}{lrrrrrrrrrrrrrrr}  
\tabletypesize                      
\footnotesize                      
\tablenum{6d} 
\tablewidth{0pt}                
\tablecaption{Abundance Ratios for M3: Zn to La \label{tab4d}} 
\tablehead{                     
\colhead{Star} &                
\colhead{[Zn/Fe]} & \colhead{$N$} &
\colhead{[Y/Fe]} & \colhead{$N$} &
\colhead{[Zr/Fe]} & \colhead{$N$} &
\colhead{[Ba/Fe]} & \colhead{$N$} &
\colhead{[La/Fe]} & \colhead{$N$} \\
\colhead{} &
\colhead{$\pm\sigma/ \sqrt{N}$} & \colhead{} & \colhead{$\pm\sigma/ \sqrt{N}$} & \colhead{
} &
\colhead{$\pm\sigma/ \sqrt{N}$} & \colhead{} & \colhead{$\pm\sigma/ \sqrt{N}$} & \colhead{
} & 
\colhead{$\pm\sigma/ \sqrt{N}$} & \colhead{} \\
\colhead{} & 
\colhead{(dex)} & \colhead{} & \colhead{(dex)} & \colhead{} &
\colhead{(dex)} & \colhead{} & \colhead{(dex)} & \colhead{} & 
\colhead{(dex)} & \colhead{} 
}
\startdata
VZ1397       &  $-$0.14 $\pm$ 0.05* & 2 & $-$0.56 $\pm$  0.07~ & 3 &  0.13 $\pm$ 0.05~ & 3
 &  0.13 $\pm$ 0.05~ & 3 &  0.07$\pm$ 0.05 & 3  \\
II-46        &    0.09 $\pm$ 0.08~ & 2 & $-$0.27 $\pm$  0.10~ & 3 & $-$0.12 $\pm$ 0.05* & 
3 &  0.12 $\pm$ 0.05~ & 3 &  0.03$\pm$ 0.08 & 3  \\
VZ1000       &  $-$0.12 $\pm$ 0.05* & 2 & $-$0.25 $\pm$  0.09~ & 4 & $-$0.11 $\pm$ 0.05* &
 3 &  0.15 $\pm$ 0.06~ & 3 &  0.01$\pm$ 0.05 & 3  \\
III-28       &  $-$0.12 $\pm$ 0.05* & 2 & $-$0.14 $\pm$  0.13~ & 4 & $-$0.03 $\pm$ 0.06~ &
 3 &  0.08 $\pm$ 0.09~ & 3 &  0.17$\pm$ 0.08 & 2  \\
IV-25        &  $-$0.10 $\pm$ 0.07~ & 2 & $-$0.25 $\pm$  0.09~ & 4 &  0.05 $\pm$ 0.05~ & 3
 &  0.15 $\pm$ 0.05* & 3 & $-$0.02$\pm$ 0.05 & 3  \\
C41303\_2217 &  $-$0.10 $\pm$ 0.05* & 2 & $-$0.15 $\pm$  0.10~ & 4 &  0.22 $\pm$ 0.10~ & 3
 &  0.16 $\pm$ 0.05~ & 3 & $-$0.02$\pm$ 0.15 & 2  \\
IV-27        &  $-$0.13 $\pm$ 0.05* & 2 & $-$0.35 $\pm$  0.09~ & 4 &     \nodata        & 
0 &  0.14 $\pm$ 0.06~ & 3 &  0.05$\pm$ 0.05 & 3  \\
III-61       &      \nodata        & 0 &    \nodata            & 0 &     \nodata         &
 0 & $-$0.09 $\pm$ 0.05* & 3 &    \nodata   & 0 \\
\enddata       
\tablenotetext{*}{The minimum value of 0.05 dex has been adopted; the nominal 
calculated value is smaller.}
\end{deluxetable}    

%
\begin{deluxetable}{lrrrrrrrrrrrrr}  
\scriptsize    
\tablenum{6e} 
\tablewidth{0pt}                
\tablecaption{Abundance Ratios for M3: Nd to Dy \label{tab4e}} 
\tablehead{                     
\colhead{Star} &                
\colhead{[Nd/Fe]} & \colhead{$N$} &
\colhead{[Eu/Fe]} & \colhead{$N$} &
\colhead{[Dy/Fe]} & \colhead{$N$} \\
\colhead{} &
\colhead{$\pm\sigma/ \sqrt{N}$} & \colhead{} & \colhead{$\pm\sigma/ \sqrt{N}$} & \colhead{
} &
\colhead{$\pm\sigma/ \sqrt{N}$} & \colhead{} \\
\colhead{} & 
\colhead{(dex)} & \colhead{} & \colhead{(dex)} & \colhead{} &
\colhead{(dex)} & \colhead{} 
}
\startdata
VZ1397       &   0.12 $\pm$ 0.07~ & 5 &  0.46 $\pm$ 0.10 & 1 & $-$0.04 $\pm$ 0.10 & 1  \\
II-46        &   0.16 $\pm$ 0.05~ & 5 &  0.51 $\pm$ 0.10 & 1 &   0.12 $\pm$ 0.10 & 1  \\
VZ1000       &   0.29 $\pm$ 0.05~ & 7 &  0.44 $\pm$ 0.10 & 1 & $-$0.09 $\pm$ 0.10 & 1  \\
III-28       &   0.15 $\pm$ 0.05* & 7 &  0.52 $\pm$ 0.10 & 1 &  0.08 $\pm$ 0.10 & 1  \\
IV-25        &   0.16 $\pm$ 0.05* & 7 &  0.46 $\pm$ 0.10 & 1 &  0.19 $\pm$ 0.10 & 1  \\
Stet189      &   0.28 $\pm$ 0.06~ & 5 &  0.51 $\pm$ 0.10 & 1 &  \nodata & 0 \\
IV-27        &   0.21 $\pm$ 0.05* & 5 &  0.55 $\pm$ 0.10 & 1 &  \nodata & 0 \\
III-61       &      \nodata        & 0 &     \nodata       & 0 &  \nodata & 0 \\
III-60       &      \nodata        & 0 &     \nodata       & 0 &  \nodata & 0 \\
C41544\_2336 & $-$0.06 $\pm$ 0.10~ & 1 &  0.53 $\pm$ 0.10 & 1 &  \nodata & 0 \\
V-30         &      \nodata        & 0 &     \nodata       & 0 &  \nodata & 0 \\
V-31         &      \nodata        & 0 &     \nodata       & 0 &  \nodata & 0 \\
C41543\_2334 &      \nodata        & 0 &     \nodata       & 0 &  \nodata & 0 \\
\enddata       
\tablenotetext{*}{The minimum value of 0.05 dex has been adopted; the nominal 
calculated value is smaller.}
\end{deluxetable}

\clearpage

%
\begin{deluxetable}{lrrrrr}                       
\tabletypesize
\small                                           
\tablenum{7}                                      
\tablewidth{0pt}                                  
\tablecaption{Sensitivity of Abundances \label{table_sens}}
\tablehead{                                       
\colhead{} &                                      
\colhead{$\Delta$(EW)} &                          
\colhead{$\Delta$(\teff)} &                        
\colhead{$\Delta$(log g)} &                        
\colhead{$\Delta$(vt)} &                         
\colhead{$\Delta$([Fe/H])} \\                        
\colhead{} &                                      
\colhead{10\%\tablenotemark{a}} &                                  
\colhead{+100 K\tablenotemark{b}} &                               
\colhead{+0.2 dex} &                             
\colhead{+0.4/$-0.2$ \kms} &                            
\colhead{+0.2 dex} \\
\colhead{} & \colhead{(dex)} &  \colhead{(dex)} & 
\colhead{(dex)} & \colhead{(dex)} & \colhead{(dex)} \nl 
}                      
\startdata                                        
O I  :  & & & & &  \\
4250/1.0/1.6 &  0.05 &$-$0.01 &  0.09 &$-$0.01/0.00 &  0.08 \\
5250/3.0/1.3 &  0.06 &$-$0.09 &  0.07 &$-$0.01/+0.01 & $-$0.01 \\
5750/3.5/2.0 &  0.06 &$-$0.08 &  0.07 &$-$0.02/+0.01 & $-$0.01 \\
Na I :  & & & & &  \\
4250/1.0/1.6 &  0.07 &  0.09 &$-$0.02 &$-$0.02/+0.02 &$-$0.01 \\
5250/3.0/1.3 &  0.06 &  0.06 &$-$0.03 &$-$0.01/+0.01 &  0.00 \\
5750/3.5/2.0 &  0.05 &  0.04 &  0.00 &$-$0.01/0.00 &  0.00 \\
\enddata 
\tablenotetext{a}{This is an underestimate of the uncertainty for
the weakest lines.}
\tablenotetext{b}{The \teff\ uncertainty is a function of \teff;  see
\S5.}                                                                           
\end{deluxetable}     
%
%

%
%
\clearpage                                                                      
\begin{deluxetable}{lrrcrrrrrr}     
\tablenum{8}                                                                    
\tablewidth{0pt}                                                                
\tablecaption{Mean Abundances and Abundance Spreads for Stars in M3
\label{table_abundsig}}                      
\tablehead{ 
\colhead{Species} & \colhead{Mean Abund.} &
\colhead{$\sigma$} & \colhead{$\sigma$(tot)\tablenotemark{a}} &
\colhead{Spread Ratio\tablenotemark{b}} & \colhead{No. of} &
\colhead{$x^2$\tablenotemark{c}} & \colhead{$p(x)$\tablenotemark{d}} & 
\colhead{Variation\tablenotemark{e}} \\
\colhead{} & \colhead{[X/Fe] (dex)} &\colhead{(dex)} &
\colhead{(dex)} & \colhead{(dex)} &
\colhead{Stars\tablenotemark{e}} & \colhead{} & \colhead{\%} & \colhead{}
}
\startdata
OI  &   0.33 &  0.20 & 0.07 &   2.26 & 12 &  56.1 & $<$0.1 & Yes \\ 
NaI & $-$0.16 &  0.20 & 0.08 &   2.32 & 12 &  55.4 & $<$0.1 & Yes \\ 
MgI &   0.40 &  0.12 & 0.17 &   0.71 & 12 &  5.5 & 90.7  & No \\
SiI &   0.27 &  0.06 & 0.07 &   0.88 & 12 &  7.6 & 75.0  & No \\
CaI &   0.10 &  0.05 & 0.12 &   0.43 & 12 &  2.1 & 99.8 & No \\
ScII&   0.13 &  0.05 & 0.12 &   0.41 & 12 &  1.7 & 99.9  & No \\
TiI &   0.17 &  0.08 & 0.12 &   0.65 & 12 &  4.2 & 95.7 & No \\
TiII&   0.27 &  0.11 & 0.15 &   0.74 & 10 &  4.6 & 86.6 & No \\
\enddata
\tablenotetext{a}{$\sigma(tot)$ is defined in \S6.2.}
\tablenotetext{b}{This is the ratio of $\sigma$
for the M3 sample stars to $\sigma$(tot).  See text.} 
\tablenotetext{c}{$x^2 = \sum (y_i-\bar{y})^2/\sigma_i^2$, where 
$\bar{y} = \sum (y_i/\sigma_i^2) / \sum(1/\sigma_i^2)$. 
A minimum $\sigma_i = \sigma(tot)$ was adopted} 
\tablenotetext{d}{$p(x)$ is the probability of exceeding $x^2$ by chance} 
\tablenotetext{e}{If $p(x) <$ 0.1\% there is abundance variation; 
if $ p(x) > 0.5\%$ there is no variation.}
\tablenotetext{e}{The number of stars in the M3 sample in which lines of
this species were detected.}
\end{deluxetable}

%
%
\begin{deluxetable}{lcrlrrcrcrcccc}  
\tabletypesize
\footnotesize                  
\tablenum{9a}                    
\tablewidth{0pt}                
\tablecaption{Abundance Ratios for M13: O to Mg \label{table_m13abund_a}} 
\tablehead{                     
\colhead{Star} &                
\colhead{[Fe/H]$_{\rm I}$}  & \colhead{$N$} &
\colhead{[Fe/H]$_{\rm II}$} & \colhead{$N$} &
\colhead{[O/Fe]} & \colhead{$N$} &
\colhead{[Na/Fe]} & \colhead{$N$} &
\colhead{[Mg/Fe]} & \colhead{$N$} \\
\colhead{} &
\colhead{$\pm\sigma/ \sqrt{N}$} & \colhead{} & \colhead{$\pm\sigma/ \sqrt{N}$} & \colhead{} &
\colhead{$\pm\sigma/ \sqrt{N}$} & \colhead{} & \colhead{$\pm\sigma/ \sqrt{N}$} & \colhead{} & 
\colhead{$\pm\sigma/ \sqrt{N}$} & \colhead{} \\
\colhead{} & 
\colhead{(dex)} & \colhead{} & \colhead{(dex)} & \colhead{} &
\colhead{(dex)} & \colhead{} & \colhead{(dex)} & \colhead{} & 
\colhead{(dex)} & \colhead{} 
}
\startdata
II-67     & $-$1.45 $\pm$ 0.05* & 110  & $-$1.14 $\pm$ 0.06 & 12 & $-$1.14 $\pm$  0.10~ &  1 &   0.32  $\pm$ 0.05~ & 3 &   0.29  $\pm$ 0.19  & 3 \\
IV-25     & $-$1.47 $\pm$ 0.05* & 128  & $-$1.35 $\pm$ 0.05 & 12 & $-$0.33 $\pm$  0.05* &  2 &   0.41  $\pm$ 0.09~ & 4 &   0.35  $\pm$ 0.21  & 3 \\
II-76     & $-$1.56 $\pm$ 0.05* & 123  & $-$1.49 $\pm$ 0.05* & 13 &   0.55  $\pm$ 0.05* & 2  & $-$0.32  $\pm$ 0.05~ & 4 &  0.49  $\pm$ 0.17  & 3 \\
III-18    & $-$1.43 $\pm$ 0.05* & 123  & $-$1.46 $\pm$ 0.05* & 13 & $-$0.35 $\pm$  0.10~ & 1 &  0.36  $\pm$ 0.06~ & 4 &  0.29  $\pm$ 0.14  & 3 \\
K188      & $-$1.45 $\pm$ 0.05* & 111  & $-$1.44 $\pm$ 0.05* & 13 &   0.33  $\pm$ 0.07~ & 2  & $-$0.08  $\pm$ 0.05* & 4 &  0.39  $\pm$ 0.17  & 3 \\
III-7     & $-$1.60 $\pm$ 0.05* &  93  & $-$1.46 $\pm$ 0.05* & 13 &   0.46  $\pm$ 0.08~ & 2  &  0.10  $\pm$ 0.05* & 2 &  0.52  $\pm$ 0.14  & 3 \\
I-18      & $-$1.49 $\pm$ 0.05* &  88  & $-$1.45 $\pm$ 0.05* & 13 &   0.17  $\pm$ 0.10~  & 1 &  0.32  $\pm$ 0.05* & 4 &  0.31  $\pm$ 0.18  & 3 \\
I-49      & $-$1.45 $\pm$ 0.05* &  87  & $-$1.52 $\pm$ 0.05* & 12 &   0.58  $\pm$ 0.05* & 2  &  0.00  $\pm$ 0.05* & 3 &  0.38  $\pm$ 0.12  & 3 \\
\enddata             			
\tablenotetext{*}{The minimum value of 0.05 dex has been adopted, the 
nominal calculated value is smaller.}
\end{deluxetable}    

%
\begin{deluxetable}{llrlrrrrrrrlr}  
\tabletypesize
\footnotesize                  
\tablenum{9b} 
\tablewidth{0pt}                
\tablecaption{Abundance Ratios for M13: Al to Ti \label{table_m13abund_b}} 
\tablehead{                     
\colhead{Star} &                
\colhead{[Al/Fe]} & \colhead{$N$} &
\colhead{[Si/Fe]} & \colhead{$N$} &
\colhead{[Ca/Fe]} & \colhead{$N$} &
\colhead{[Sc/Fe]} & \colhead{$N$} &
\colhead{[Ti/Fe]} & \colhead{$N$} \\
\colhead{} &
\colhead{$\pm\sigma/ \sqrt{N}$} & \colhead{} & \colhead{$\pm\sigma/ \sqrt{N}$} & \colhead{} &
\colhead{$\pm\sigma/ \sqrt{N}$} & \colhead{} & \colhead{$\pm\sigma/ \sqrt{N}$} & \colhead{} & 
\colhead{$\pm\sigma/ \sqrt{N}$} & \colhead{} \\
\colhead{} & 
\colhead{(dex)} & \colhead{} & \colhead{(dex)} & \colhead{} &
\colhead{(dex)} & \colhead{} & \colhead{(dex)} & \colhead{} & 
\colhead{(dex)} & \colhead{} 
}
\startdata	
II-67     &  0.64  $\pm$ 0.10  &  1 &  0.56 $\pm$ 0.05  & 9  & $-$0.10 $\pm$ 0.05* &  13  &   0.31  $\pm$  0.06~ & 7  &  $-$0.13 $\pm$  0.05*& 28 \\
IV-25     &  0.85  $\pm$ 0.10  &  1 &  0.44 $\pm$ 0.07  & 8  &   0.04  $\pm$ 0.05* &  13  &   0.27  $\pm$  0.05* &  7 &    0.04  $\pm$ 0.05*& 27 \\
II-76     & $\leq$ 0.59        &  1 &  0.41 $\pm$ 0.05  & 8  &   0.03  $\pm$ 0.05* &  13  &   0.20  $\pm$  0.05* &  7 &    0.07  $\pm$ 0.05*& 24 \\
III-18    &  0.74  $\pm$ 0.10  &  1 &  0.36 $\pm$ 0.05* & 8  &   0.10  $\pm$ 0.05* &  13  &   0.21  $\pm$  0.05* &  7 &    0.09  $\pm$ 0.05*& 22 \\
K188      & $\leq$ 0.57        &  1 &  0.27 $\pm$ 0.05* & 9  &   0.11  $\pm$ 0.05* &  13  &   0.16  $\pm$  0.05* &  7 &    0.13  $\pm$ 0.05*& 15 \\
III-7     & $\leq$ 0.70        &  1 &  0.37 $\pm$ 0.05* & 8  &   0.16  $\pm$ 0.05* &  13  &   0.19  $\pm$  0.05* &  7 &    0.04  $\pm$ 0.05*& 11 \\
I-18      & $\leq$ 0.76        &  1 &  0.39 $\pm$ 0.05* & 7  &   0.15  $\pm$ 0.05* &  13  &   0.21  $\pm$  0.05* &  7 &    0.16  $\pm$ 0.05*& 10 \\
I-49      & $\leq$ 0.67        &  1 &  0.32 $\pm$ 0.05* & 7  &   0.15  $\pm$ 0.05* &  13  &   0.13  $\pm$  0.06~ &  7 &    0.15  $\pm$ 0.05~ & 10 \\
\enddata       
\tablenotetext{*}{The minimum value of 0.05 dex has been adopted, 
the nominal calculated value is smaller.}
\end{deluxetable}

%
\begin{deluxetable}{lrrrrrrrrrrr}  
\tablenum{9c} 
\tabletypesize
\footnotesize                  
\tablewidth{0pt}                
\tablecaption{Abundance Ratios for M13: V to Ni \label{table_m13abund_c}} 
\tablehead{                     
\colhead{Star} & 
\colhead{[V/Fe]} & \colhead{$N$} &
\colhead{[Cr/Fe]} & \colhead{$N$} &
\colhead{[Mn/Fe]} & \colhead{$N$} &
\colhead{[Co/Fe]} & \colhead{$N$} &
\colhead{[NiFe]} & \colhead{$N$} \\
\colhead{} &
\colhead{$\pm\sigma/ \sqrt{N}$} & \colhead{} & \colhead{$\pm\sigma/ \sqrt{N}$} & \colhead{} &
\colhead{$\pm\sigma/ \sqrt{N}$} & \colhead{} & \colhead{$\pm\sigma/ \sqrt{N}$} & \colhead{} & 
\colhead{$\pm\sigma/ \sqrt{N}$} & \colhead{} \\
\colhead{} & 
\colhead{(dex)} & \colhead{} & \colhead{(dex)} & \colhead{} &
\colhead{(dex)} & \colhead{} & \colhead{(dex)} & \colhead{} & 
\colhead{(dex)} & \colhead{} 
}
\startdata
II-67     &  $-$0.33  $\pm$ 0.05~ &  9 &    0.04 $\pm$ 0.08~ & 10 & $-$0.20  $\pm$ 0.10~  &  4  &  $-$0.03 $\pm$ 0.09~ &  4 & $-$0.09 $\pm$ 0.05* & 19  \\ 
IV-25     &  $-$0.14  $\pm$ 0.05* &  9 &    0.10 $\pm$ 0.07~ &  9 & $-$0.26  $\pm$ 0.13~  &  4  &     0.00 $\pm$ 0.05~ &  4  & $-$0.06 $\pm$ 0.05* & 18  \\
II-76     &  $-$0.19  $\pm$ 0.05~ &  8 &  $-$0.11 $\pm$  0.07~ & 7 & $-$0.36 $\pm$  0.05~  &  4  &  $-$0.03 $\pm$ 0.05* &  4 & $-$0.12 $\pm$ 0.05* & 18 \\
III-18    &  $-$0.11  $\pm$ 0.05* &  8 &  $-$0.01 $\pm$  0.05~ & 7 & $-$0.25 $\pm$  0.08~  &  4  &  $-$0.02 $\pm$ 0.06~ &  4 & $-$0.09 $\pm$ 0.05* & 18 \\
K188      &  $-$0.16  $\pm$ 0.05* &  7 &  $-$0.04 $\pm$  0.05~ & 7 & $-$0.16 $\pm$  0.06~  &  4  &  $-$0.05 $\pm$ 0.07~ &  3 & $-$0.12 $\pm$ 0.05* & 15 \\
III-7     &  $-$0.09  $\pm$ 0.10~ &  1 &  $-$0.06 $\pm$  0.06~ & 7 & $-$0.14 $\pm$  0.05~  &  4  &  $-$0.02 $\pm$ 0.08~ &  2 & $-$0.19 $\pm$ 0.05  & 11 \\
I-18      &    0.05  $\pm$ 0.10~ &  1 &  $-$0.12  $\pm$ 0.07~ & 5 & $-$0.25  $\pm$ 0.09~  &  4  &  $-$0.07 $\pm$ 0.10~ &  1 & $-$0.10 $\pm$ 0.08  & 11 \\
I-49      &  $-$0.17  $\pm$ 0.10~ &  1 &  $-$0.02 $\pm$  0.05~ & 5 & $-$0.25 $\pm$  0.07~  &  3  &   0.08 $\pm$ 0.10~ &  1 &  $-$0.08 $\pm$ 0.05* & 11 \\
\enddata                                          
\tablenotetext{*}{The minimum value of 0.05 dex has been adopted, the 
nominal calculated value is smaller.}
\end{deluxetable}

%
\begin{deluxetable}{llrrrrrrrrrrrrrr}  
\tabletypesize
\footnotesize                  
\tablenum{9d} 
\tablewidth{0pt}                
\tablecaption{Abundance Ratios for M13: Cu to Ba \label{table_m13abund_d}} 
\tablehead{                     
\colhead{Star} &                
\colhead{[Cu/Fe]} & \colhead{$N$} &
\colhead{[Zn/Fe]} & \colhead{$N$} &
\colhead{[Y/Fe]} & \colhead{$N$} &
\colhead{[Zr/Fe]} & \colhead{$N$} &
\colhead{[Ba/Fe]} & \colhead{$N$} \\
\colhead{} &
\colhead{$\pm\sigma/ \sqrt{N}$} & \colhead{} & \colhead{$\pm\sigma/ \sqrt{N}$} & \colhead{} &
\colhead{$\pm\sigma/ \sqrt{N}$} & \colhead{} & \colhead{$\pm\sigma/ \sqrt{N}$} & \colhead{} & 
\colhead{$\pm\sigma/ \sqrt{N}$} & \colhead{} \\
\colhead{} & 
\colhead{(dex)} & \colhead{} & \colhead{(dex)} & \colhead{} &
\colhead{(dex)} & \colhead{} & \colhead{(dex)} & \colhead{} & 
\colhead{(dex)} & \colhead{} 
}
\startdata
II-67     &$-$0.55  $\pm$  0.13  & 2 &  0.17  $\pm$  0.14~ &  2  &$-$0.08  $\pm$  0.11~ & 5  &$-$0.37  $\pm$  0.05~ & 3  & 0.43  $\pm$ 0.05* & 3 \\ 
IV-25     &$-$0.62  $\pm$  0.06  & 2 &  0.02  $\pm$  0.05* &  2  &$-$0.18  $\pm$  0.07~ & 5  &$-$0.12  $\pm$  0.05* & 3  & 0.28  $\pm$ 0.05~ & 3 \\
II-76     &$-$0.62  $\pm$  0.12  & 2 &  0.04  $\pm$  0.05* &  2  &$-$0.26  $\pm$  0.05* & 4  &  0.06  $\pm$  0.06~ & 3  & 0.38  $\pm$ 0.06~ & 3 \\
III-18    &$-$0.61  $\pm$  0.13  & 2 &$-$0.05  $\pm$  0.05* &  2  &$-$0.26  $\pm$  0.06~ & 5  &  0.05  $\pm$  0.07~ & 2  & 0.36  $\pm$ 0.05~ & 3 \\
K188      &$-$0.70  $\pm$  0.05* & 2 &  0.08  $\pm$  0.05* &  2  &$-$0.13  $\pm$  0.05~ & 4  &       \nodata        & 0  & 0.44  $\pm$ 0.05* & 3 \\
III-7     &$-$0.73  $\pm$  0.05* & 2 &  0.16  $\pm$  0.05* &  2  &$-$0.21  $\pm$  0.05* & 5  &       \nodata        & 0  & 0.44  $\pm$ 0.05~ & 3 \\
I-18      &$-$0.61  $\pm$  0.08  & 2 &  0.09  $\pm$  0.05* &  2  &$-$0.25  $\pm$  0.05~ & 4  &       \nodata        & 0  & 0.37  $\pm$ 0.06~ & 3 \\
I-49      &$-$0.68  $\pm$  0.11  & 2 &  0.06  $\pm$  0.05* &  2  &$-$0.33  $\pm$  0.09~ & 4  &       \nodata        & 0  & 0.28  $\pm$ 0.05~ & 3 \\
\enddata                                          
\tablenotetext{*}{The minimum value of 0.05 dex has been adopted, 
the nominal calculated value is smaller.}
\end{deluxetable}    

%
\begin{deluxetable}{lrrlrlrrrrrrrr}  
\tabletypesize
\footnotesize                  
\tablenum{9e} 
\tablewidth{0pt}                
\tablecaption{Abundance Ratios for M13: La to Dy \label{table_m13abund_e}} 
\tablehead{                     
\colhead{Star} &                
\colhead{[La/Fe]} & \colhead{$N$} &
\colhead{[Nd/Fe]} & \colhead{$N$} &
\colhead{[Eu/Fe]} & \colhead{$N$} &
\colhead{[Dy/Fe]} & \colhead{$N$} \\
\colhead{} &
\colhead{$\pm\sigma/ \sqrt{N}$} & \colhead{} & \colhead{$\pm\sigma/ \sqrt{N}$} & \colhead{} &
\colhead{$\pm\sigma/ \sqrt{N}$} & \colhead{} \\
\colhead{} & 
\colhead{(dex)} & \colhead{} & \colhead{(dex)} & \colhead{} &
\colhead{(dex)} & \colhead{} 
}
\startdata
II-67     &  0.12  $\pm$  0.06~ & 3  &  0.41  $\pm$  0.05  & 7  &  0.56  $\pm$  0.10  & 1  &$-$0.03 $\pm$  0.10  & 1 \\
IV-25     &  0.20  $\pm$  0.10~ & 3  &  0.40  $\pm$  0.05* & 7  &  0.59  $\pm$  0.10  & 1  &  0.09  $\pm$  0.10  & 1 \\
II-76     &  0.11  $\pm$  0.10~ & 3  &  0.21  $\pm$  0.05* & 7  &  0.30  $\pm$  0.10  & 1  &  0.10  $\pm$  0.10  & 1 \\
III-18    &  0.09  $\pm$  0.09~ & 3  &  0.31  $\pm$  0.06  & 7  &  0.58  $\pm$  0.10  & 1  &  0.22  $\pm$  0.10  & 1 \\
K188      &  0.09  $\pm$  0.09~ & 3  &  0.27  $\pm$  0.05* & 7  &  0.67  $\pm$  0.10  & 1  &       \nodata        & 0 \\
III-7     &  0.21  $\pm$  0.05~ & 3  &  0.24  $\pm$  0.07  & 6  &  0.71  $\pm$  0.10  & 1  &       \nodata        & 0 \\
I-18      &  0.11  $\pm$  0.10~ & 2  &  0.12  $\pm$  0.07  & 6  &  0.67  $\pm$  0.10  & 1  &       \nodata        & 0 \\
I-49      &  0.06  $\pm$  0.14~ & 2  &  0.23  $\pm$  0.05* & 7  &  0.52  $\pm$  0.10  & 1  &       \nodata        & 0 \\
\enddata                                          
\tablenotetext{*}{The minimum value of 0.05 dex has been adopted, the nominal 
calculated value is smaller.}
\end{deluxetable}

%
%
\clearpage                                                                      
\begin{deluxetable}{lrrcrrrrrr}     
\tablenum{10}
\tablewidth{0pt}                                                                
\tablecaption{Mean Abundances and Abundance Spreads for Stars in M13
\label{table_abundsig_m13}}                      
\tablehead{ 
\colhead{Species} & \colhead{Mean Abund.} &
\colhead{$\sigma$} & \colhead{$\sigma$(tot)\tablenotemark{a}} &
\colhead{Spread Ratio\tablenotemark{b}} & \colhead{No. of} &
\colhead{$x^2$\tablenotemark{c}} & \colhead{$p(x)$\tablenotemark{d}} & 
\colhead{Variation\tablenotemark{e}} \\
\colhead{} & \colhead{[X/Fe] (dex)} &\colhead{(dex)} &
\colhead{(dex)} & \colhead{(dex)} &
\colhead{Stars\tablenotemark{f}} & \colhead{} & \colhead{\%} & \colhead{}
}
\startdata
OI  &   0.18 &  0.37 & 0.10 &   3.70 & 24 & 319 & $<$0.1 & Yes \\ 
NaI &   0.14 &  0.22 & 0.08 &   2.75 & 25 & 174 & $<$0.1 & Yes \\ 
MgI &   0.24 &  0.15 & 0.18 &   0.83 & 21 & 14.2 & 81.9  & No \\
AlI &   0.74\tablenotemark{g} &  0.11 & \nodata & \nodata & 3 &  \nodata 
 &  \nodata  &   \nodata \\
SiI &   0.31 &  0.11 & 0.08 &   1.37 & 21 & 36.9 & 1.9  & No \\
CaI &   0.12 &  0.07 & 0.12 &   0.58 & 25 &  9.1 & 99.7 & No \\
ScII&   0.12 &  0.11 & 0.12 &   0.92 & 25 & 19.2 & 74.4  & No \\
TiI &   0.16 &  0.11 & 0.13 &   0.85 & 21 & 15.4 & 75.4 & No \\
\enddata
\tablenotetext{a}{$\sigma(tot)$ is defined in \S6.2.}
\tablenotetext{b}{This is the ratio of $\sigma$
for the M13 sample stars to $\sigma$(tot).  See text.} 
\tablenotetext{c}{$x^2 = \sum (y_i-\bar{y})^2/\sigma_i^2$, where 
$\bar{y} = \sum (y_i/\sigma_i^2) / \sum(1/\sigma_i^2)$. 
A minimum $\sigma_i$ = $\sigma$(tot) was adopted} 
\tablenotetext{d}{$p(x)$ is the probability of exceding $x^2$ by chance} 
\tablenotetext{e}{If $p(x) <$ 0.1\% there is abundance variation; 
if $ p(x) > 0.5\%$ there is no variation.}
\tablenotetext{f}{The number of stars in the M13 sample in which lines of
this species were detected.}
\tablenotetext{g}{[Al/Fe] is based on the weak and rarely used
line at 5557~\AA, detected in only three of the most luminous stars of M13.}
\end{deluxetable}

%
%
\clearpage                                                                      
\begin{deluxetable}{lccccll}     
\tablenum{11}           
\tablewidth{0pt}                                                                
\tabletypesize
\small
\tablecaption{Comparison with Atmospheric Parameters From the Literature
\label{table_teff_comp} } 
\tablehead{
\colhead{} & \multicolumn{2}{c}{This work} & \colhead{} & 
\multicolumn{3}{c}{Literature} \\
\cline{2-3} \cline{5-7} \\
\colhead{ID} & \colhead{\teff} & \colhead{log $g$}  & \colhead{} &
\colhead{\teff} & \colhead{log $g$} & \colhead{Reference} 
}
\startdata
\multicolumn{7}{c}{\bf M 3} \\
VZ 1397 & 3985   &  0.40  & & 3950 &  0.40  & Sneden et al. (2004), photometric \\
        &        &        & & 3925 &  0.10  & Sneden et al. (2004), spectroscopic \\
        &        &        & & 3950 &  0.40  & Kraft et al. (1992) \\
        &        &        & & 4000 &  0.6   & Cohen (1978) \\
II-46   & 3998   &  0.40  & & 4050 &  0.40  & Sneden et al. (2004) \\
        &        &        & & 4000 &  0.60  & Kraft et al. (1992) \\
        &        &        & & 4000 &  0.6   & Cohen (1978) \\
\enddata
\end{deluxetable}

\clearpage

\begin{deluxetable}{l ll cc r}     
\tablenum{12}           
\tablewidth{0pt}                                                                
\tabletypesize
\small
\tablecaption{Comparison of Deduced Abundance Ratios
with Sneden et al. (2004)\label{table_compare}}    
\tablehead{ 
\colhead{Species} & \multicolumn{2}{c}{This work}
& \multicolumn{2}{c}{Sneden et al. (2004)} & \colhead{Us -- Sneden} \\
\colhead{} & \colhead{[X/Fe]} &
\colhead{$\sigma_{N}$\tablenotemark{a}} & 
\colhead{[X/Fe]} &
\colhead{$\sigma_{N}$\tablenotemark{a}} & 
\colhead{$\Delta$([X/Fe])} \\
\colhead{} & \colhead{(dex)} & \colhead{(dex)} &
\colhead{(dex)} & \colhead{(dex)} & \colhead{(dex)}
}
\startdata
\multicolumn{6}{c}{\bf M 3} \\
OI   &    +0.33 & 0.05 &     +0.15 & 0.03 &  +0.18 \\ 
NaI  &   $-$0.16 & 0.06 &    +0.01 & 0.04 &  $-0.17$ \\ 
MgI  &    +0.40 & 0.03 &     +0.22 & 0.03 &  +0.18 \\
SiI  &    +0.27 & 0.02 &     +0.30 & 0.01 &  $-0.03$ \\
CaI  &    +0.10 & 0.01 &     +0.23 & 0.01 &  $-0.13$ \\
ScII &    +0.13 & 0.01 &     $-$0.11 & 0.02 &  +0.24 \\
TiI  &    +0.17 & 0.02 &     +0.21 & 0.01 &  $-0.04$\\
\enddata
\tablenotetext{a}{$\sigma_{N}$ is the uncertainty in the mean value
of [X/Fe] for the cluster, i.e. $\sigma/\sqrt{N_{stars}}$}
\tablenotetext{b}{These values are [Fe/H].}
\end{deluxetable}

%
%
\begin{deluxetable}{lrrrrrrrrrrrrrrrrr}     
\tabletypesize
\small
\tablenum{13a}           
\tablewidth{0pt}                                                                
\tablecaption{Star by Star Comparison of Abundances for M3\tablenotemark{a}
\label{table_ab_comp_m3} } 
\tablehead{
\colhead{Species}  & \colhead{VZ 1397} & \colhead{II-46} & \colhead{VZ1000} &
\colhead{$\sigma$}
}
\startdata
{[O/Fe]}   &   0.21 & $-$0.13 &  0.02  &    0.17 \\
{[Na/Fe]}  &   0.09 &   0.14  &  0.17  &   0.04 \\
{[Mg/Fe]}  &   0.09 & \nodata  & \nodata &   \nodata \\
{[Si/Fe]}  &   0.03 &   0.10  &  0.09  &   0.04 \\
{[Ca/Fe]}  & $-$0.04 & $-$0.04 &  0.01  &   0.03 \\
{[Sc/Fe]}  & $-$0.12 &   0.09  &  0.19 &   0.16 \\
{[Ti/Fe]}  &    0.03 & $-$0.09 &  0.05 &   0.08 \\
{[V/Fe]}   &    0.15 & $-$0.10 & $-$0.04 &  0.13 \\
{[Mn/Fe]}  &    0.09 &   0.10 & $-$0.05 &  0.08 \\
{[Fe/H]}  &    $-0.06$ &  -0.16 & $-$0.08 &  0.05 \\
{[Ni/Fe]}  &    0.08 &   0.12 &   0.00 &   0.06 \\
{[Ba/Fe]}  &    0.21 &   0.24  &  0.15 &   0.05 \\
{[La/Fe]}  &    0.16 & \nodata  & \nodata &  \nodata \\
{[Eu/Fe]}  &    0.07 &   0.24 &  $-$0.20 & 0.22 \\
\enddata
\tablenotetext{a}{[X/Fe](this work) -- [X/Fe](Sneden \etal\ 2004) -- (mean $\Delta$ 
from Table~\ref{table_compare}).} 
\end{deluxetable}

%
%
\begin{deluxetable}{lrrrrrrrrrrrrrrrrr}     
\tabletypesize
\small
\tablenum{13b}           
\tablewidth{0pt}                                                                
\tablecaption{Star by Star Comparison of Abundances for M13\tablenotemark{a}
\label{table_ab_comp_m13} } 
\tablehead{
\colhead{Species}  & \colhead{II-67} & \colhead{IV-25} & \colhead{II-76} &
\colhead{III-18} & \colhead{K188} & \colhead{$\sigma$}
}
\startdata
{[O/Fe]}   &   0.17 &  0.44  &  $-$0.06  & $-$0.40  & $-$0.28 & 0.34 \\
{[Na/Fe]}  &  $-0.03$ &  $-0.04$  &  0.02  &  $-0.01$  & $-0.02$ & 0.02 \\
{[Mg/Fe]}  &   0.06 &  0.06  &  \nodata &  \nodata  & $-0.16$ & 0.13 \\
{[Al/Fe]}  &  $-0.51$ &  $-0.35$  & \nodata &  \nodata  & \nodata & 0.11 \\
{[Fe/H]}   &   0.05 &   0.04  & $-0.04$ &  $-0.04$  & 0.14  & 0.07 \\
{[Ba/Fe]}  &    0.29 &  0.14  & \nodata &  \nodata  & 0.14 &  0.09 \\
{[La/Fe]}  &    0.20 &  0.30  &  \nodata &  \nodata  & $-0.32$ & 0.33 \\
{[Eu/Fe]}  &    0.09 &  0.17  &  \nodata &  \nodata  & $-0.10$ & 0.14 \\
\enddata
\tablenotetext{a}{[X/Fe](this work) -- [X/Fe](Sneden \etal\ 2004) -- (mean $\Delta$  
from Table~\ref{table_compare}).} 
\end{deluxetable}

\clearpage

%
%

\clearpage

\end{document}